\documentclass[  reprint,  amsmath,amssymb,  aps, pra, floatfix]{revtex4-2}
\usepackage{graphicx}
\usepackage{dcolumn}
\usepackage{bm}
\usepackage[utf8]{inputenc}
\usepackage[T1]{fontenc}
\usepackage{physics}
\usepackage{amsmath}
\usepackage{amsfonts}
\usepackage{amssymb}
\usepackage{mathrsfs}
\usepackage{placeins}
\usepackage{float}
\makeatletter
\let\newfloat\newfloat@ltx
\makeatother
\usepackage{algorithm}
\usepackage{algpseudocode}
\usepackage{color}
\usepackage{xcolor}
\usepackage[most]{tcolorbox}
\usepackage{tikz}
\usepackage{relsize}
\usepackage[normalem]{ulem}
\usepackage{parskip}
\usepackage{xr}
\usepackage{lineno}
\usepackage[utf8]{inputenc}
\usepackage{tikz}
\usepackage{amsmath}
\usepackage{amsfonts}
\usepackage{subcaption}
\usepackage{graphicx}

\usetikzlibrary{shapes,arrows.meta,positioning,calc,fit,backgrounds}

\tikzset{
  block/.style = {rectangle, draw, fill=blue!10, text centered, rounded corners, minimum height=3em, text width=8em},
  large_panel/.style = {rectangle, draw, fill=gray!10, rounded corners, inner sep=25pt}, 
  highlight_block/.style = {rectangle, draw=orange, fill=orange!10, rounded corners, inner sep=10pt, align=center},
  step_block/.style = {rectangle, draw, fill=green!10, minimum size=2em},
  inner_block/.style = {rectangle, draw, fill=white, text width=11em, align=center, rounded corners, minimum height=2.5em},
  math_block/.style = {inner_block, fill=yellow!10, text width=16em},
  line/.style = {draw, -Latex, thick},
  dotted_line/.style = {draw, -Latex, thick, dashed}
}

\begin{document}

\title{
How well can Diffusion Models learn Lagrangian-Tracer Statistics in Non-reciprocal Turbulence?}

\author{Pratyush Jha\textsuperscript{\#}}
\email{pratyushjha@iisc.ac.in}

\author{Biswajit Maji\textsuperscript{\#}}
\email{biswajitmaji@iisc.ac.in}

\author{Rahul Pandit}
\email{rahul@iisc.ac.in}

\affiliation{Centre for Condensed Matter Theory, Department of Physics, Indian Institute of Science, Bangalore 560012, India}
\thanks{\textsuperscript{\#}These authors contributed equally to this work.}
 \begin{abstract}
    Recent advances in generative artificial intelligence have led to significant potential applications in conventional fluid flows, including those that are turbulent. Can these methods be carried over to studies of novel types of turbulence, such as turbulence induced by non-reciprocity in binary-fluid systems? To answer this question, we analyze the statistics of Lagrangian-tracer particles in non-reciprocal binary-fluid turbulence, which has been  studied recently in the non-reciprocal Cahn-Hilliard-Navier-Stokes (NRCHNS). We obtain our \textit{ground-truth} data via extensive pseudospectral direct numerical simulations (DNSs) of the two-dimensionsl (2D) NRCHNS model. Our study yields a variety of intriguing results for probability distribution functions (PDFs) for particle accelerations and velocity-component PDFs; the latter turn out to be bimodal, \textit{completely unlike their 2D-fluid-turbulence counterparts}. We relate this bimodality to lane-type structures in Eulerian-velocity components. Furthermore, we characterize Lagrangian multiscaling  via Lagrangian velocity increments, their structure functions and flatnesses, and multiscaling exponent ratios, for the first time in non-reciprocal hydrodynamics. Finally, we use generative diffusion models to obtain synthetic Lagrangian trajectories for the NRCHNS system, assess how effectively they can emulate the Lagrangian statistics that we obtain from our DNSs, and highlight open challenges in the application of generative artificial intelligence in non-reciprocal systems.

\end{abstract}

\date{\today}%

\maketitle


\section{Introduction} 
\label{sec:intro}

Turbulence continues to pose grand-challenge problems for scientists and engineers~\cite{frisch1985singularity,meneveau1991multifractal,frisch1995turbulence,pope2001turbulent,falkovich2006lessons,boffetta2008twenty,pandit2009statistical,boffetta2012two,pandit2017overview,benzi2023lectures,sreenivasan2025turbulence}, computational physicists~\cite{canuto2007spectral,kim2024early,yeung2025small}, and specialists in artificial intelligence (AI)~\cite{beck2021perspective,hadi2024machine,sahibzada2025ai,fang2026breakthroughs}. In particular, the application of generative-artificial intelligence (GenAI) to challenging problems in science, in general, and the physics of turbulent flows, in particular, continues to move apace. One fertile field for such applications has been the development of stochastic models, which yield synthetic turbulence, and are capable, \textit{inter alia}, of obtaining trajectories of Lagrangian-tracer particles and their statistical properties (see, e.g., Refs.~\cite{benzi1993random,arneodo1998random,bacry2001multifractal,bacry2003log,wilczek2016non,verma2020first}). Recent advances in studies of such Lagrangian statistics have adopted  Generative Diffusion Models, whose origins lie in work on Denoising Diffusion Probabilistic Models (DDPMs)~\cite{ho2020denoising,dickstein2015deep}. Specifically, DDPMs have been used to obtain tracer trajectories and their statistical properties in three-dimensional (3D), statistically homogeneous and isotropic turbulence in the incompressible Navier-Stokes (NS) equations~\cite{Biferale_DM_2024,li2024generative}; similar methods have been extended to investigate the statistics of bubbly flows~\cite{narula2026gappy}.

Meanwhile, there has been a renaissance in studies of systems in which  effective non-reciprocity of interactions occurs in a large variety of nonequilibrium systems~\cite{sompolinsky1986temporal,hong2011kuramoto,montbrio2018kuramoto,menzel2013traveling,ivlev2015statistical,kryuchkov2018dissipative,you2020nonreciprocity,gompper20202020,saha2020scalar,hosaka2021nonreciprocal,yasuda2021nonreciprocality,bowick2022symmetry,gupta2022active,poncet2022soft,clerk2022introduction,dinelli2023non,frohoff2023non,frohoff2023nonreciprocal,ryu2023dynamics,tucci2024nonreciprocal,rana2024defect,markovich2024nonreciprocity,mandal2024molecular,reisenbauer2024non,guislain2024collective,al2025non,garces2025phase,avni2025nonreciprocal,pisegna2025can,weis2025generalized,te2025metareview,sahoo2025nonreciprocal}. The non-reciprocal research frontier is extensive; and it is advancing rapidly. Non-reciprocity has been shown to have important consequences in, e.g., neuronal networks and excitation-inhibition models~\cite{sompolinsky1986temporal,hong2011kuramoto,montbrio2018kuramoto}, non-Hermitian models~\cite{miri2019exceptional,clerk2022introduction,guislain2024collective}, active-matter systems~\cite{gompper20202020,poncet2022soft,gupta2022active,bowick2022symmetry,dinelli2023non,te2023microscopic,markovich2024nonreciprocity,te2025metareview,al2025non}, phase transitions~\cite{kryuchkov2018dissipative,fruchart2021non}, and biological~\cite{halatek2018self} and chemical~\cite{soto2014self,nasouri2020exact,yasuda2021nonreciprocality,mandal2024molecular,liu2024self} systems. Recently, we have uncovered the intriguing statistical properties of \textit{non-reciprocal turbulence}, in the Eulerian framework, using a minimal, two-dimensional (2D) binary-fluid model, in which unequal off-diagonal elements of the diffusion tensor lead to non-reciprocity~\cite{maji2026non}; this is the Non-reciprocal Cahn-Hilliard-Navier-Stokes (NRCHNS) model.

We now pose two important questions: What are the statistical properties of Lagrangian tracers that are advected by non-reciprocal turbulence? And can these properties be replicated by a suitably trained diffusion model [cf. Refs.~\cite{Biferale_DM_2024,li2024generative} for 3D fluid turbulence]? Our findings are based on a synthesis of two rapidly growing frontiers in computational science and physics, namely, generative AI and non-reciprocal systems. 

To address the first question, we use the Non-Reciprocal Cahn-Hilliard-Navier-Stokes (NRCHNS) model~\cite{maji2026non} to demonstrate, that, in the nonequilibrium statistically steady state (NESS) of 2D non-reciprocal or NRCHNS turbulence, the statistical properties of Lagrangian tracers
are similar to, \textit{but distinctly different from,} their counterparts in 2D fluid turbulence. We answer the second question by showing that diffusion models, which have been used recently for Lagrangian-tracer statistics in 3D fluid turbulence~\cite{Biferale_DM_2024,li2024generative}, can be employed, \textit{mutatis mutandis}, for 2D NRCHNS turbulence, too. Before we present the details of our studies, we give a qualitative overview of our principal results.

 
We begin with an investigation, based on extensive direct numerical simulations (DNSs), of the statistical properties of Lagrangian-tracer particles in non-reciprocal binary-fluid flows, in the NRCHNS model. Our study brings out several remarkable results that have not been anticipated hitherto. We find that the probability distribution functions (PDFs) for parallel and perpendicular components of particle accelerations are qualitatively similar to their fluid-turbulence counterparts~\cite{Biferale_DM_2024}: they have near-Gaussian central peaks, but these give way to fat tails. The velocity-component PDFs and JPDFs  are bimodal, \textit{completely unlike their 2D-fluid-turbulence counterparts}~\cite{perlekar2009statistically,pandit2017overview}; this bimodality is related to lane-type structures~\cite{reeves2021emergence} in pseudocolor plots of the Eulerian-velocity components for the NRCHNS system. Next, we characterize multiscaling in the Lagrangian framework via Lagrangian velocity increments, their structure functions and flatnesses, and multiscaling exponent ratios. Although such Lagrangian studies have been conducted for conventional fluid turbulence and bacterial turbulence~\cite{biferale2004multifractal,arneodo2008universal,benzi2010inertial,kiran2025onset,pandit2025particles}, they have not been explored, heretofore, in non-reciprocal hydrodynamics. Finally, we present the first study of irreversibility in this non-reciprocal system by computing the asymmetry of PDFs of Lagrangian energy increments $W(\tau)$ and the Lagrangian power $p_L(t)$ [cf. Refs.~\cite{xu2014flight,bhatnagar2018heavy,kiran2023irreversibility,shukla2023inertial} for fluid, bacterial, and superfluid turbulence~\cite{shukla2023inertial}]. 

We then use generative diffusion models~\cite{li2024generative} to obtain synthetic Lagrangian trajectories for the NRCHNS system and assess, thereby, how effectively they can emulate the Lagrangian statistics that we obtain from our DNSs. To the best of our knowledge, this is the first time that such generative AI models have been used for a non-reciprocal system~\footnote{When evaluating the efficacy of these models, it is critical to distinguish the physical time-evolution of the system from the data generation process in diffusion models. Sampling synthetic data from diffusion models occurs via a predefined stochastic differential equation, rather than by autoregressively integrating the exact physical trajectories. Consequently, the generative process is not hindered by the fundamentally non-conservative and non-reciprocal dynamics of the NRCHNS equations.}. 
In particular, we demonstrate that two models (called DDPM-800 and DDIM-100 below) yield synthetic Lagrangian trajectories whose statistical properties are very close to those that we obtain from our DNSs. We provide compelling evidence for this agreement via extensive benchmarking against our DNS results for acceleration, velocity-component, and curvature PDFs, and for Lagrangian velocity increments, their structure functions and flatnesses, and multiscaling exponent ratios, and, finally, signatures of irreversibility. We also find that these diffusion models exhibit slight inaccuracies for low non-reciprocity; this indicates a potential performance gap in learning highly subtle flow features. Last, but not least, we explore a type of transfer learning, which allows us to refine models, trained at one value of the non-reciprocity, so that they can be employed at another value of the non-reciprocity. 

The remainder of this paper is organized as follows: In Sec.~\ref{sec:MandM} we describe the models we use and the methods that we employ to study them. Section~\ref{sec:results} presents our statistical analysis of Lagrangian-tracer trajectories and their diffusion-model counterparts. Concluding remarks and a discussion of our results are given in Sec.~\ref{sec:Discussions}.

\section{Models and Methods}
\label{sec:MandM}
We begin in Sec.~\ref{subsec:NRT} with an overview of the non-reciprocal Cahn-Hilliard-Navier-Stokes (NRCHNS) model that has been used recently~\cite{maji2026non} to study non-reciprocal binary-fluid turbulence. We obtain Lagrangian-tracer trajectories as described in Sec.~\ref{subsec:lagrangian}.
Initial conditions are given in Sec.~\ref{subsec:init}. We give a brief description of our DNS of the NRCHNS model in Sec.~\ref{subsec:DNS}. The Diffusion Models that we use are described in detail in Sec.~\ref{subsec:DMOver}. 

\subsection{Non-reciprocal Turbulence} 
\label{subsec:NRT}


To prepare the ground for our work, we recall that the spatiotemporal evolution of binary mixtures is governed by the Cahn-Hilliard (CH) partial differentail equations (PDEs) [see, e.g., Refs.~\cite{cahn1958free,cahn1959free,cahn1961spinodal,bray2002theory,puri2009kinetics}]. These PDEs use  a scalar order-parameter field, say $\varphi$, to distinguish between $A$-rich and $B$-rich components, which are the two coexisting phases; $\varphi$ is positive (negative) in $A$-rich ($B$-rich) regions. To model two-component \textit{fluid} mixtures, we must couple the CH and Navier-Stokes (NS) PDEs~\cite{hohenberg1977theory,puri2009kinetics,padhan2025cahn}. For \textit{active fluids}, a rapidly growing area~\cite{alert2022active,pandit2025particles}, we must include active-fluid terms [see, e.g., Refs.~\cite{padhan2023activity,padhan2024novel,padhan2025cahn}]. Here, we consider  non-reciprocity-induced activity, for which we turn to the 2D NRCHNS PDEs that have been used, principally, at low Reynolds numbers~\cite{menzel2013traveling,you2020nonreciprocity,saha2020scalar,frohoff2023non,frohoff2023nonreciprocal,brauns2024nonreciprocal,blom2026dynamic}. A recent study~\cite{maji2026non} has uncovered binary-fluid turbulence in the following 2D NRCHNS system, which we employ here:
\begin{eqnarray}
   \pdv{\phi}{t} + \bm{u}\cdot\nabla\phi & = & M_1\nabla^2[ - \frac{3}{2}\sigma_1\epsilon_1\nabla^2\phi \nonumber \\
    &+& \frac{3}{4}\frac{\sigma_1}{\epsilon_1}(-\phi + \phi^3) + D_{12}\psi] \,;\label{eq:phi} \\
\pdv{\psi}{t} + \bm{u}\cdot\nabla\psi &=& M_2\nabla^2[- \frac{3}{2}\sigma_2\epsilon_2\nabla^2\psi \nonumber \\
&+& \frac{3}{4}\frac{\sigma_2}{\epsilon_2}(-\psi + \psi^3) + D_{21}\phi] \,;\label{eq:psi} \\
\pdv{\omega}{t} + \bm{u}\cdot\nabla\omega &=& \nu\nabla^2\omega  - \frac{3}{2}\sigma_1\epsilon_1\curl{(\nabla^2\phi\nabla\phi)}\nonumber \\  
&-& \frac{3}{2}\sigma_2\epsilon_2\curl{(\nabla^2\psi\nabla\psi)}   -  
 \alpha\omega\,;\label{eq:omega}\\
\nabla \cdot \bm u &=& 0\,.\label{eq:incom} 
\end{eqnarray}
This system of PDEs comprises two CH PDEs for the interacting scalar fields, $\phi$ and $\psi$, which are also coupled to an incompressible 2D NS fluid with bottom friction $\alpha$, kinematic viscosity $\nu$, velocity $\bm u$, and the vorticity $\bm \omega \equiv (\nabla \times \bm u) = \omega \hat{z}$, which points out of the $xy$ plane in the $\hat{z}$ direction;  $\sigma_1$ and $\sigma_2$ are surface-tension coefficients; and $\epsilon_1$ and $\epsilon_2$ are the widths for the $\phi$ and $\psi$ interfaces, respectively~\cite{padhan2025cahn,maji2026non}. For simplicity we set $\alpha = 0$, $\sigma_1=\sigma_2$,  $\epsilon_1=\epsilon_2$,
and the mobilities $M_1=M_2$. We identify the diffusivities for the scalar fields $\phi$ and $\psi$ as $D_{11}\equiv -M_1\frac{3}{4}\frac{\sigma_1}{\epsilon_1}$ and $D_{22}\equiv -M_2\frac{3}{4}\frac{\sigma_2}{\epsilon_2}$, respectively [see, e.g., Refs.~\cite{saha2020scalar,frohoff2023non,frohoff2023nonreciprocal,brauns2024nonreciprocal,maji2026non}]; their negative signs lead, in their corresponding CH models, to the well-known, anti-diffusive spinodal instability~\cite{puri2009kinetics,padhan2025cahn},
which leads to the early stages of phase separation. For non-reciprocity we must have $D_{21}= -D_{12}$ for off-diagonal elements of the diffusion tensor; we take $D_{21}= -D_{12}$~\footnote{Reference~\cite{brauns2024nonreciprocal} calls this the \textit{anti-reciprocal limit}.}; this suffices to obtain 2D NRCHNS turbulence~\cite{maji2026non}, where the non-dimensional non-reciprocity parameter is $\mathcal{D}_{12}\equiv |D_{12}|/|D_{11}|$. It is worth recalling that, for a non-reciprocal CH model, the terms with the coefficients $D_{12}$ and $D_{21}$, which are responsible for the non-reciprocal nature of this model, do not follow from a functional derivative of a free-energy functional like $\mathcal{F}$ [see, e.g., Refs.~\cite{saha2020scalar,frohoff2023non,frohoff2023nonreciprocal,brauns2024nonreciprocal}].

 
 A recent study~\cite{maji2026non} has used pseudospectral direct numerical simulations (DNSs) to obtain the NESSs of the 2D NRCHNS model~\eqref{eq:phi}-\eqref{eq:incom} and their statistical properties in the Eulerian framework for different values of parameters of $\nu$ and 
 $\mathcal{D}_{12}$. In brief, at low Reynolds number $Re$, this model displays travelling waves~\cite{saha2020scalar,frohoff2023non,frohoff2023nonreciprocal,brauns2024nonreciprocal,maji2026non}; as we decrease $\nu$ or increase $\mathcal{D}_{12}$, 2D NRCHNS turbulence sets in.
 This turbulence displays an inverse cascade of energy with fluid energy spectrum $E(k)\sim k^{-5/3}$, where $k$ is the wavenumber; this is reminiscent of conventional, forced, 2D fluid turbulence~\cite{boffetta2012two,pandit2017overview}; but there are important differences between 2D fluid and NRCHNS turbulence; in particular, the latter exhibits the non-reciprocal flux~\cite{rana2024defect,maji2026non}
\begin{eqnarray}
\bm{J} \equiv \langle \phi \nabla \psi - \psi \nabla \phi \rangle_s\,,
\label{eq:fluxJ}
\end{eqnarray}
where the subscript $s$ indicates the spatial average. The modulus $J(t)\equiv |\bm{J}|$ achieves a steady, finite value at low $Re$; as $Re$ increases, so does the intensity of 2D NRCHNS turbulence, $J(t)$ fluctuates about a mean value ($  > 0$); this mean value 
decreases with increasing $Re$. Details of the statistical properties of 2D NRCHNS turbulence, in the Eulerian framework, are given in Ref.~\cite{maji2026non}.

\subsection{Trajectories of Lagrangian tracers}
\label{subsec:lagrangian}

We now examine 2D NRCHNS turbulence in the Lagrangian framework, for which purpose we introduce $N_L$ Lagrangian tracer particles into our 2D NRCHNS system, after it has reached a NESS. The position $\bm r(t)$ at time $t$ of a given tracer, which has the position $\bm r_0$ at time $t_0$, is
\begin{eqnarray}
    \frac{d\bm r(t)}{dt} = \bm{v}(\bm r, t|\bm r_0, t_0) = \bm u(\bm r, t) \;, \label{eq:tracer}
\end{eqnarray}
where the Lagrangian velocity $\bm v$ follows from the Eulerian velocity field $\bm u$~\cite{benzi2010inertial,verma2020first,bihari2025interfaces}. 

\subsection{Initial conditions}
\label{subsec:init}

To obtain the NESSs for the NRCHNS model~\eqref{eq:phi}-\eqref{eq:incom}, in the Eulerian framework, we follow Ref.~\cite{maji2026non} and use an initial condition in which $\omega(x, y, t = 0) = 0$ and $\phi(x, y, t = 0)$ and
$\psi(x, y, t = 0)$ are zero-mean,  random numbers, independent and identically distributed, uniformly on the interval $[-0.1, 0.1]$, at each spatial point $(x,y)$. Once the NESS has been established, we introduce $N_L$ Lagrangian tracer particles
at random locations in our simulation domain.

\subsection{Direct Numerical Simulations}
\label{subsec:DNS}

We employ the Fourier pseudospectral DNS of Ref.~\cite{maji2026non} to solve Eqs.~\eqref{eq:phi}-\eqref{eq:incom}; this is appropriate for a square simulation domain, with side 
$2\pi$, doubly periodic boundary conditions, $\mathfrak{N}^2$ collocation points, and the $1/2$ rule for dealiasing [see, e.g., Refs.~\cite{canuto2007spectral,padhan2025cahn,maji2026non}]. 
Note that spatial derivatives can be  evaluated simply in Fourier space; by contrast, products are evaluated conveniently in physical space. The semi-implicit ETDRK-2 method~\cite{cox2002exponential} works well
for the time integration. We use a CUDA C code on an A100 Nvidia GPU. The data  we present here use:
$\mathfrak{N} = 1024;\, M_1 = M_2 = 0.0001;\, \epsilon_1 = \epsilon_2 = 0.01839;\,
\sigma_1 = \sigma_2 = 3;$ and $\alpha = 0$; 
non-dimensional parameters are listed in Sec.~\ref{subsec:appendix:non_dim_parameters} in the Appendix.

Note that Lagrangian particles can move off the Eulerian grid (in any numerical scheme); we use bilinear interpolation at off-grid points and a first-order Euler scheme for the integration of Eq.~(\ref{eq:tracer}) [see, e.g., Refs.~\cite{benzi2010inertial,verma2020first,bihari2025interfaces}], with equal time steps ($\delta_L t=10^{-4})$.


\FloatBarrier

\subsection{Diffusion Model: Task Definition and Overview}
\label{subsec:DMOver}
Diffusion models (DMs) are a class of generative models that learn data distributions by reversing a 
step-by-step noising process~\cite{dickstein2015deep, ho2020denoising} in
two stages: a forward process that systematically corrupts the data with Gaussian noise until it reaches an isotropic noise distribution; and then a learnable reverse process (called \textit{denoising}) that progressively removes this noise to reconstruct the original data distribution~\cite{dickstein2015deep, ho2020denoising}. 

We train three distinct diffusion models corresponding to the datasets $\{\mathfrak{D}_i\mid i\in\{1, 2, 3\}\}$ (with $\mathcal{D}_{12}= 30,\, 40,\,$ and $45$ for $i=1,\,2,$ and $3$, respectively). Each one of these datasets consists of $\lvert\mathfrak{D}_i \rvert = N_L$ Lagrangian-particle trajectories, generated via our DNS [Sec.~\ref{subsec:DNS}], after the NESS is achieved (we choose $N_L = 131,072$). We now train models $\mathfrak{M}_i$, with trainable parameter sets $\theta_i$, to obtain synthetic Lagrangian trajectories (with the appropriate 
value of $\mathcal{D}_{12}$) and benchmark their statistical properties against their DNS counterparts.

Formally, for the $j^{\rm th}$ Lagrangian trajectory, belonging to dataset $\mathfrak{D}_i$, we define the following time-ordered set of 2D velocity vectors at times $t_l = l \delta t$, with the integers $l \in \{0, \ldots,(l_{max}-1)\}$ [we choose $l_{max}=2000$]:
\begin{equation}
\mathcal{V}^{(i,j)}_0=\left\{\mathbf{V}^{(i,j)}(t_l)\mid  l\in\{0, \dots, (l_{max}-1)\} \right\}\,,
\end{equation}
where $\mathbf{V}^{(i,j)}(t_l)=\left(\bm{v}_x^{(i,j)}(t_l), \bm{v}_y^{(i,j)}(t_l)\right)$, the subscripts $x$ and $y$ denote the Cartesian components of the 
Lagrangian velocity $\bm{v}$ [Eq.~\eqref{eq:tracer}], and $j\in\{1, \dots, \lvert\mathfrak{D}_i \rvert\}$.

\newsavebox{\ddpmbox}
\sbox{\ddpmbox}{%
  \begin{minipage}{23em} 
    \centering
    {\LARGE \textbf{ONE DENOISING\\[0.1em] OPERATION}}\\[0.4em]
    {\Large \textbf{(STEP $n$ to $n-1$)}}\\[0.8em]
    \hrule\vspace{1em}
    \begin{tikzpicture}[
        node distance=0.8cm, auto, every node/.style={font=\Large},
        base/.style={rectangle, rounded corners=8pt, draw=black!50, line width=1.2pt, align=center, inner sep=10pt, fill=white, text width=20em},
        split_node/.style={rectangle, rounded corners=8pt, draw=black!50, line width=1.2pt, align=center, inner sep=10pt, fill=white, text width=8em},
        model/.style={rectangle, rounded corners=12pt, draw=blue!60!black, line width=1.8pt, align=center, inner sep=12pt, fill=blue!5, text width=12em},
        math/.style={rectangle, rounded corners=8pt, draw=purple!60!black, line width=1.2pt, align=center, inner sep=10pt, fill=purple!5, text width=20em},
        arrow/.style={->, >=stealth, thick, draw=black!60, rounded corners=6pt}
    ]
        \node [split_node] (in_x) {\textbf{Input Data}\\[0.2em]$\mathcal{V}_n$};
        \node [split_node, right=1.cm of in_x] (in_t) {\textbf{Timestep}\\[0.2em]$n$};
        
        \path (in_x) -- (in_t) coordinate[midway] (mid_in);
        
        \node [model, below=2.2cm of mid_in] (unet) {\includegraphics[width=1.5cm]{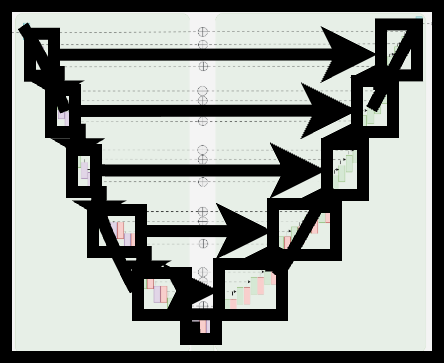}\\[0.3em]\textbf{U-Net}};
        
        \node [base, below=0.8cm of unet] (pred) {\textbf{Predicted Noise} $\epsilon_\Theta$};
        
        \node [math, below=0.8cm of pred] (est) {\textbf{Step 1: Estimate Mean ${\mu}_\Theta$:}\\[0.6em] ${\mu}_\Theta(\mathcal{V}_n,n)=\frac{1}{\sqrt{\alpha_n}}\left(\mathcal{V}_n-\frac{\beta_n}{\sqrt{1-\bar{\alpha}_n}}\epsilon_\Theta(\mathcal{V}_n,n)\right)$ [Eq.~\eqref{eq:muTheta}]};
        
        \node [base, below=0.8cm of est] (rep) {\textbf{Reparameterization}\\[0.2em]Add scaled Gaussian noise to transition to $\mathcal{V}_{n-1}$};
        
        \node [math, below=0.8cm of rep] (calc) {\textbf{Step 2: Calculate $\mathcal{V}_{n-1}$:}\\[0.6em] $\displaystyle \mathcal{V}_{n-1} = {\mu}_\Theta(\mathcal{V}_n,n) + \sqrt{{\beta}_n} z$ [Eq.~\eqref{eq:DDPM_Vn-1_to_Vn}]}; 
        
        \node [base, below=0.8cm of calc] (out) {\textbf{Output Data}\\$\mathcal{V}_{n-1}$};

        \draw [arrow] (in_x.south) |- ([yshift=1.0cm]unet.north) -- (unet.north);
        \draw [arrow] (in_t.south) |- ([yshift=1.0cm]unet.north) -- (unet.north);
        
        \draw [arrow] (unet.south) -- (pred.north);
        \draw [arrow] (pred.south) -- (est.north);
        \draw [arrow] (est.south) -- (rep.north);
        \draw [arrow] (rep.south) -- (calc.north);
        \draw [arrow] (calc.south) -- (out.north);
    \end{tikzpicture}
  \end{minipage}%
}

\newsavebox{\ddimbox}
\sbox{\ddimbox}{%
  \begin{minipage}{23em} 
    \centering
    {\LARGE \textbf{ONE DENOISING\\[0.1em] OPERATION}}\\[0.4em]
    {\Large \textbf{(STEP $n$ to $n-K$)}}\\[0.8em]
    \hrule\vspace{1em}
    \begin{tikzpicture}[
        node distance=0.8cm, auto, every node/.style={font=\Large},
        base/.style={rectangle, rounded corners=8pt, draw=black!50, line width=1.2pt, align=center, inner sep=10pt, fill=white, text width=20em},
        split_node/.style={rectangle, rounded corners=8pt, draw=black!50, line width=1.2pt, align=center, inner sep=10pt, fill=white, text width=8em},
        model/.style={rectangle, rounded corners=12pt, draw=blue!60!black, line width=1.8pt, align=center, inner sep=12pt, fill=blue!5, text width=12em},
        math/.style={rectangle, rounded corners=8pt, draw=purple!60!black, line width=1.2pt, align=center, inner sep=10pt, fill=purple!5, text width=20em},
        arrow/.style={->, >=stealth, thick, draw=black!60, rounded corners=6pt}
    ]
        \node [split_node] (in_x) {\textbf{Input Data}\\[0.2em]$\mathcal{V}_n$};
        \node [split_node, right=1.cm of in_x] (in_t) {\textbf{Timestep}\\[0.2em]$n$};
        
        \path (in_x) -- (in_t) coordinate[midway] (mid_in);
        
        \node [model, below=2.2cm of mid_in] (unet) {\includegraphics[width=1.5cm]{chapter6_figures/unet_icon.png}\\[0.3em]\textbf{U-Net}};
        
        \node [base, below=0.8cm of unet] (pred) {\textbf{Predicted Noise} $\epsilon_\Theta$};
        
        \node [math, below=0.8cm of pred] (est) {\textbf{Step 1: Estimate Clean Data $\hat{\mathcal{V}}_0$:}\\[0.6em] $\displaystyle \hat{\mathcal{V}}_0 = \frac{\mathcal{V}_n - \sqrt{1-\bar{\alpha}_n}\epsilon_\Theta}{\sqrt{\bar{\alpha}_n}}$ [Eq.\eqref{eq:V0DDIM}]};
        
        \node [base, below=0.8cm of est] (rep) {\textbf{Accelerated Jump}\\Bypass intermediate steps; jump directly to $\mathcal{V}_{n-K}$};
        
        \node [math, below=0.8cm of rep] (calc) {\textbf{Step 2: Calculate $\mathcal{V}_{n-K}$:}\\[0.6em] $\displaystyle \mathcal{V}_{n-K} = \sqrt{\bar{\alpha}_{n-K}}\hat{\mathcal{V}}_0 + \sqrt{1-\bar{\alpha}_{n-K}}\epsilon_\Theta$ [Eq.~\eqref{eq:DDIM_Vn_to_Vn-K}]};
        
        \node [base, below=0.8cm of calc] (out) {\textbf{Output Data}\\$\mathcal{V}_{n-K}$};

        \draw [arrow] (in_x.south) |- ([yshift=1.0cm]unet.north) -- (unet.north);
        \draw [arrow] (in_t.south) |- ([yshift=1.0cm]unet.north) -- (unet.north);
        
        \draw [arrow] (unet.south) -- (pred.north);
        \draw [arrow] (pred.south) -- (est.north);
        \draw [arrow] (est.south) -- (rep.north);
        \draw [arrow] (rep.south) -- (calc.north);
        \draw [arrow] (calc.south) -- (out.north);
    \end{tikzpicture}
  \end{minipage}%
}

\begin{figure*}[t] 
  \centering
  
  \tikzset{
    every node/.style={font=\Large},
    base/.style={rectangle, rounded corners=8pt, draw=black!40, thick, align=center, inner sep=10pt, fill=white, text width=20em},
    clean/.style={rectangle, rounded corners=8pt, draw=green!60!black, line width=1.8pt, align=center, inner sep=12pt, fill=green!5, text width=8em},
    noise/.style={rectangle, rounded corners=8pt, draw=red!60!black, line width=1.8pt, align=center, inner sep=12pt, fill=red!5, text width=10em},
    mid/.style={rectangle, rounded corners=8pt, draw=orange!70!black, thick, align=center, inner sep=12pt, fill=orange!5, text width=8em},
    unet/.style={rectangle, rounded corners=12pt, draw=blue!70!black, line width=2pt, align=center, inner sep=16pt, fill=blue!5, text width=14em},
    loss/.style={rectangle, rounded corners=10pt, draw=magenta!80!black, line width=2pt, align=center, inner sep=14pt, fill=magenta!5, text width=22em},
    equation/.style={rectangle, rounded corners=6pt, draw=black!30, fill=white, inner sep=8pt},
    arrow/.style={->, >=stealth, thick, draw=black!70, rounded corners=8pt},
    backarrow/.style={->, >=stealth, dashed, very thick, draw=magenta!90!black, rounded corners=8pt},
    panel_bg/.style={rectangle, draw=black!15, thick, rounded corners=16pt, fill=black!2, inner sep=24pt}
  }

  \begin{subfigure}[t]{0.48\textwidth}
    \vspace{0pt} 
    \centering
    \resizebox{\linewidth}{!}{
    \begin{tikzpicture}
      
      \node (outer_top) at (0, 6) {};
      \node (outer_bot) at (0, -34) {}; 

      \node [font=\huge\bfseries, text=black!85, scale=1.2] (p1_title) at (0, 5) {PART 1: NOISE PREDICTION TRAINING};
      \node [font=\LARGE\bfseries, scale=1.4] (forward_title) at (0, 3.4) {Forward Diffusion Process};
      
      \node [clean] (v0) at (-6, -1) {\textbf{Clean Data}\\$\mathcal{V}_0$};
      \node [mid] (v1) at (-2, -1) {\textbf{Noisy}\\$\mathcal{V}_1$};
      \node [font=\LARGE] (dots) at (0.5, -1) {$\dots$};
      \node [mid] (vn) at (3, -1) {\textbf{Noisy}\\$\mathcal{V}_n$};
      \node [noise, text width=8em] (vN) at (7, -1) {\textbf{Pure Noise}\\$\mathcal{V}_N$};
      
      \draw [arrow] (v0) -- node[above]{} (v1);
      \draw [arrow] (v1) -- (dots);
      \draw [arrow] (dots) -- (vn);
      \draw [arrow] (vn) -- (vN);


      \node (arc_top) at (0, 1.5) {};
      \begin{scope}[on background layer]
        \node [panel_bg, fit=(v0) (vN) (arc_top)] (forward_box) {};
      \end{scope}

      \node [font=\LARGE\bfseries, scale=1.4] (train_title) at (0, -6) {At a randomized training step};

      \node [noise] (sample_eps) at (-4.5, -10) {\textbf{Sample Noise}\\$\epsilon \sim \mathcal{N}(0,I)$};
      \node [clean] (train_v0) at (4.5, -10) {\textbf{Clean Data}\\$\mathcal{V}_0$};
      
      \node [circle, draw=orange!80!black, fill=orange!10, very thick, inner sep=6pt, font=\Large\bfseries] (plus) at (0, -13.5) {$+$};
      
      \draw [arrow] (sample_eps.south) |- (plus.west);
      \draw [arrow] (train_v0.south) |- (plus.east);

      \node [mid, text width=16em] (noisy_in) at (0, -18) {\textbf{Generated Noisy Input}\\$(\mathcal{V}_n, n)$};
      
      \draw [arrow] (plus.south) -- node[right, xshift=0.2cm, fill=white] {\begin{tikzpicture}\node[equation]{$\mathcal{V}_n = \sqrt{\bar{\alpha}_n}\mathcal{V}_0 + \sqrt{1-\bar{\alpha}_n}\epsilon$ [Eq.~\eqref{eq:V0_to_Vn}]};\end{tikzpicture}} (noisy_in.north);

      \node [unet] (unet_node) at (0, -23) {\includegraphics[width=1.5cm]{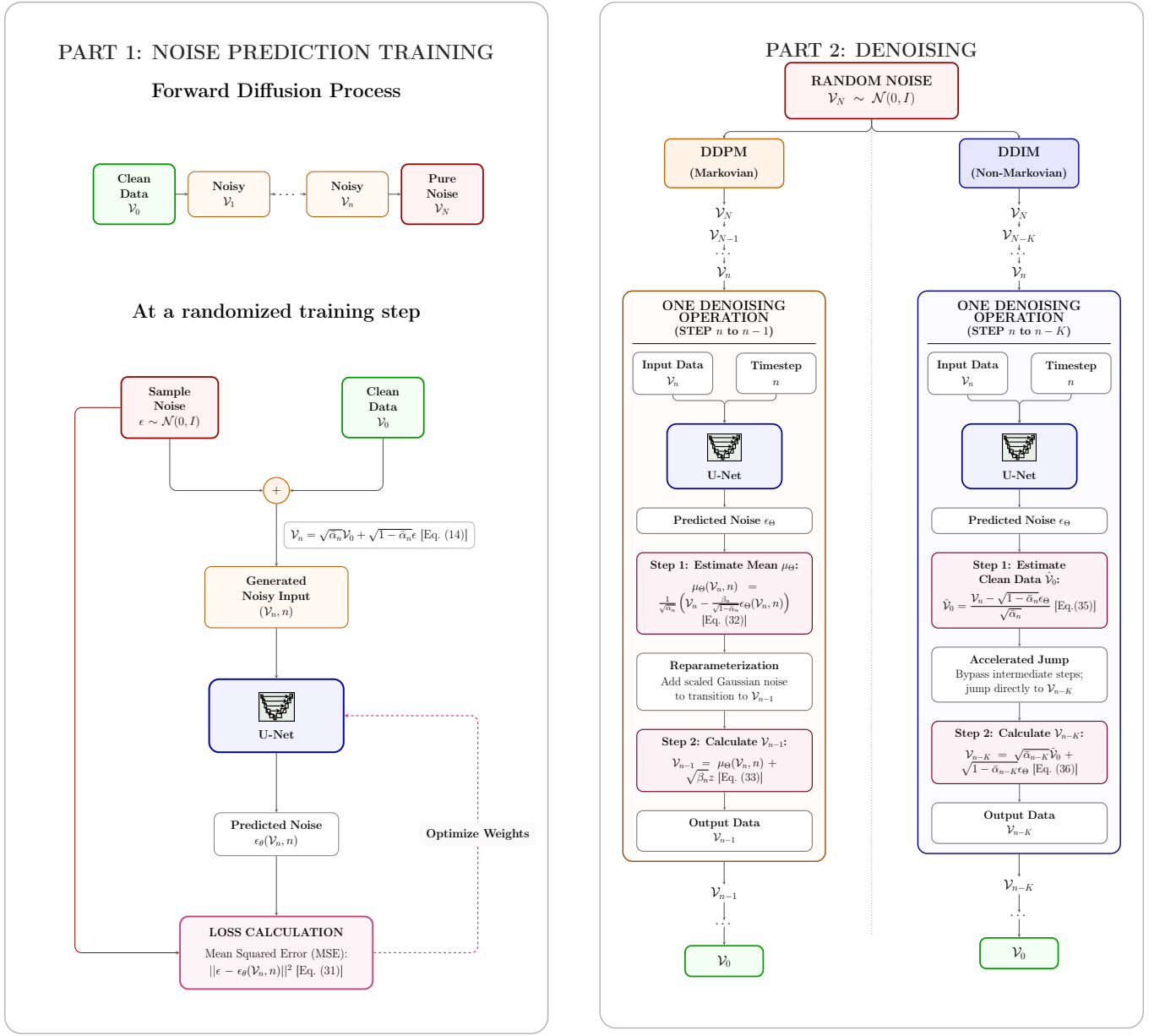}\\[0.2em]\textbf{U-Net}};
      \draw [arrow] (noisy_in.south) -- (unet_node.north);

      \node [base, draw=black!50, text width=14em] (pred_eps) at (0, -28) {\textbf{Predicted Noise}\\$\epsilon_\theta(\mathcal{V}_n, n)$};
      \draw [arrow] (unet_node.south) -- (pred_eps.north);

      \node [loss] (loss_node) at (0, -33) {\textbf{LOSS CALCULATION}\\[0.5em] Mean Squared Error (MSE):\\[0.2em] $||\epsilon - \epsilon_\theta(\mathcal{V}_n, n)||^2$  [Eq.~\eqref{eq:simple_loss}]};
      
      \draw [arrow] (pred_eps.south) -- (loss_node.north);
      
      \draw [arrow, draw=red!60!black] (sample_eps.west) -- (-8.5, -10) -- (-8.5, -33) -- (loss_node.west);

      \draw [backarrow] (loss_node.east) -- (8.5, -33) -- node[midway, fill=black!2, inner sep=8pt, font=\Large\bfseries] {Optimize Weights} (8.5, -23) -- (unet_node.east);

      \node (train_left) at (-9.5, -10) {};
      \node (train_right) at (9.5, -10) {};
      \begin{scope}[on background layer]
        \node [panel_bg, fit=(sample_eps) (train_v0) (loss_node) (train_left) (train_right)] (train_box) {};
      \end{scope}

      \begin{scope}[on background layer]
        \node [rectangle, draw=black!30, line width=1.5pt, rounded corners=20pt, fill=white, inner sep=28pt, fit=(outer_top) (forward_box) (train_box) (outer_bot)] {};
      \end{scope}

    \end{tikzpicture}
    }
  \end{subfigure}
  \hfill
  \begin{subfigure}[t]{0.48\textwidth}
    \vspace{0pt} 
    \centering
    \resizebox{\linewidth}{!}{
    \begin{tikzpicture}
      
      \node (outer_top) at (0, 6) {};
      \node (outer_bot) at (0, -37) {}; 

      \node [font=\huge\bfseries, scale=1.2, text=black!85] (p2_title) at (0, 5) {PART 2: DENOISING};

      \node [noise, text width=20em, font=\LARGE, inner sep=16pt] (r_noise) at (0, 3.2) {\textbf{RANDOM NOISE}\\ $\mathcal{V}_N \sim \mathcal{N}(0,I)$};

      \coordinate (split_pt) at (0, 1.4);

      \begin{scope}[shift={(-6.5, 0)}]
        \node [rectangle, rounded corners=8pt, draw=orange!80!black, line width=1.5pt, fill=orange!10, text width=14em, align=center, inner sep=10pt, font=\LARGE\bfseries] (ddpm_title) at (0, 0) {DDPM\\[0.2em] \Large (Markovian)};
        
        \node [font=\LARGE] (d1) at (0, -2.2) {$\mathcal{V}_N$};
        \node [font=\LARGE] (d2) at (0, -3.2) {$\mathcal{V}_{N-1}$};
        \node [font=\LARGE] (d3) at (0, -4.0) {$\dots$};
        \node [font=\LARGE] (d4) at (0, -4.8) {$\mathcal{V}_n$};
        
        \node [rectangle, draw=orange!60!black, line width=1.5pt, fill=orange!2, rounded corners=12pt, inner sep=12pt, anchor=north] (ddpm_box) at (0, -5.6) {\usebox{\ddpmbox}};
        
        \node [font=\LARGE, below=1.0cm of ddpm_box] (d5) {$\mathcal{V}_{n-1}$};
        \node [font=\LARGE, below=0.8cm of d5] (d6) {$\dots$};
        \node [clean, font=\LARGE, below=0.8cm of d6] (d7) {$\mathcal{V}_0$};

        \draw [arrow] (ddpm_title) -- (d1); \draw [arrow] (d1) -- (d2); \draw [arrow] (d2) -- (d3); \draw [arrow] (d3) -- (d4);
        \draw [arrow] (d4) -- (ddpm_box.north); \draw [arrow] (ddpm_box.south) -- (d5); \draw [arrow] (d5) -- (d6); \draw [arrow] (d6) -- (d7);
      \end{scope}

      \begin{scope}[shift={(6.5, 0)}]
        \node [rectangle, rounded corners=8pt, draw=blue!80!black, line width=1.5pt, fill=blue!10, text width=14em, align=center, inner sep=10pt, font=\LARGE\bfseries] (ddim_title) at (0, 0) {DDIM\\[0.2em] \Large (Non-Markovian)};
        
        \node [font=\LARGE] (i1) at (0, -2.2) {$\mathcal{V}_N$};
        \node [font=\LARGE] (i2) at (0, -3.2) {$\mathcal{V}_{N-K}$};
        \node [font=\LARGE] (i3) at (0, -4.0) {$\dots$};
        \node [font=\LARGE] (i4) at (0, -4.8) {$\mathcal{V}_n$};
        
        \node [rectangle, draw=blue!60!black, line width=1.5pt, fill=blue!2, rounded corners=12pt, inner sep=12pt, anchor=north] (ddim_box) at (0, -5.6) {\usebox{\ddimbox}};
        
        \node [font=\LARGE, below=1.0cm of ddim_box] (i5) {$\mathcal{V}_{n-K}$};
        \node [font=\LARGE, below=0.8cm of i5] (i6) {$\dots$};
        \node [clean, font=\LARGE, below=0.8cm of i6] (i7) {$\mathcal{V}_0$};

        \draw [arrow] (ddim_title) -- (i1); \draw [arrow] (i1) -- (i2); \draw [arrow] (i2) -- (i3); \draw [arrow] (i3) -- (i4);
        \draw [arrow] (i4) -- (ddim_box.north); \draw [arrow] (ddim_box.south) -- (i5); \draw [arrow] (i5) -- (i6); \draw [arrow] (i6) -- (i7);
      \end{scope}

      \draw [arrow] (r_noise.south) -- (split_pt) -| (ddpm_title.north);
      \draw [arrow] (r_noise.south) -- (split_pt) -| (ddim_title.north);

      \draw [dotted, very thick, draw=black!30] (split_pt) -- (0, -34);

      \begin{scope}[on background layer]
        \node [rectangle, draw=black!30, line width=1.5pt, rounded corners=20pt, fill=white, inner sep=28pt, fit=(outer_top) (r_noise) (ddpm_box) (ddim_box) (d7) (i7) (outer_bot)] {};
      \end{scope}

    \end{tikzpicture}
    }
  \end{subfigure}
  \caption{Operational flows in the diffusion models we use. (Left) During training, a clean sample $\mathcal{V}_0$ is corrupted with random noise $\epsilon$ to create a noisy state $\mathcal{V}_n$. A U-Net learns to predict this added noise, optimized via Mean Squared Error (MSE) loss. (Right) During inference, the trained U-Net iteratively denoises a pure noise sample $\mathcal{V}_N$. The standard DDPM uses a Markovian process to step sequentially from $\mathcal{V}_n$ to $\mathcal{V}_{n-1}$, whereas the accelerated DDIM formulation leverages a non-Markovian deterministic jump from $\mathcal{V}_n$ directly to $\mathcal{V}_{n-K}$, significantly reducing the required sampling steps (see text).}
  \label{fig:flow_chart}
\end{figure*}

\begin{figure*}[htp]
    \centering
    \includegraphics[width=1.0\linewidth]{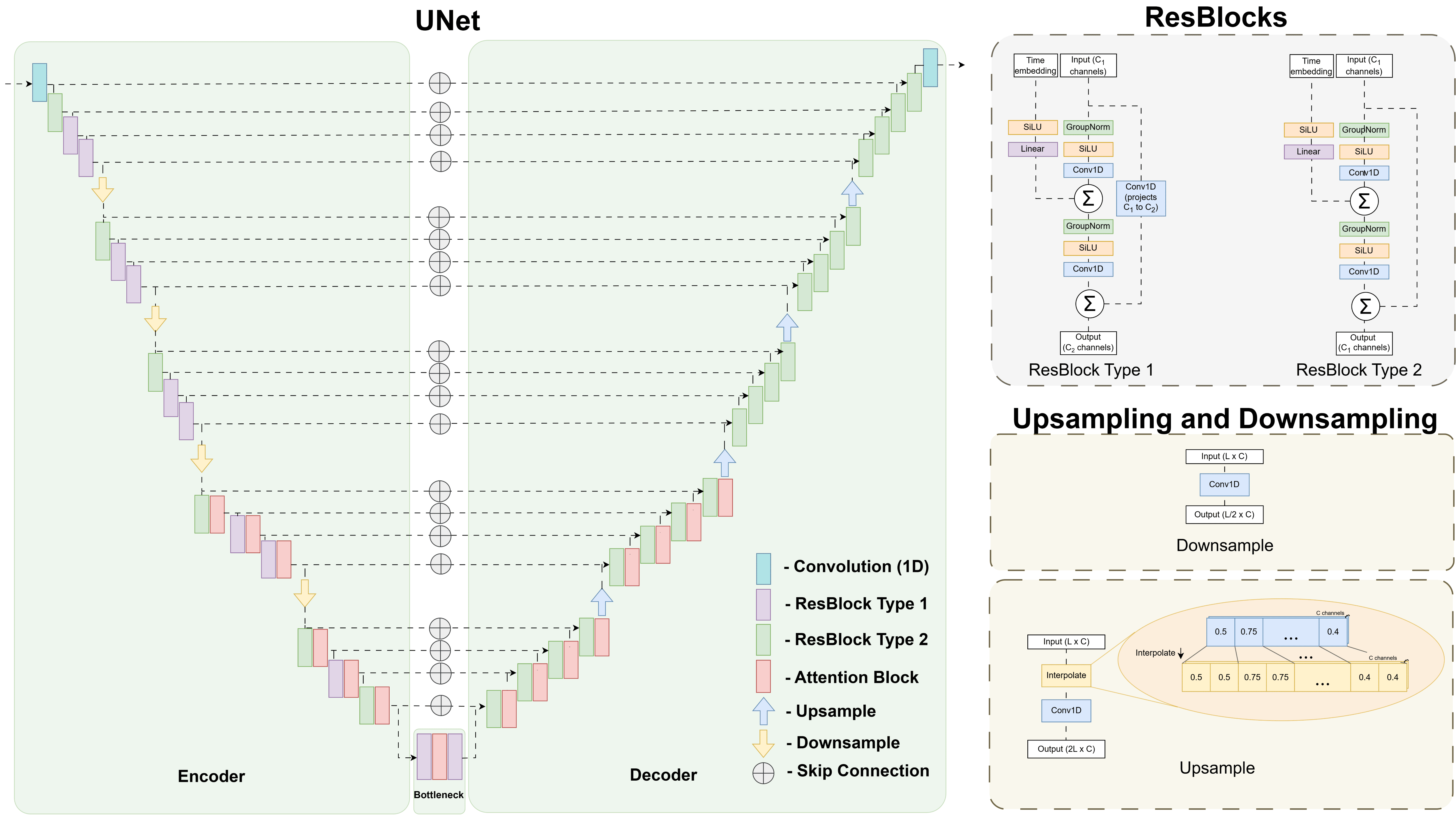}
    \caption{The UNet architecture used to learn the diffusion process, along with the ResBlocks, Upsampling and Downsampling blocks that build the UNet. The UNet architecture is trained as the denoiser to remove noise during the sampling process. ResBlock Type 1 changes the number of channels in the data pass, whereas ResBlock Type 2 maintains the number of channels. \textcircled{$\Sigma$} denotes an element-wise sum, Conv1D refers to a one-dimensional convolution layer, and SiLU (Sigmoid Linear Unit: $\mathrm{SiLU}(x)=x\cdot\frac{1}{1+e^{-x}}$) refers to the activation function.}

    \label{fig:unet}
\end{figure*}

\subsubsection{Diffusion Model: Details}
\label{subsubsec:DMdet}

\textit{Forward diffusion:} Let the ground-truth trajectory data distribution function of the NRCHNS system be $\pi_{data}$, which depends on the data-set index $i$; we suppress this index henceforth for notational convenience. In the diffusion-model framework, $n$ labels discrete time in the $N$-step forward-noising (or forward-diffusion) process; this is a discrete sequence of continuous random variables $\mathcal{V}_{n} \in \mathbb{R}^{l_{max}\times 2}$, with $0\leq n\leq N$. We choose $\mathcal{V}_0$ to be distributed according to  $\pi_{data}$ and denote by $q(\mathcal{V}_n)$ the PDF of $\mathcal{V}_{n}$; formally we write:
\begin{eqnarray}
\mathcal{V}_0 &\sim& q(\mathcal{V}_0) = \pi_{data}\,;\nonumber \\
\mathcal{V}_n &\sim& q(\mathcal{V}_n)\,.
\label{eq:V0}
\end{eqnarray}
The forward-diffusion process adds noise to $\mathcal{V}_0$, over $N$ discrete steps, by introducing Gaussian noise at each discrete time $n$ (and delta-correlated in $n$), thus producing the Markovian sequence $\mathcal{V}_1,\dots,\mathcal{V}_N$. The noise variance $\beta_n\in(0, 1)$, at each step $n$, is a predefined noise schedule, with $\beta_1<\beta_2<\ldots<\beta_N$,  chosen such that, after $N$-steps, $\mathcal{V}_N$ is \textit{pure noise}, i.e.,
\begin{equation}
q(\mathcal{V}_N)=\mathcal{N} (\mathbf{0},\mathbf{I})\,,
\end{equation}
where $\mathcal{N} (\mathbf{0},\mathbf{I})$ denotes a zero-mean normal distribution with the unit covariance matrix $\mathbf{I}$. Furthermore, because we consider a Markov process,
the conditional probability distribution factorizes as follows:
\begin{equation}
q_N(\mathcal{V}_1, \mathcal{V}_2, \dots, \mathcal{V}_N|\mathcal{V}_0)
=\prod_{n=1}^Nq_{(n-1)\to n}(\mathcal{V}_n|\mathcal{V}_{n-1})\,;
\end{equation}
we often employ the shorthand notation $q_N(\mathcal{V}_{1:N}|\mathcal{V}_0)\equiv q_N(\mathcal{V}_1, \mathcal{V}_2, \dots, \mathcal{V}_N|\mathcal{V}_0)$~\footnote{The subscripts on $q$ are often not shown because it is assumed implicitly that they can be surmised from the arguments of $q$.}. Explicitly, the one-step transition probability distribution, from step $n-1$ to $n$, is
\begin{equation}
q_{(n-1)\to n}(\mathcal{V}_n|\mathcal{V}_{n-1})=\mathcal{N}\left(\sqrt{1-\beta_n}\mathcal{V}_{n-1},\beta_n\mathbf{I}\right)\,.
\end{equation}
Therefore, we can obtain
\begin{eqnarray}
\mathcal{V}_n&=&\sqrt{1-\beta_n}\mathcal{V}_{n-1}+\sqrt{\beta_n}z\,,\nonumber \\
{\rm where}\quad z&\sim&\mathcal{N}(\mathbf{0},\mathbf{I})\,.
\label{eq:Vn-1_to_Vn}
\end{eqnarray}
We define $\alpha_n\equiv (1-\beta_n)$ and its cumulative product $\bar{\alpha}_n\equiv\prod_{i=1}^n\alpha_i$. The sum of independent Gaussian-distributed random variables remains Gaussian-distributed. Therefore, we can directly reach an arbitrary step $n$ of the forward process, from the initial state $\mathcal{V}_0$, by using Eq.~\eqref{eq:Vn-1_to_Vn} recursively. The resulting $n$-step transition probability distribution is
\begin{equation}
q_{0\rightarrow n}(\mathcal{V}_n|\mathcal{V}_0)=\mathcal{N}\left(\sqrt{\bar{\alpha}_n}\mathcal{V}_0,(1-\bar{\alpha}_n)\mathbf{I}\right)\,,
\end{equation}
whence we obtain
\begin{eqnarray}
\mathcal{V}_n&=&\sqrt{\bar{\alpha}_n}\mathcal{V}_0+\sqrt{1-\bar{\alpha}_n}\epsilon\,,\nonumber \\
{\rm where}\quad \epsilon &\sim& \mathcal{N}(\mathbf{0},\mathbf{I})\,.
\label{eq:V0_to_Vn}
\end{eqnarray}

\textit{Backward diffusion:} We now define the \textit{reverse denoising sequence} (or backward-diffusion process), which generates $\mathcal{V}_{n-1}$ from $\mathcal{V}_n$ via a sequence of denoising steps, starting from the pure Gaussian noise $\mathcal{V}_N\sim p(\mathcal{V}_n)=\mathcal{N}(\bm 0, \bm I)$ that has the Markov property. We denote the reverse one-step transition probability distribution functions as 
\begin{equation}
p_{n\rightarrow (n-1)}(\mathcal{V}_{n-1}\mid\mathcal{V}_n)\,. 
\end{equation}
The forward-noise schedule $\beta_n$ is chosen to be sufficiently small so that $p_{n\rightarrow (n-1)}(\mathcal{V}_{n-1}|\mathcal{V}_n)$ preserves the Gaussian structural form~\cite{feller1949theory}. 

At this juncture, we recall that a U-Net~\cite{ronneberger2015unet} is a convolutional neural network (CNN) that consists of an encoder-decoder architecture with skip connections. The encoder compresses the input to capture macroscopic, long-range structural correlations in the input data, and the skip connections bypass the bottleneck to preserve localized, high-frequency details; details of the U-Net architecture that we use are discussed in Sec.~\ref{subsubsec:net_arch_noise_schedule}.

The computation of the true reverse transition probability distribution $p_{n\rightarrow (n-1)}(\mathcal{V}_{n-1}|\mathcal{V}_n)$ is intractable; therefore, we approximate it using a U-Net parameterized by learnable parameters, denoted g
enerically by $\theta$, for each denoising step $n$. The Markov property still holds for the reverse process, and the probability distribution for the whole sequence, \textit{as approximated by the U-Net}, is:
\begin{equation}
p_{\theta,N}(\mathcal{V}_{0:N})=p_\theta(\mathcal{V}_N)\prod_{n=1}^Np_{\theta,n\rightarrow(n-1)}(\mathcal{V}_{n-1}|\mathcal{V}_n)\,;
\end{equation}
here, each generative reverse step is a Gaussian that can, in general, be modeled as:
\begin{equation}
p_{\theta,n\rightarrow(n-1)}(\mathcal{V}_{n-1}|\mathcal{V}_n)=\mathcal{N}\left(\mu_\theta(\mathcal{V}_n,n),\Sigma_\theta(\mathcal{V}_n,n)\right)\,,
\label{eq:ptheta1}
\end{equation}
where $\mu_\theta$ and  $\Sigma_\theta$ are the mean and the standard deviation, respectively.

Our objective is to train the U-Net model to perform the reverse denoising sequence in such a way that the synthetic data, thus generated, have the same statistical properties as those of the original distribution $\pi_{data}$. The training scheme is illustrated in the flow chart of Fig.~\ref{fig:flow_chart}.

\subsubsection{Loss Function}
\label{subsubsec:Loss}

The training of the U-net involves maximizing the log-likelihood of the ground-truth data. This maximization is equivalent to the minimization of the cross-entropy loss $L_{\text{CE}}$ between the DNS-data distribution and the learned probability distribution:
\begin{align}
L_{\text{CE}} &\equiv-\mathbb{E}_{\mathcal{V}_0\sim q(\mathcal{V}_0)}\log(p_{\theta}(\mathcal{V}_0)) \nonumber \\
&=-\mathbb{E}_{\mathcal{V}_0\sim q(\mathcal{V}_0)}\log\left(\int p_{\theta,N+1}(\mathcal{V}_{0:N})d\mathcal{V}_{1:N}\right)\,,
\end{align}
where $\mathbb{E}$ denotes the expectation value.
A direct computation of this integral, over all possible latent pathways, is numerically prohibitive. Hence, the optimization targets the \textit{Variational Lower Bound} (VLB) of 
\begin{figure*}[htp]
    \centering
    \includegraphics[width=1.0\linewidth]{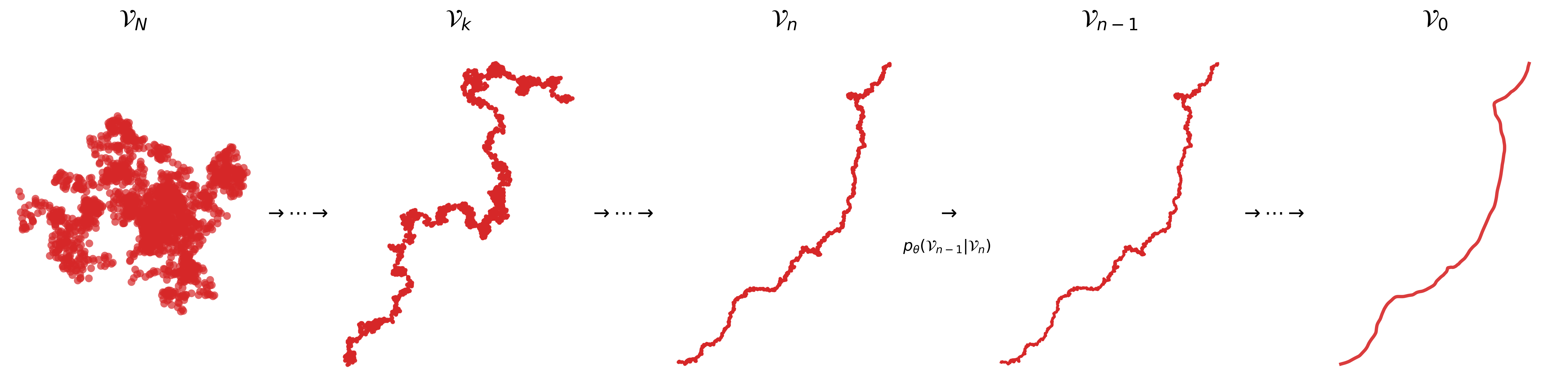}
    \caption{A schematic representation of steps in the sampling part in our diffusion models. We begin by sampling a trajectory from the standard Gaussian prior distribution, denoted as $\mathcal{V}_N$. The model then reverses the diffusion process by progressively denoising the sequence through intermediate latent states $(\mathcal{V}_N,\dots,\mathcal{V}_k,\dots, \mathcal{V}_n, \mathcal{V}_{n-1},\dots)$. Each generative reverse step is approximated by the U-Net-parameterized transition probability $p_{\theta,n\rightarrow(n-1)}(\mathcal{V}_{n-1}\vert{\mathcal{V}_n})$. This iterative denoising process ultimately yields the synthetic Lagrangian trajectory $\mathcal{V}_0$.}
    \label{fig:denoising_trajectory}
\end{figure*}
the cross-entropy loss [here and henceforth we suppress the subscripts on $p$ and $q$, as is common in the literature in this area, because these subscripts can be surmised from the arguments of the relevant distributions]:

\begin{eqnarray}
L_{\text{CE}}&\leq&\mathbb{E}_{q(\mathcal{V}_0)}\mathbb{E}_{p(\mathcal{V}_{1:N}|\mathcal{V}_0)}\left[\log\frac{q(\mathcal{V}_{1:N}|\mathcal{V}_0)}{p_{\theta}(\mathcal{V}_{0:N})}\right]\nonumber \\
&\equiv& L_{\text{VLB}}\,.
\end{eqnarray}
This bound can be expressed as a sum of $D_{\rm KL}$, the Kullback-Leibler (KL) divergences~\footnote{This divergence measures how the probability distribution, say $\mathcal{Q}$, differs from a reference distribution, say $\mathcal{P}$; for the discrete and continuous forms we write, respectively, $D_{\rm KL}(\mathcal{Q}\parallel \mathcal{P}) = \sum_x \mathcal{P}(x) \log [\mathcal{P}(x)/\mathcal{Q}(x)]$ and $D_{\rm KL}(\mathcal{Q}\parallel \mathcal{P}) = \int^{\infty}_{-\infty} \mathcal{P}(x) \log [\mathcal{P}(x)/\mathcal{Q}(x)] dx$; or, equivalently, $D_{\rm KL}(\mathcal{Q}\parallel \mathcal{P})=\mathbb{E}_{x\sim \mathcal{P}(x)}\Bigg[ \log [\mathcal{P}(x)/\mathcal{Q}(x)]\Bigg]$ [Ref.\cite{Goodfellow2016} page 72 discusses this in detail].} for the individual reverse steps; this isolates the transitions into mathematically manageable terms:
\begin{align}
L_{\text{VLB}} &= \mathbb{E}_{q(\mathcal{V}_0)}\Bigg[ L_N +  \sum_{n>1}^N L_{n-1} -\log p_{\theta} (\mathcal{V}_0|\mathcal{V}_1)\Bigg]\,,
\label{eq:VLB}
\end{align}
where 
\begin{align}
    L_N&\equiv& D_{\text{KL}}\big(&q(\mathcal{V}_N|\mathcal{V}_0)\parallel p_\theta(\mathcal{V}_{N})\big), \\
    L_{n-1}&\equiv& D_{\text{KL}}\big(&q(\mathcal{V}_{n-1}|\mathcal{V}_n,\mathcal{V}_0)\parallel p_{\theta}(\mathcal{V}_{n-1}|\mathcal{V}_n)\big)\,.\quad
    \label{eq:DKL}
\end{align}

The term $L_N$ contains no trainable parameters, as $p_\theta(\mathcal{V}_N)=\mathcal{N}(\bm 0, \bm I)$; therefore, it can be discarded from the loss calculation during training. For the remaining terms with $L_{n-1}$, the inverse conditional probability $q(\mathcal{V}_{n-1}|\mathcal{V}_n,\mathcal{V}_0)$ becomes analytically tractable, via the Bayes theorem, because it is explicitly conditioned on the ground-truth data:
\begin{equation}
q(\mathcal{V}_{n-1}|\mathcal{V}_n,\mathcal{V}_0)=q(\mathcal{V}_n|\mathcal{V}_{n-1},\mathcal{V}_0)\frac{q(\mathcal{V}_{n-1}|\mathcal{V}_0)}{q(\mathcal{V}_n|\mathcal{V}_0)}\,.
\end{equation} 
We then substitute the known forward-transition PDFs into this expression, in the conditioned reverse-step PDF, to obtain the following Gaussian distribution:
\begin{equation}
q(\mathcal{V}_{n-1}|\mathcal{V}_n,\mathcal{V}_0)=\mathcal{N}(\tilde{\mu}_n(\mathcal{V}_n,\mathcal{V}_0),\tilde{\beta}_n\mathbf{I})\,,
\end{equation}
where
\begin{eqnarray}
\tilde{\mu}_n(\mathcal{V}_n,\mathcal{V}_0)&\equiv&\frac{\sqrt{\bar{\alpha}_{n-1}}\beta_n}{1-\bar{\alpha}_n}\mathcal{V}_0 \nonumber\\
&&+\frac{\sqrt{\alpha_n}(1-\bar{\alpha}_{n-1})}{1-\bar{\alpha}_n}\mathcal{V}_n\,; \nonumber \\
\tilde{\beta}_n&\equiv&\frac{1-\bar{\alpha}_{n-1}}{1-\bar{\alpha}_n}\beta_n\,.
\end{eqnarray}

If we fix the standard deviation of the network's predicted reverse step to align with the forward step, i.e., we choose 
\begin{equation}
\Sigma_\theta(\mathcal{V}_n,n)=\beta_n\mathbf{I}\,,
\label{eq:ptheta2}
\end{equation}
then the KL divergences~\eqref{eq:DKL} 
 simplify to the mean-squared error (MSE) 
\begin{eqnarray}
L_{n-1}=\mathbb{E}_{\mathcal{V}_0\sim q(\mathcal{V}_0)}\left[\frac{1}{2\beta_n}\|\tilde{\mu}_n(\mathcal{V}_n,\mathcal{V}_0)-\mu_\theta(\mathcal{V}_n,n)\|^2\right]. 
\label{eq:Euclid}
\end{eqnarray}
We can express $\mathcal{V}_0$ as a function of $\mathcal{V}_n$ and $\epsilon$, the noise added to $\mathcal{V}_0$ to construct $\mathcal{V}_n$, using the inverted relation from the forward-sampling relation~\eqref{eq:V0_to_Vn}, to get
\begin{equation}
\label{eq:mean}
\tilde{\mu}(\mathcal{V}_n,\epsilon)=\frac{1}{\sqrt{\alpha_n}}\left(\mathcal{V}_n-\frac{\beta_n}{\sqrt{1-\bar{\alpha}_n}}\epsilon\right)\,.
\end{equation}
If we replace $\epsilon$ by $\epsilon_\theta(\mathcal{V}_n,n)$, which we obtain from the U-Net while it is being trained, we get 
\begin{equation}
\mu_\theta(\mathcal{V}_n,n)=\frac{1}{\sqrt{\alpha_n}}\left(\mathcal{V}_n-\frac{\beta_n}{\sqrt{1-\bar{\alpha}_n}}\epsilon_\theta(\mathcal{V}_n,n)\right)\,.
\label{eq:reparametrization}
\end{equation}
Now we substitute Eqs.~\eqref{eq:mean} and \eqref{eq:reparametrization} into Eq.~\eqref{eq:Euclid} to obtain 
\begin{equation}
L_{n-1}=\mathbb{E}_{q(\mathcal{V}_0),\epsilon}\left[\frac{\beta_n}{2\alpha_n(1-\bar{\alpha}_n)}\|\epsilon-\epsilon_\theta(\mathcal{V}_n,n)\|^2\right]\,. 
\label{eq:lnm1penultimate}
\end{equation}
Thus, we have a noise-matching task, in which the model is trained to reduce the MSE between the actual noise injected and the noise predicted by the model.
The step-dependent weighting coefficient is often omitted to improve training stability, as in Ref.~\cite{ho2020denoising}, which we follow, so we minimize the simple unweighted, loss function given below:
\begin{equation}
L_{n-1}^{\text{simple}}\equiv\mathbb{E}_{q(\mathcal{V}_0),\epsilon}\left[\|\epsilon-\epsilon_\theta(\mathcal{V}_n,n)\|^2\right]\,.
\label{eq:simple_loss}
\end{equation}

\subsubsection{Sampling}

We now utilize the trained U-Net to generate synthetic trajectories by reversing the diffusion process. Specifically, we initialize this generation by sampling a trajectory from the prior distribution, $\mathcal{V}_N \sim \mathcal{N}(\mathbf{0}, \mathbf{I})$, and progressively denoise it, as shown in Fig.~\ref{fig:denoising_trajectory}. There are two common formulations of this denoising process:  the Denoising Diffusion Probabilistic Model (DDPM) and the Denoising Diffusion Implicit Model (DDIM); the former imposes a Markovian constraint on the reverse steps, but the latter does not.
\\
\paragraph{Denoising Diffusion Probabilistic Model (DDPM) Sampler:}
The standard DDPM framework explicitly relies on the Markovian assumption for both the forward and backward processes, i.e., the generation step follows the exact reverse-process transition probability that is parameterized during training [Eqs.~\eqref{eq:ptheta1} and \eqref{eq:ptheta2}]. The final output from the  U-Net is the predicted noise $\epsilon_{\Theta}(\mathcal{V}_n, n)$, where $\Theta$ denotes the fully trained value of $\theta$, so we can express the estimated mean of the reverse step as
\begin{equation}
{\mu}_\Theta(\mathcal{V}_n,n)=\frac{1}{\sqrt{\alpha_n}}\left(\mathcal{V}_n-\frac{\beta_n}{\sqrt{1-\bar{\alpha}_n}}\epsilon_\Theta(\mathcal{V}_n,n)\right)\,.
\label{eq:muTheta}
\end{equation}
To sample the preceding state $\mathcal{V}_{n-1}$, we draw from the Gaussian distribution $p_\theta(\mathcal{V}_{n-1}|\mathcal{V}_n)$ by taking the predicted mean and injecting the stochastic noise component with the fixed variance ${\beta}_n$:
\begin{eqnarray}
\mathcal{V}_{n-1} &=& {\mu}_\Theta(\mathcal{V}_n,n)+ \sqrt{\beta_n}\mathbf{z}\,,
\label{eq:DDPM_Vn-1_to_Vn}
\end{eqnarray}
where $\mathbf{z}\sim\mathcal{N}(\mathbf{0}, \mathbf{I})$ for $n>1$. For the final step at $n=1$, no noise is added ($\mathbf{z}=\mathbf{0}$); this yields U-Net result for the synthetic Lagrangian-particle trajectory $\mathcal{V}_0$. We note that DDPM enforces the Markov-chain property, so generating a high-fidelity trajectory requires sequentially evaluating the network for all $N$ steps, which is computationally expensive.
\\
\paragraph{Denoising Diffusion Implicit Model (DDIM) Sampler:}
Although the diffusion model is trained by optimizing the VLB, under the strict Markovian assumption, Ref.~\cite{song2021denoising} has demonstrated that the same VLB works as a surrogate objective function for a family of \textit{non-Markovian} inference distributions, so we can use the same trained U-Net, as in DDPM, with the trained parameters $\Theta$. This holds provided that $q_{0\rightarrow n}(\mathcal{V}_n|\mathcal{V}_0)$ remain strictly Gaussian [Eq.~\eqref{eq:V0_to_Vn}]. Then, an application of the Bayes theorem yields~\cite{song2021denoising} the  reverse-transition probability that parameterized by the variance $\sigma_n$:
\begin{align}
&{q_\sigma(\mathcal{V}_{n-1}\vert{\mathcal{V}_n, \mathcal{V}_0)}} = \nonumber\\ 
&\mathcal{N }\Bigg(\sqrt{\frac{1-\bar{\alpha}_{n-1}-\sigma_n^2}{1-\bar{\alpha}_n}}\mathcal{V}_n+ \nonumber\\ 
&\left(\sqrt{\bar{\alpha}_{n-1}}-\sqrt{\bar{\alpha}_n}\sqrt{\frac{1-\bar{\alpha}_{n-1}-\sigma_n^2}{1-\bar{\alpha}_n}}\right)\mathcal{V}_0, \sigma_n^2\mathbf{I}\Bigg)\,.
\end{align}
By inverting the reparameterization used in Eq.~\eqref{eq:reparametrization}, we obtain the following U-Net approximation of the denoised trajectory:
\begin{equation}
\hat{\mathcal{V}}_{0}=\frac{1}{\sqrt{\bar{\alpha}_n}}\left(\mathcal{V}_n-\sqrt{1-\bar{\alpha}_n}\epsilon_\Theta(\mathcal{V}_n, n)\right)\,;
\label{eq:V0DDIM}
\end{equation}
Denoising Diffusion Implicit Models (DDIM) emerge as a special case of this generalized process when $\sigma_n= 0$ for all steps $n$, i.e., the stochasticity of the reverse process vanishes entirely, resulting in a deterministic mapping from the initial latent noise $\mathcal{V}_N$ to the synthetic DDIM Lagrangian-particle trajectory $\hat{\mathcal{V}}_{0}$. 
Thus, we can directly jump to the state $n-K$ to get the DDIM sampling update:
\begin{align}
\mathcal{V}_{n-K} &= \sqrt{\bar{\alpha}_{n-K}}\left(\frac{\mathcal{V}_n-\sqrt{1-\bar{\alpha}_n}\epsilon_\Theta(\mathcal{V}_n, n)}{\sqrt{\bar{\alpha}_n}}\right) \nonumber \\
&\quad + \sqrt{1-\bar{\alpha}_{n-K}}\epsilon_\Theta(\mathcal{V}_n, n)\,,
\label{eq:DDIM_Vn_to_Vn-K}
\end{align}
where we have used Eq.~\eqref{eq:V0DDIM}. 
    The crucial advantage of DDIM, relative to DDPM,
    is that the reverse diffusion can be performed over a much smaller subset of denoising steps $M < N$, by taking jumps of length $K$  [Eq.~\eqref{eq:DDIM_Vn_to_Vn-K} and the flow chart in Part II of Fig.~\ref{fig:flow_chart}], so there is a considerable gain in computational efficiency.

\begin{figure*}[htp]
    \centering
    \includegraphics[width=1.0\linewidth]{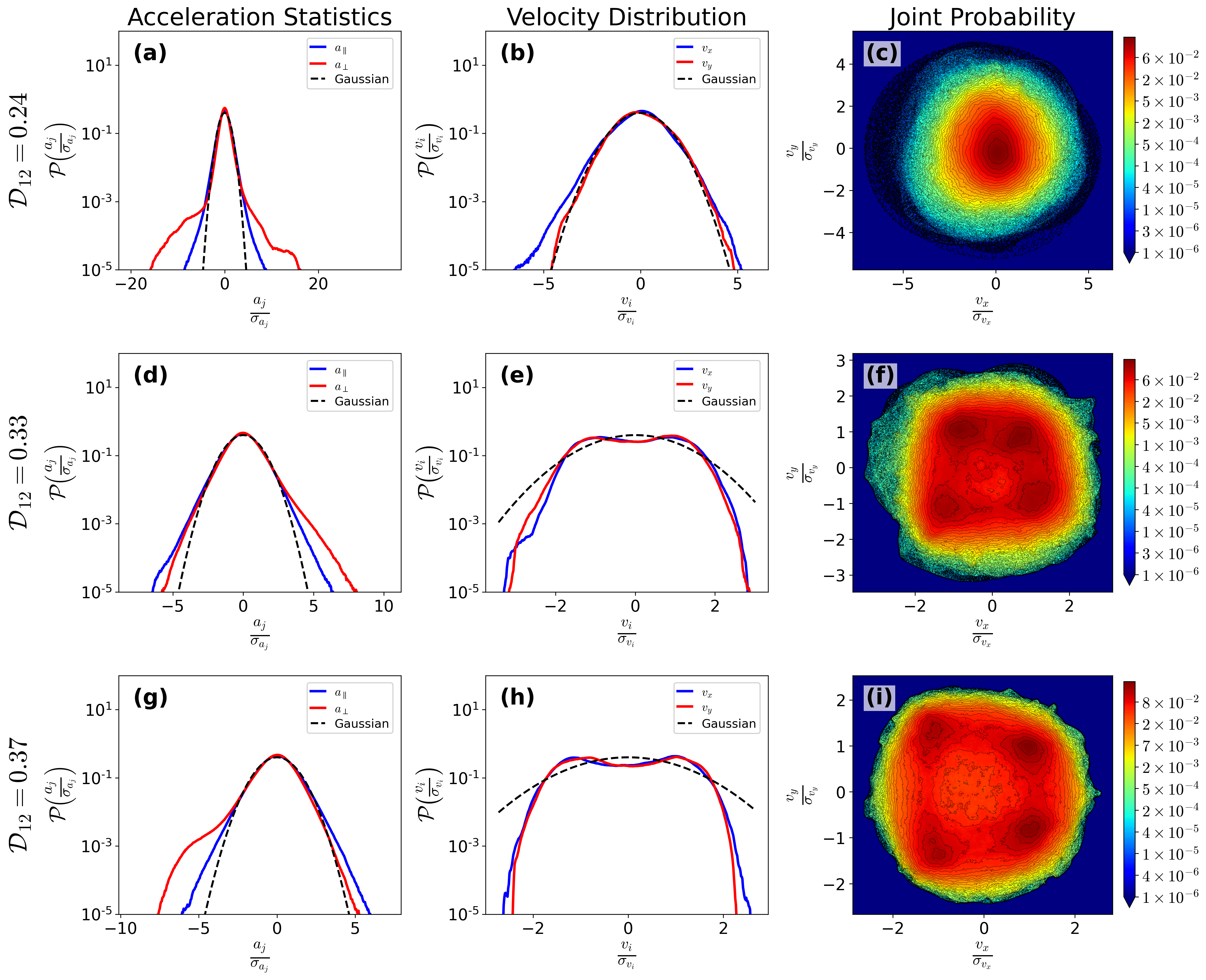}
    \caption{Probability density functions (PDFs) of acceleration components ($a_\parallel$ and $a_\perp$) and velocity ($v_x$ and $v_y$), alongside their joint probability distributions for the non-reciprocity parameters $\mathcal{D}_{12} \in \{0.24, 0.33, 0.37\}$. The distributions illustrate that the acceleration components and the velocity components show significant deviations from Gaussian distribution; in particular, the latter are bimodal for $\mathcal{D}_{12}=0.33$ and $\mathcal{D}_{12}=0.37$ [cf. their Eulerian counterparts in Figs.~\ref{fig:subplot_pdf} (c) and (d), which also show bimodal PDFs that are associated with lane-type structures in pseudocolor plots of the Eulerian-velocity components [Figs.~\ref{fig:subplot_pdf} (a) and (b)].}
    \label{fig:vel_accel_stats}
\end{figure*}

\begin{figure*}[htp]
    \centering
    \includegraphics[width=1.0\linewidth]{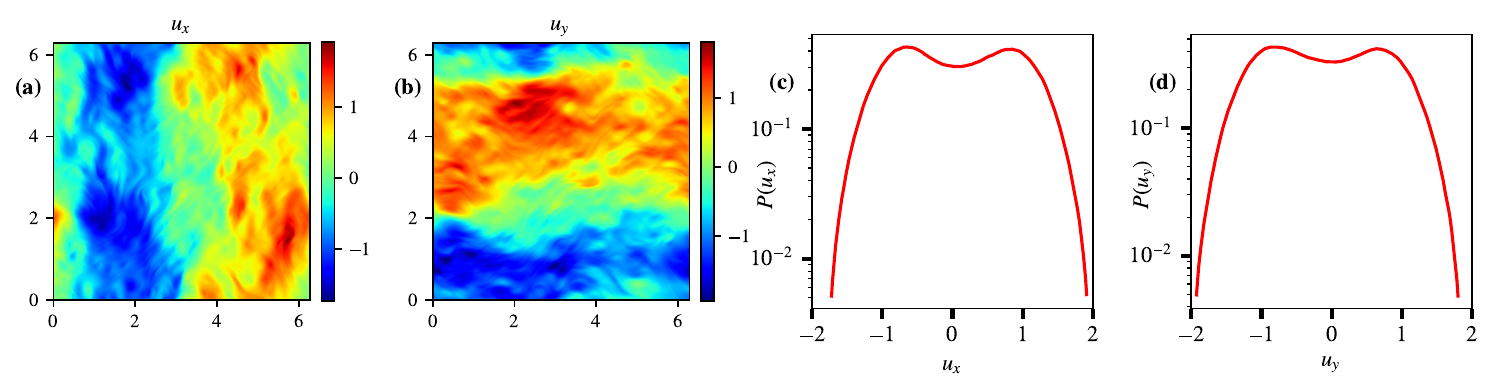}
    \caption{(a) and (b) Pseudocolor plots of the Eulerian velocity components $u_x$ and $u_y$, respectively, for $\mathcal{D}_{12}=0.33$, showing the emergence of lane-like structures. (c) and (d) Probability density functions (PDFs) of $u_x$ and $u_y$, respectively, exhibiting a bimodal distribution consistent with lane formation.}
    \label{fig:subplot_pdf}
\end{figure*}

\begin{figure*}[htp]
    \centering
    
    \includegraphics[width=1.0\linewidth]{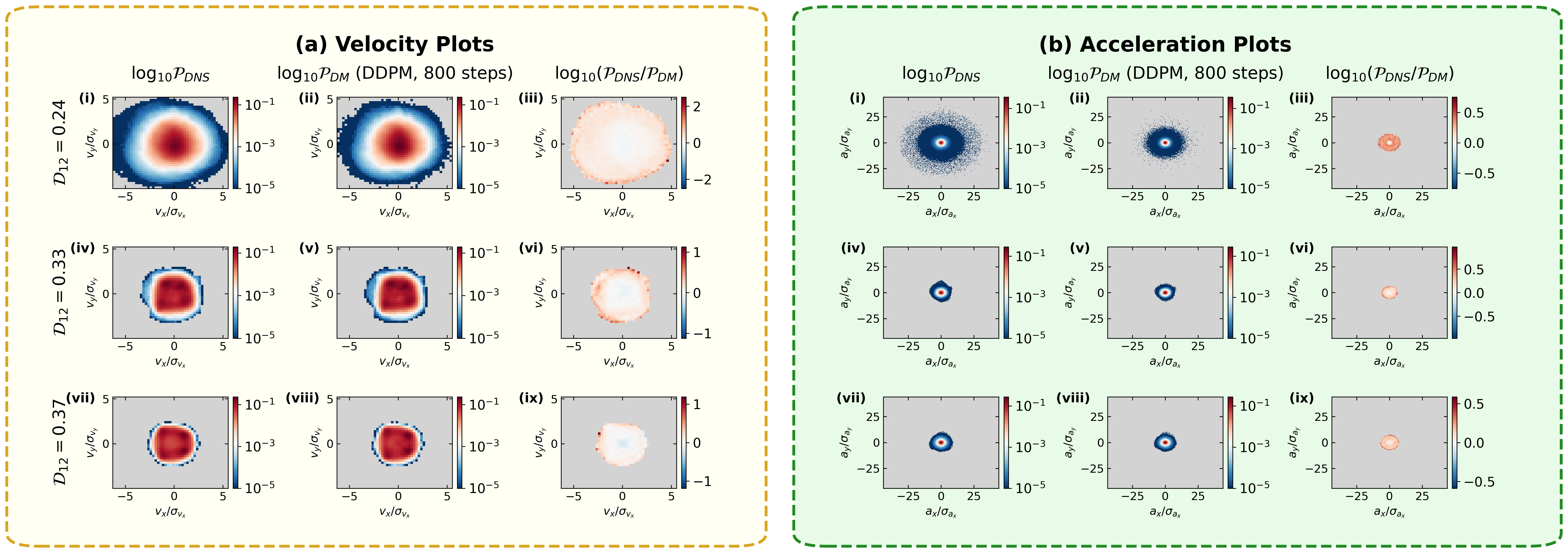}
    \caption{\textbf{(a) - Left} Comparison of the joint probability density functions (JPDFs) of normalized velocity components $(v_x/\sigma_{v_x}, v_y/\sigma_{v_y})$ between the ground-truth DNS data and synthetic trajectories generated by the diffusion model (DDPM sampler with 800 denoising steps: DDPM-800).
    \textbf{(b) - Right} Comparison of the JPDFs of normalized acceleration components $(a_x/\sigma_{a_x}, a_y/\sigma_{a_y})$ from DNS data versus diffusion model (DDPM-800). The residuals, plotted as $\log_{10}(\mathcal{P}_{DNS}/\mathcal{P}_{DM})$, provide a quantitative assessment of the model's fidelity in reliably reproducing the target distribution.}
    \label{fig:vel_accel_stats_DDPM_800}
\end{figure*}

\subsubsection{U-Net Architecture and Training Schedule}
\label{subsubsec:net_arch_noise_schedule}

The U-Net we employ is similar to the U-Nets in Refs.~\cite{dhariwal2021diffusionUnet,Biferale_DM_2024}. Our U-Net, adapted, \textit{mutatis mutandis}, to process the 2D Lagrangian-particle trajectories of our NRCHNS system, is the core noise-estimation network that yields $\epsilon_\Theta$ for high-accuracy particle-trajectory generation. The U-Net architecture, along with the Residual Blocks (ResBlocks), Upsampling and Downsampling, are shown schematically in Fig.~\ref{fig:unet}.

The encoder and decoder pathways in this U-Net consist of five levels each. As the data pass through these levels, the temporal resolution is sequentially halved, via downsampling layers, in the encoder, and doubled, via upsampling layers, in the decoder. Each level consists of three ResBlocks~\cite{He2016ResNet} in the encoder, and four ResBlocks in the decoder. The Bottleneck of the U-Net consists of an Attention Block~\cite{vaswani2017attention} sandwiched between two ResBlocks. The two levels closest to the bottleneck in both the encoder and the decoder contain an Attention Block with each ResBlock. Each Attention Block is a multi-head attention~\cite{vaswani2017attention}, with four attention heads, to enable the U-Net to capture dependencies at specific critical temporal regions within the trajectory sequences. [See Appendix~\ref{subsec:appendix:unet_math} for details.]

During the training phase, we sample mini-batches of DNS ground-truth trajectories and assign a uniformly distributed random diffusion step to each sample; we use a mini-batch size of $64$. The U-Net is then optimized by minimizing the simplified loss objective $L_{n-1}^{\text{simple}}$ [Eq.~\eqref{eq:simple_loss}] for each mini-batch; we train three diffusion models independently, one for each dataset $\mathfrak{D}_i$, with $i=1,\,2$, and $3$, for $3.5\times10^5$ training iterations. The batch-averaged loss [Eq.~\eqref{eq:simple_loss}] reaches a plateau rapidly, but this does not guarantee complete training of $\epsilon_\theta$. To verify if the U-Net has been trained adequately, we check every $10,000$ steps if the statistics of synthetic trajectories matches the statistics from our DNS ground-truth.

To optimize both the training efficiency and the generative sampling speed, we use the non-linear noise schedule proposed in Ref.~\cite{li2024generative}; this is referred to as the \textit{tanh6-1} schedule. In the \textit{tanh6-1} schedule, $\bar{\alpha}_n$ has the following functional form:
\begin{equation}
\bar{\alpha}_n=\frac{-\tanh(7n/N - 6) + \tanh(1)}{-\tanh(-6) + \tanh(1)}\,,
\end{equation}
with $\beta_{\text{min}}=\beta_1=10^{-4}$ and $\beta_{\text{max}}=\beta_N=0.02$.
This tailored schedule allows us to achieve high-quality sample generation with only $N=800$ total diffusion steps in the DDPM sampler.

\subsubsection{Training and Sampling Time:}
Training a single U-Net on any dataset $\mathfrak{D}_i$ for $3.5\times10^5$ training iterations takes approximately $34$ hours on one A100 GPU. Performing 100 steps of denoising takes approximately $6.5$ minutes per $2048$ trajectories. Our  results [Sec.~\ref{sec:results}]
are based on $102,400$ samples for $\mathcal{D}_{12}=0.24$, $\mathcal{D}_{12}=0.33$, and $\mathcal{D}_{12}=0.37$, for every sampling technique that we use [sampled from the respective trained U-Nets].

\section{Results}
\label{sec:results}

We present our results for PDFs of velocity and acceleration components in Sec.~\ref{subsec:vel_accel_pdfs}. In Sec.~\ref{subsec:curvature_analysis} we turn to PDFs of the trajectory curvature $\kappa$. Then, in Sec.~\ref{subsec:structure_flatness_zeta}, we give PDFs of velocity increments along with velocity structure functions, the associated flatness, and the local slopes of the structure functions. Finally, we analyse the irreversibility of 2D NRCHNS turbulence in Sec.~\ref{subsec:irrev_analysis}. We display results from our DNS and then compare them with their counterparts from the different sampling techniques that we use.

\begin{figure*}[htp]
    \centering
    \includegraphics[width=1.0\linewidth]{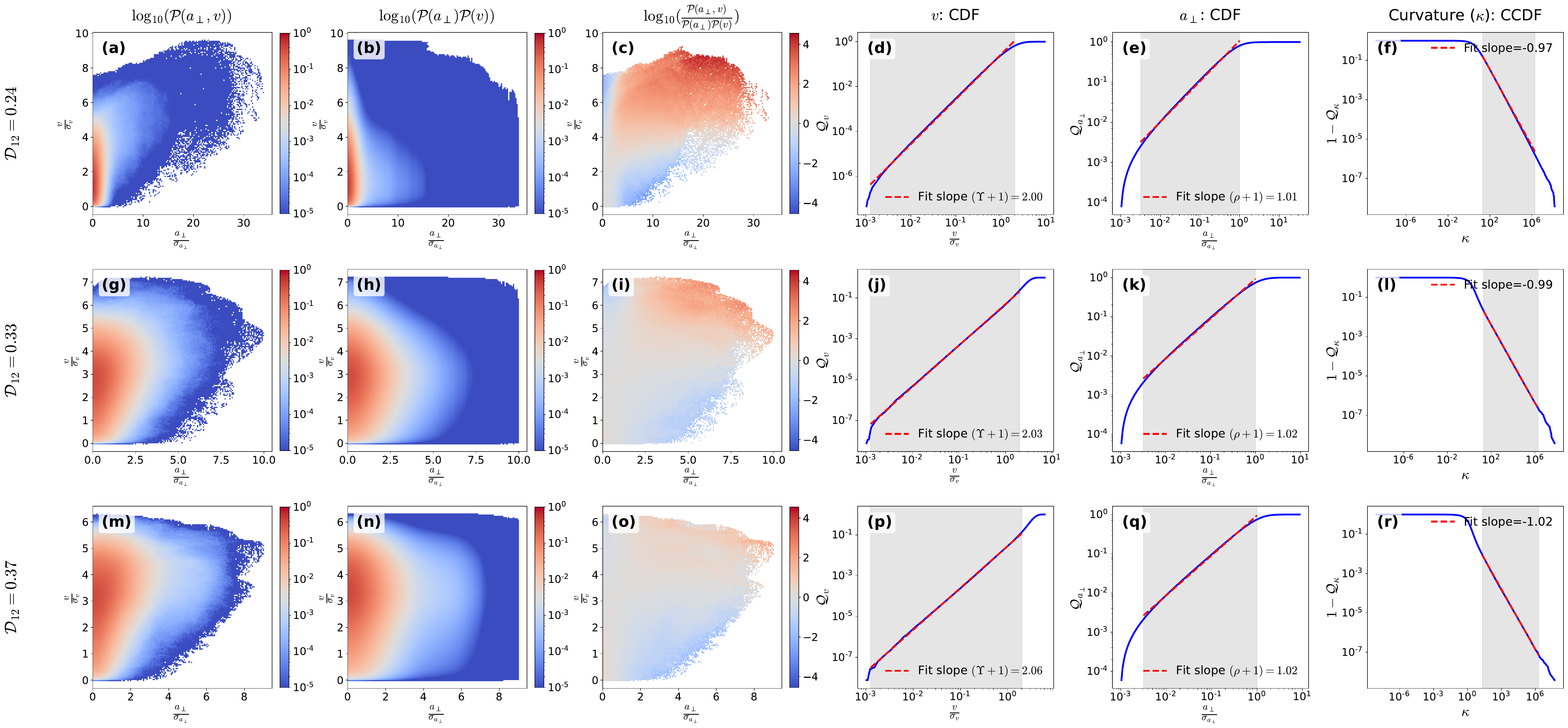}
    \caption{Plots (a), (g), and (m) display the vc  joint probability density function $\log_{10}(\mathcal{P}(a_\perp,v))$ of the velocity magnitude ($v$) and perpendicular acceleration ($a_\perp$). Plots (b), (h), and (n) show the factorized approximation $\log_{10}(\mathcal{P}(a_\perp)\mathcal{P}(v))$ under the assumption of statistical independence. Plots (c), (i), and (o) present the residual $\log_{10}(\frac{\mathcal{P}(a_\perp,u)}{\mathcal{P}(a_\perp)\mathcal{P}(v)})$ between the joint and factorized distributions. Plots (d), (j), and (p) display the cumulative distribution function ($\mathcal{Q}_v$) for the normalized velocity, where $\sigma_v$ is the standard deviation of $v$. Plots (e), (k), and (q) show the CDF ($\mathcal{Q}_a$) for the normalized perpendicular acceleration, with $\sigma_{a_\perp}$ representing its standard deviation. Plots (f), (l), and (r) plot the complementary cumulative distribution function $(1-\mathcal{Q}_\kappa)$ for trajectory curvature ($\kappa$). Red dashed lines indicate the linear fits within the shaded asymptotic regimes. }
    \label{fig:residual_plot}
\end{figure*}

\begin{figure*}[htp]
    \centering
    \includegraphics[width=1.0\linewidth]{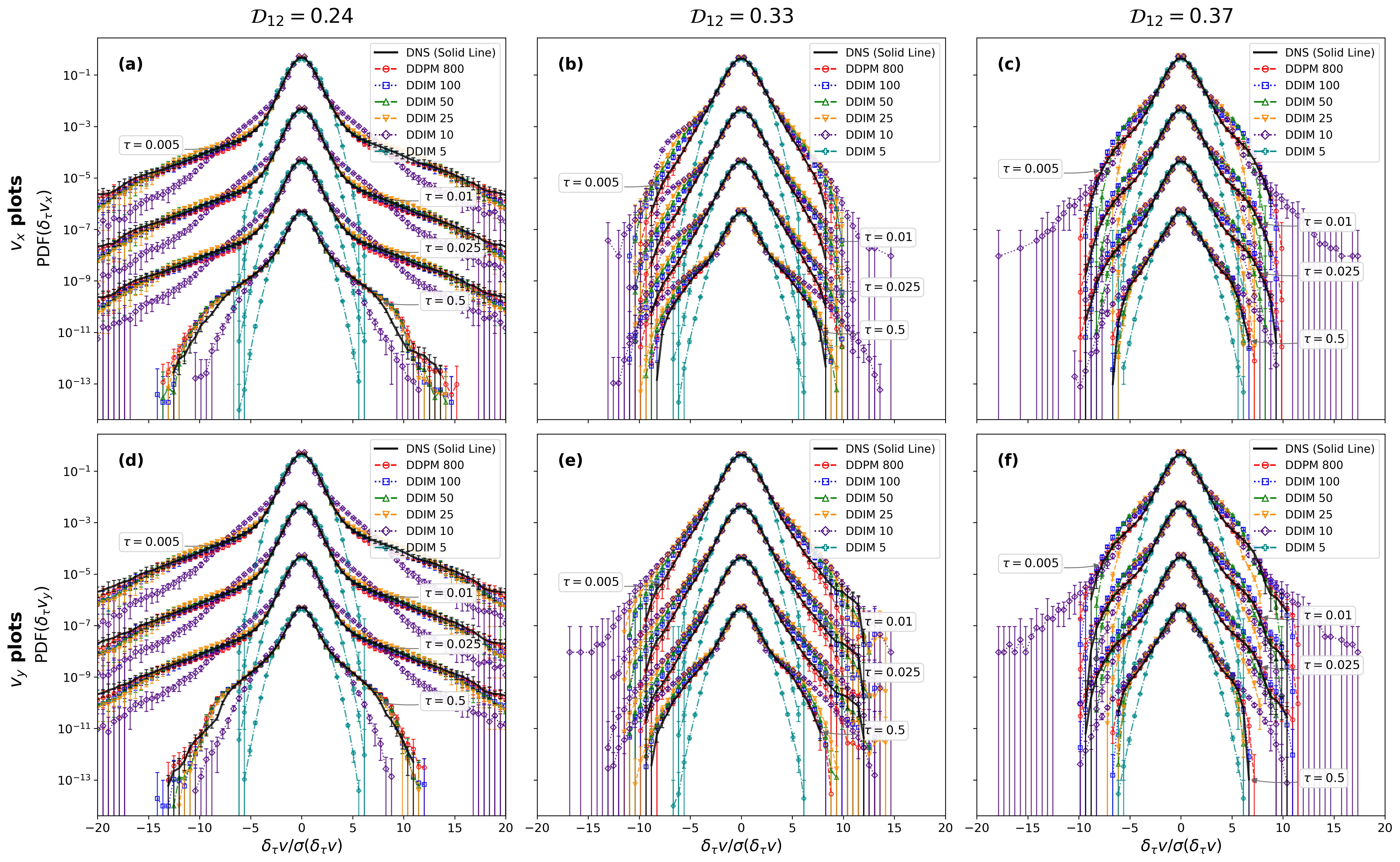}
    \caption{Probability density functions (PDFs) of Lagrangian velocity increments $\delta_{\tau}v_{i}(t)$ for the $x$ and $y$ components. The comparison demonstrates that DDPM sampling with 800 steps and DDIM sampling with 100 steps successfully capture the non-Gaussian tails associated with intermittency in the turbulent flow across all analyzed $\mathcal{D}_{12}$ regimes. DDIM with 50 and 25 steps also follow the DNS curve closely, but with slight deviations. DDIM with 10 steps shows greater deviation, whereas DDIM with 5 steps does not follow the DNS plots at all. Error bars represent the minimum and maximum values obtained after dividing the dataset into $10$ smaller sub-batches.}
    \label{fig:vel_inc_pdf}
\end{figure*}

\begin{figure*}[htp]
    \centering
    
    \begin{subfigure}[b]{0.48\textwidth}
        \centering
        \begin{tikzpicture}
            \node[draw=green!50, dotted, line width=1.5pt, rounded corners=8pt, inner sep=5pt] (boxA)
            {\includegraphics[width=1\linewidth, height=7cm]{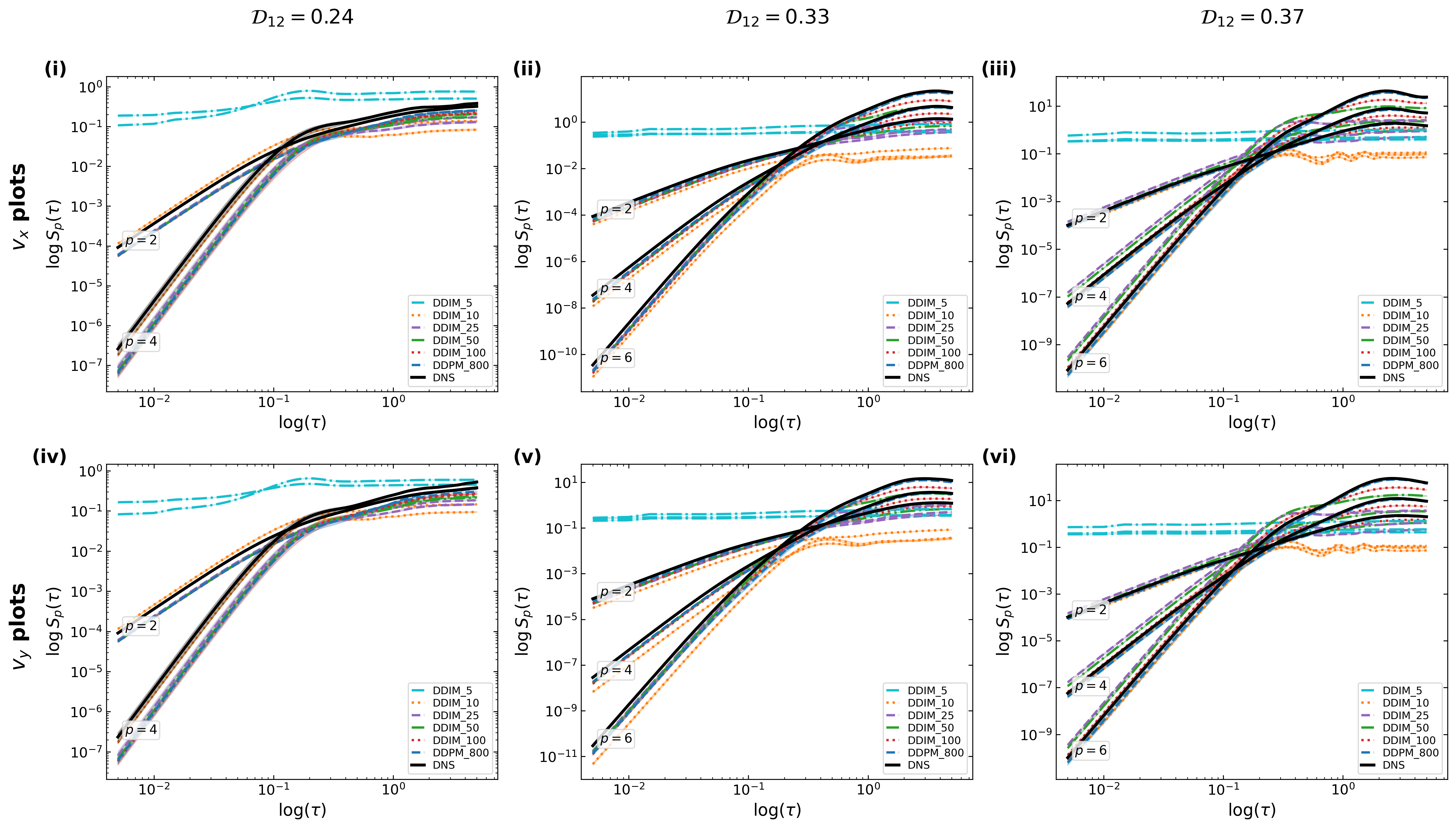}};
            
            \node[anchor=north west, fill=white, inner sep=3pt, font=\bfseries] at ([xshift=5pt, yshift=-5pt]boxA.north west) {(a)};
        \end{tikzpicture}
    \end{subfigure}
    \hfill 
    \begin{subfigure}[b]{0.48\textwidth}
        \centering
        \begin{tikzpicture}
            \node[draw=orange, dotted, line width=1.5pt, rounded corners=8pt, inner sep=5pt] (boxB)
            {\includegraphics[width=1\linewidth, height=7cm]{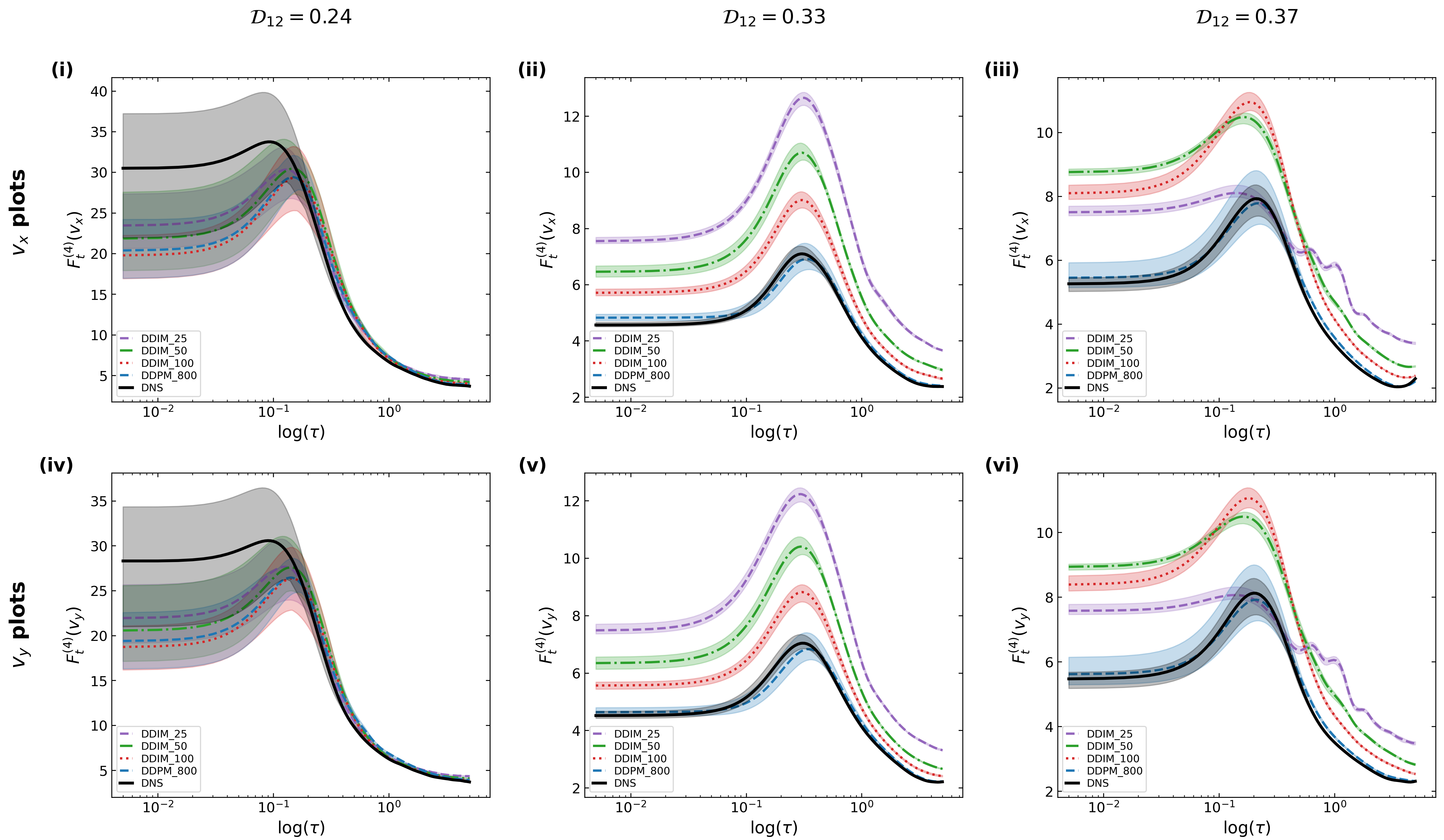}};
            
            \node[anchor=north west, fill=white, inner sep=3pt, font=\bfseries] at ([xshift=5pt, yshift=-5pt]boxB.north west) {(b)};
        \end{tikzpicture}
    \end{subfigure}

    \vspace{1.5em} 

    \begin{subfigure}[b]{0.85\textwidth}
        \centering
        \begin{tikzpicture}
            \node[draw=cyan!60, dotted, line width=1.5pt, rounded corners=8pt, inner sep=5pt] (boxC)
            {\includegraphics[width=1\linewidth]{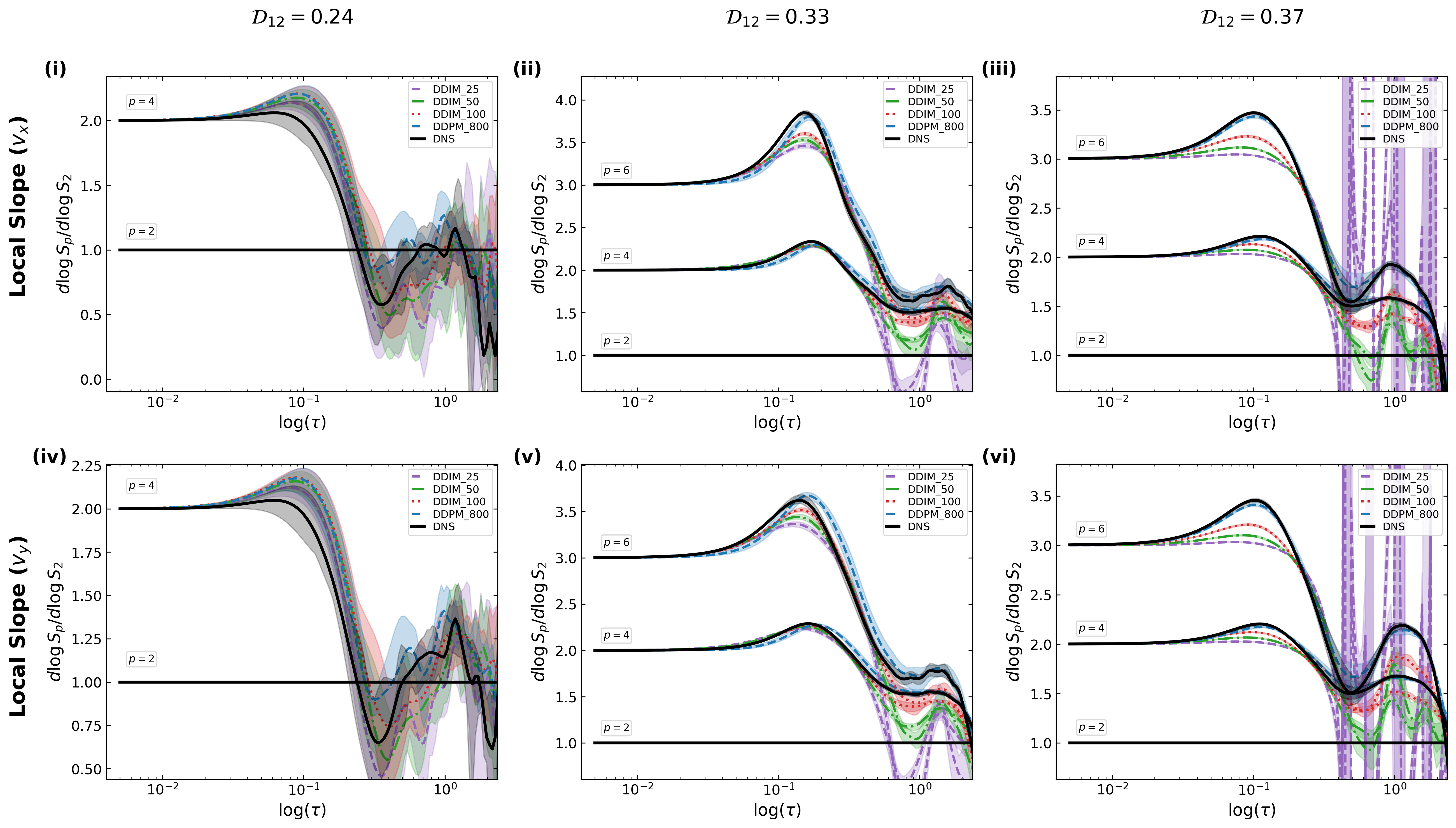}};
            
            \node[anchor=north west, fill=white, inner sep=3pt, font=\bfseries] at ([xshift=5pt, yshift=-5pt]boxC.north west) {(c)};
        \end{tikzpicture}
    \end{subfigure}

    \caption{\textbf{(a)} Lagrangian structure functions $S_{\tau}^{(p)}$ of order $p=2, 4,$ and $6$ for spatial components $v_x$ and $v_y$ across the three non-reciprocity regimes ($\mathcal{D}_{12} \in \{0.24, 0.33, 0.37\}$). \textbf{(b)} (i)-(vi) Log-log plots of the flatnesses for increments of $v_x$ and $v_y$. \textbf{(c)} (i)-(vi) Semi-log plots of the local slope $d\log S_p/d\log S_2$ for $p=2$ and $4$ for $\mathcal{D}_{12}=0.24$, and for $p=2$, $4$ and $6$ for $\mathcal{D}_{12}=0.33$, and $0.37$. 
    The shaded regions in each plot represent the minimum and maximum values obtained after dividing the dataset into $10$ smaller sub-batches.}
    \label{fig:structure_fn_flatness_local_slope}
\end{figure*}

\subsection{Velocity and Acceleration PDFs}
\label{subsec:vel_accel_pdfs}
In Fig.~\ref{fig:vel_accel_stats} we present PDFs of the parallel and perpendicular components of Lagrangian-particle acceleration and PDFs and joint PDFs of the Cartesian components of their velocities, in the nonequilibrium statistically steady states of the NRCHNS system, for the non-reciprocity parameter $\mathcal{D}_{12}=0.24$ [Figs.~\ref{fig:vel_accel_stats} (a), (b), and (c)], $\mathcal{D}_{12}=0.33$ [Figs.~\ref{fig:vel_accel_stats} (d), (e), and (f)], and $\mathcal{D}_{12}=0.37$ [Figs.~\ref{fig:vel_accel_stats} (g), (h), and (i)].

For $a_\parallel$ and $a_\perp$, the local acceleration components parallel and perpendicular to the velocity vector, which we normalize by their respective standard deviations $\sigma_{a_\perp}$ and $\sigma_{a_\parallel}$, we obtain the PDFs $\mathcal{P}_{a_j}(a_j/\sigma_{a_j})$, where $j\in\{\parallel, \perp\}$. We present semi-log plots of these PDFs in Figs.~\ref{fig:vel_accel_stats} (a), (d), and (h); these are qualitatively similar to their fluid-turbulence counterparts~\cite{Biferale_DM_2024}, insofar as they are close to Guassian PDFs near their peaks, but they have fat tails (i.e., they are leptokurtic).

The velocity-component PDFs [Figs.~\ref{fig:vel_accel_stats} (b), (e), and (h)] and JPDFs [Figs.~\ref{fig:vel_accel_stats} (c), (f), and (i)] are distinctly non-Gaussian, unlike their 2D-fluid-turbulence counterparts~\cite{perlekar2009statistically,pandit2017overview}. In particular, for $\mathcal{D}_{12}=0.33$ and $\mathcal{D}_{12}=0.37$ these PDFs are distinctly bimodal. This bimodality is associated with lane-type structures~\cite{reeves2021emergence} in pseudocolor plots of the Eulerian-velocity components [Figs.~\ref{fig:subplot_pdf} (a) and (b)]  The Eulerian counterparts of the Lagrangian PDFs are given in Figs.~\ref{fig:subplot_pdf} (c) and (d), these also show bimodal PDFs that follow naturally from the lane-type structures mentioned above, Not surprisingly, the JPDFs of these velocity components [Figs.~\ref{fig:vel_accel_stats} (c), (f), and (i)] display four peaks for $\mathcal{D}_{12}=0.33$ and $\mathcal{D}_{12}=0.37$. We note also that the statistically stationary non-reciprocal turbulent state has no net mean flow, so $\langle \mathbf{u}\rangle \simeq 0$; the angular brackets denote the average over this state. Furthermore, we quantify the degree of anisotropy via the tensor
$a_{ij}\equiv\frac{\langle u_i u_j\rangle}{\langle u_k u_k\rangle} -\frac{1}{2}\delta_{ij}$; in the statistically stationary state, we find that all components of $a_{ij}\simeq 10^{-3}$, indicating that, when we average over the nonequilibrium state, it is statistically isotropic by this measure.. 

We now compare the Lagrangian  PDFs and JPDFs from our ground-truth DNSs [Fig.~\ref{fig:vel_accel_stats}] with those that we have derived from the synthetic Lagrangian trajectories that we obtain from our trained diffusion models. This comparison of JPDFs is critical for a robust verification of the ability of our diffusion models to capture such statistics. The residuals, $\log_{10}(\mathcal{P}_{DNS}/\mathcal{P}_{DM})$, provide a quantitative measure for such verification; here, the subscript $DM$ denotes, generically, a diffusion model (such as the DDPM and DDIM that we use). 
In Fig.~\ref{fig:vel_accel_stats_DDPM_800}, we present the JPDFs of the normalized Cartesian components of the velocities [Fig.~\ref{fig:vel_accel_stats_DDPM_800}(a):(i)-(ix)] and accelerations [Fig.~\ref{fig:vel_accel_stats_DDPM_800}(b):(i)-(ix)] for the Lagrangian-particle trajectories in the NESSs for the NRCHNS system. The plots are organized by the value of the non-reciprocity parameter: $\mathcal{D}_{12}=0.24$ in the top rows [Figs.~\ref{fig:vel_accel_stats_DDPM_800}(a) and (b):(i)-(iii)], $\mathcal{D}_{12}=0.33$ in the middle rows [Figs.~\ref{fig:vel_accel_stats_DDPM_800}(a) and (b):(iv)-(vi)], and $\mathcal{D}_{12}=0.37$ in the bottom rows [Figs.~\ref{fig:vel_accel_stats_DDPM_800}(a) and (b):(vii)-(ix)]. Within each row, labelled by $\mathcal{D}_{12}$, we compare the ground-truth DNS data [Figs.~\ref{fig:vel_accel_stats_DDPM_800}(a) and (b): (i), (iv), (vii)] with their counterparst for synthetic trajectories generated by DDPM with 800 steps [Figs.~\ref{fig:vel_accel_stats_DDPM_800}(a) and (b): (ii), (v), (viii)]. The residuals between the two models [Figs.~\ref{fig:vel_accel_stats_DDPM_800}(a) and (b): (iii), (vi), (ix)] demonstrate that the DDPM-800 works well for these PDFs and JPDFs.


To explore the trade-off between the speed of computation and sampling accuracy in our 2D NRCHNS system, we have evaluated the deterministic DDIM sampler in a range of denoising steps ($\{5,10,25,50,100\}$). DDIM with 100 steps performs slightly worse than DDPM-$800$. If we reduce the DDIM step count below 100, we find noticeable deviations from the ground-truth distributions. The DDIM sampler struggles to match the anisotropic 4-peak structure of the velocity JPDFs when the denoising steps are reduced; instead, particularly at step sizes of 50 and below, it shows a strong Gaussian-type peak around $(0, 0)$ that is absent in the DNS ground-truth for $\mathcal{D}_{12}=0.33$ and $\mathcal{D}_{12}=0.37$. However, for $\mathcal{D}_{12}=0.24$, the residual is not as prominent, until step-sizes of 25, because of the Gaussian-type JPDF. 


Complete plots detailing the DDIM sampling, for all tested step sizes, are provided in the Appendix~\ref{subsec:appendix:vel_accel_jpdf}.

\subsection{Curvature Statistics}
\label{subsec:curvature_analysis}

The local curvature $\kappa$ of a Lagrangian-particle trajectory [in any type of flow, in general, and 2D NRCHNS flows, in particular] is
\begin{equation}
    \kappa = \frac{a_\perp}{v^2}\,,\label{eq:kappa_def}
\end{equation}
where $a_\perp$ is the magnitude of the component of the acceleration perpendicular to the trajectory and $v$ is the velocity magnitude. Thus, $\mathcal{P}_\kappa(\kappa)$, the PDF of the curvature,  can be calculated by integrating over the JPDF $\mathcal{P}_j(a_\perp, v)$:
\begin{eqnarray}
    \mathcal{P}_\kappa(\kappa) = \int_{0}^{\infty} dv \int_{-\infty}^{\infty} &da_\perp \mathcal{P}_j(a_\perp, v) \nonumber \\
    & \times \delta\left(\kappa - \frac{a_\perp}{v^2}\right)\,.
    \label{eq:p_kappa_integral}
\end{eqnarray}

In Fig.~\ref{fig:residual_plot}, we present the JPDF $\mathcal{P}_j(a_\perp,v)$ for the Lagrangian-particle trajectories in the NESSs for the NRCHNS system. The plots are organized by the value of the non-reciprocity parameter: $\mathcal{D}_{12}=0.24$ in the top row [Figs.~\ref{fig:residual_plot}(a)-(f)], $\mathcal{D}_{12}=0.33$ in the middle row [Figs.~\ref{fig:residual_plot}(g)-(l)], and $\mathcal{D}_{12}=0.37$ in the bottom row [Figs.~\ref{fig:residual_plot}(m)-(r)]. Within each row, labelled by $\mathcal{D}_{12}$, we show the JPDF of the velocity magnitude $v$ and the perpendicular acceleration $a_\perp$ [Figs.~\ref{fig:residual_plot}(a), (g), (m)] and compare it with the factorized approximation $\mathcal{P}(a_\perp)\mathcal{P}(v)$ under the assumption of statistical independence [Figs.~\ref{fig:residual_plot}(b), (h), (n)]. The residuals between the full JPDF and the factorized distribution [Figs.~\ref{fig:residual_plot}(c), (i), (o)] provide a measure of the correlation between the arguments of the JPDFs. Furthermore, the remaining columns display the cumulative distribution functions (CDFs) for the normalized velocity magnitude $v/\sigma_v$ [Figs.~\ref{fig:residual_plot}(d), (j), (p)] and the normalized perpendicular acceleration $a_\perp/\sigma_{a_\perp}$ [Figs.~\ref{fig:residual_plot}(e), (k), (q)], alongside the complementary cumulative distribution function (CCDF) for the trajectory curvature $\kappa$ [Figs.~\ref{fig:residual_plot}(f), (l), (r)]. The dashed lines indicate linear fits characterizing the scaling behaviors within the gray-shaded regimes.

We can evaluate the PDF~\eqref{eq:p_kappa_integral} approximately if we assume statistical independence of $a_\perp$ and $v$, namely,
\begin{eqnarray}
    \mathcal{P}_j(a_\perp, v) &\approx& \mathcal{P}_{a_\perp}(a_\perp)\mathcal{P}_v(v)\,, \label{eq:independence}
\end{eqnarray}
where $\mathcal{P}_{a_\perp}(a_\perp)$ and $\mathcal{P}_v(v)$ denote, respectively, the marginal PDFs of $a_\perp$ and $v$.
Figure~\ref{fig:residual_plot} shows the empirical verification of this assumption in the 2D NRCHNS system by examining the residual between the full joint distribution $\log_{10}(\mathcal{P}_j(a_\perp, v))$ and the approximation $\log_{10}(\mathcal{P}_{a_\perp}(a_\perp)\mathcal{P}_v(v))$.

Equation~\eqref{eq:independence} yields the asymptotic result 
\begin{eqnarray}
    \mathcal{P}_\kappa(\kappa) &\approx& \kappa^{-2}\,, \quad \text{as} \quad \kappa \to \infty \,,\label{eq:p_kappa_scale}
\end{eqnarray}
if we make the Ans\"atze
\begin{eqnarray}
    \mathcal{P}_{a_\perp}(a_\perp) &\approx& a_\perp^\rho h(a_\perp)\quad {\rm and} \label{eq:p_an_gen} \\
    \mathcal{P}_v(v) &\approx& v^\Upsilon g(v)\,; \label{eq:p_v_gen}
\end{eqnarray}
our 2D NRCHNS DNS results are consistent with $\rho=0$ and $\Upsilon=1$, as is evident from Figs.~\ref{fig:residual_plot} (d), (e), (j), (k), (p), and (q) [see the Appendix~\ref{subsec:appendix:kappa_derivation} for details]. 
As a consequence of Eq.~\eqref{eq:p_kappa_scale}, the Complementary Cumulative Distribution Function (CCDF), $1-\mathcal{Q}_\kappa(\kappa)$, obtained by integrating $\mathcal{P}_\kappa(\kappa)$ from $\kappa$ to $\infty$, exhibits the following power-law scaling:
\begin{eqnarray}
    1-\mathcal{Q}_\kappa(\kappa) &=& \int_{\kappa}^{\infty} \mathcal{P}_\kappa(\kappa') d\kappa'\,; \nonumber \\
    &\sim& \int_{\kappa}^{\infty} (\kappa')^{-2} d\kappa' \sim \kappa^{-1}\,. \label{eq:cdf_scale}
\end{eqnarray}
The scaling forms~\eqref{eq:p_an_gen}, \eqref{eq:p_v_gen}, and \eqref{eq:cdf_scale} are in good agreement with our DNS ground truth over several decades [see the gray-shaded regions in Figs.~\ref{fig:residual_plot} (d)-(f), (j)-(l), and (p)-(r)]. The scaling form for $\mathcal{Q}_\kappa(\kappa)$ is similar to its fluid-turbulence counterpart~\cite{xu2007curvature,scagliarini2011geometric,bhatnagar2016deviation}.

We also compare $\mathcal{Q}_{v}$, $\mathcal{Q}_{a_\perp}$ and $1-\mathcal{Q}_{\kappa}$, obtained from our DNS data, with the ones obtained from DDPM with 800 steps and DDIM with 100, 50, 25, 10, and 5 steps. We observe that only the results from DDPM-800 follow the DNS data closely; results from the DDIM samplers become progressively worse with decreasing number of denoising steps. However, the slopes that we get from linear fits to the log-log plots are consistent across all the sampling schemes. [See the Appendix~\ref{subsec:appendix:curvature_comparison} for these plots.]

Curvature provides a geometrical signature of Lagrangian-tracer dynamics. In a turbulent state, in general, and NRCHNS turbulence, in particular, tracer particles do not simply accelerate in the direction of their motion; they frequently experience strong transverse accelerations that bend their trajectories. The curvature statistics show that most trajectories have moderate curvature, whereas a small fraction of trajectories exhibit extremely sharp bending. These rare, high-curvature events are associated with coherent structures, such as vortices or the lanes that appear in non-reciprocal dynamics. Such events play an important role in high-order Lagrangian statistics, which we study via high-order moments
of velocity increments in Sec.~\ref{subsec:structure_flatness_zeta}.

\begin{figure*}[htp]
    \centering
    \begin{tikzpicture}
        \node[anchor=south west,inner sep=0] (image) at (0,0) {\includegraphics[width=1.0\linewidth]{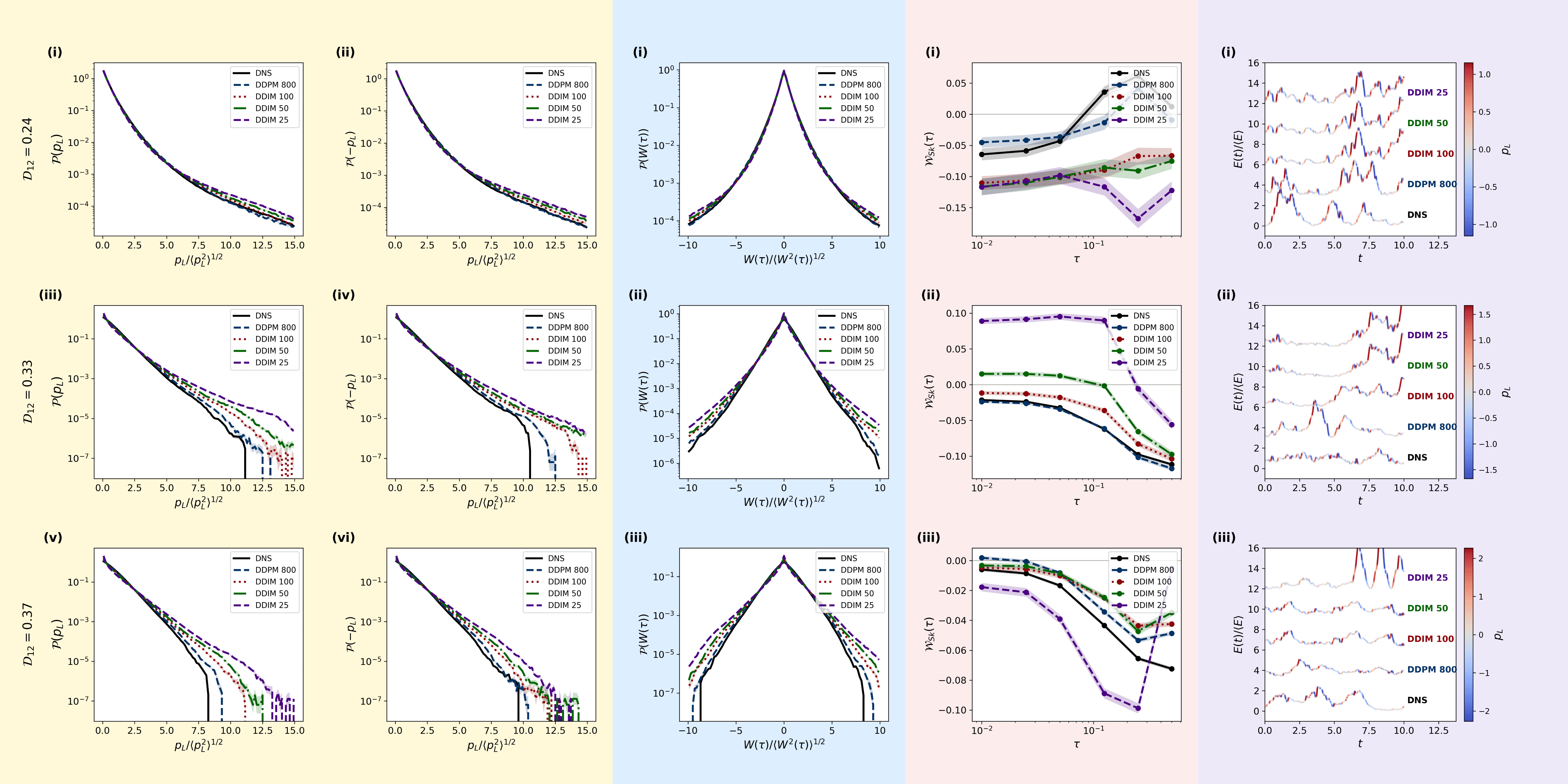}};
        
        \begin{scope}[x={(image.south east)},y={(image.north west)}, every node/.style={font=\tiny}]
            
            \node[anchor=north west] at (0.001, 0.99) {\textbf{(a)}};
            \node[anchor=north west] at (0.386, 0.99) {\textbf{(b)}};
            \node[anchor=north west] at (0.57, 0.99) {\textbf{(c)}};
            \node[anchor=north west] at (0.76, 0.99) {\textbf{(d)}};

        \end{scope}
    \end{tikzpicture}
    \caption{Statistical evaluation of the energy fluctuations in the NRCHNS system, comparing a baseline Direct Numerical Simulation (DNS, solid black line) against diffusion model. The figure is divided into four groups: \textbf{(a)} the PDFs $\mathcal{P}(p_L)$ and $\mathcal{P}(-p_L)$ of power $p_L$, which represents the rate of change of energy; \textbf{(b)} show the energy increment PDFs $\mathcal{P}(W(\tau))$ for an energy increment $W(\tau)$ at a time delay of $\tau = 0.05$; \textbf{(c)} skewness of energy increments $\mathcal{W}_{Sk}(\tau)$ across various time delays $\tau$, used to measure the asymmetry in how particles gain or lose energy; and \textbf{(d)} sample trajectories of the normalized energy $E(t)/\langle E \rangle$ over time $t$, each separated by 3 units on the vertical axis. The shaded regions in \textbf{(a)}, \textbf{(b)} and \textbf{(c)} represent the standard error of the mean, estimated by partitioning each dataset into 100 independent sub-batches.}
    \label{fig:irreversibility}
\end{figure*}
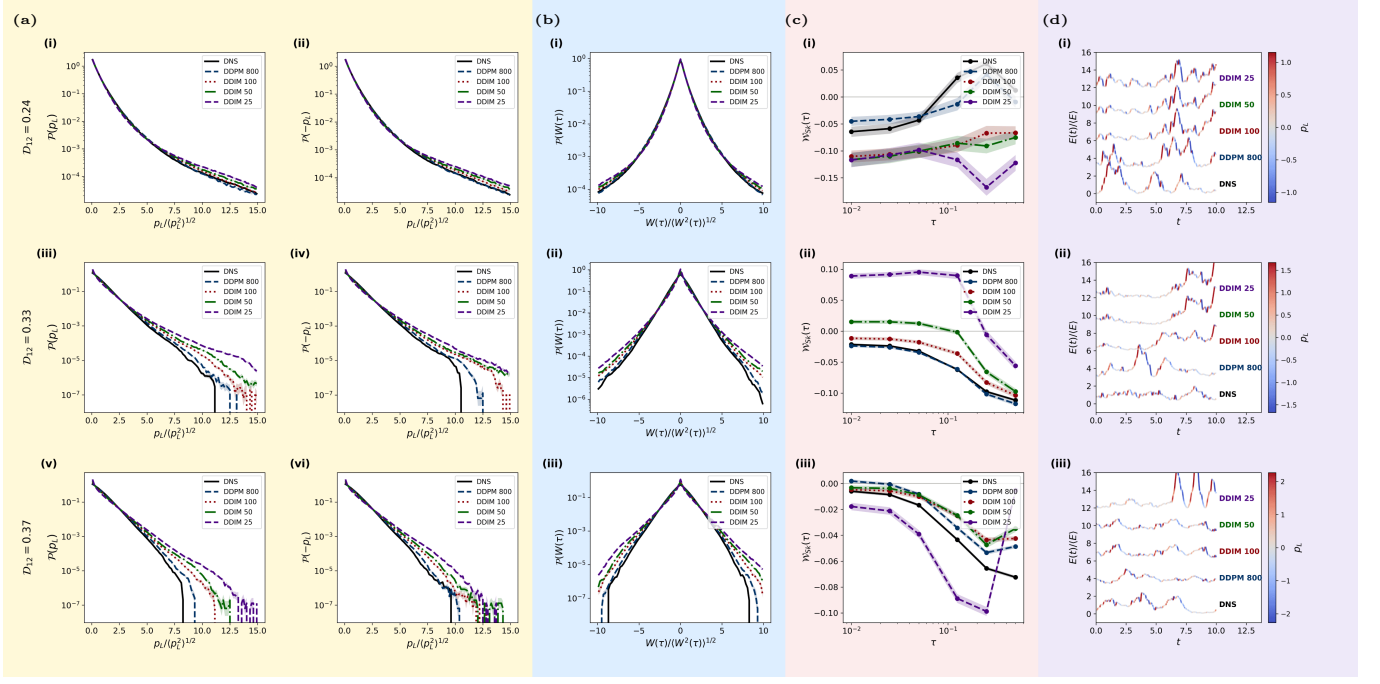

\subsection{Lagrangian Velocity Increments, Structure Functions, Flatness, and the Local Scaling Exponents}
\label{subsec:structure_flatness_zeta}

\paragraph*{Velocity Increments:}
We first define the Lagrangian-velocity increments 
\begin{equation}
    \delta_\tau v_i(t)=v_i(t+\tau)-v_i(t)\,,
\end{equation}
where $i\in\{x, y\}$. It is instructive to obtain the PDFs of these increments for the NRCHNS system. 
The solid lines in Fig.~\ref{fig:vel_inc_pdf}, from our DNSs, show that the tails of these PDFs are more pronounced for $\mathcal{D}_{12}=0.24$ (first column) than they are for $\mathcal{D}_{12}=0.33$ (second column) and $\mathcal{D}_{12}=0.37$ (third column); the top and bottom rows present, respectively, data for the $x$ and $y$ components of the velocity increments.

We then evaluate the generative fidelity of the trained diffusion models by examining their ability to capture these PDFs. Our investigations show that DDPM sampling, with $800$ steps, and DDIM sampling, with $100$, $50$, and $25$ steps, capture the non-Gaussian tails in the velocity increment PDFs; but the deviations from their DNS counterparts, especially in the tails of the PDFs, increase as the number of sampling steps decreases. These deviations become very large if the sampling steps are drastically reduced to $10$ and $5$ with a DDIM sampler. Hence, we do not include the DDIM sampler with $10$ and $5$ denoising steps in our analyses.\\

\paragraph*{Structure Functions:} The order-$p$ Lagrangian-velocity structure functions are
\begin{equation}
    S_\tau^{(p)}(v_i)=\langle(\delta_\tau v_i)^p\rangle \,,
\end{equation}
where $i \in \{x, y\}$. In Fig.~\ref{fig:structure_fn_flatness_local_slope} (a) we give log-log plots 
of $S_\tau^{(p)}$ vs $\tau$; the solid lines indicate our DNS results. We present 
data for $p=2$ and $4$, for $\mathcal{D}_{12}=0.24$,
[Figs.~\ref{fig:structure_fn_flatness_local_slope}(a): (i), (iv)], 
and for $p=2, 4,$ and $6$ for $\mathcal{D}_{12}=0.33$ [Figs.~\ref{fig:structure_fn_flatness_local_slope}(a): (ii), (v)] and $\mathcal{D}_{12}=0.37$ [Figs.~\ref{fig:structure_fn_flatness_local_slope}(a): (iii), (vi)] for DNS. These plots also contain data that we have obtained for these structure functions from our DDPM and DDIM. 

The synthetic trajectories, which we generate using the full 800-step DDPM process, demonstrate excellent agreement with those from our DNS ground truth for $\mathcal{D}_{12}=0.33$ and $0.37$; there are slight deviations for $\mathcal{D}_{12} = 0.24$. 
We can accelerate the sampling process using the DDIM; this is robust for $100$, $50$ and $25$ sampling steps (with slight deviations).
\\
\paragraph*{Flatness:} We obtain the flatnesses
\begin{equation}
    F_\tau^{(4)}(v_i)=S_\tau^{(4)}(v_i)/[S_\tau^{(2)}(v_i)]^2\,,
\end{equation}
where $i\in{x,y}$; we present lin-log plots of $F_\tau^{(4)}(v_i)$ in Fig.~\ref{fig:structure_fn_flatness_local_slope}(b); solid lines are used for data from our DNSs for $\mathcal{D}_{12}=0.24$ (first column), $\mathcal{D}_{12}=0.33$ (second column), and $\mathcal{D}_{12}=0.37$ (third column).
At intermediate values of $\tau$, these flatnesses exhibit maxima, which are more prominent for $\mathcal{D}_{12}=0.33$ and $\mathcal{D}_{12}=0.37$
than for $\mathcal{D}_{12}=0.24$.

The plots in Fig.~\ref{fig:structure_fn_flatness_local_slope}(b) also show the flatnesses that we obtain from synthetic trajectories, which we generate using our diffusion models; we compare $F_\tau^{(4)}(v_i)$ with their DNS counterparts. We find that DDPM-800 reconstructs the peak in the flatnesses accurately for $\mathcal{D}_{12}=0.33$ and $\mathcal{D}_{12}=0.37$; by contrast, the errors in DDIM sampling (with denoising steps 100, 50, and 25) are prominent.
Furthermore, the DDPM-800 $F_\tau^{(4)}(v_i)$  exhibits a greater deviation from its DNS counterpart, for $\mathcal{D}_{12}=0.24$, especially as $\tau\rightarrow0$. \footnote{Counterintuitively, the deviation from the DNS graph reduces with a decrease in the number of denoising steps in the DDIM sampler. Reducing the number of denoising steps is known to introduce error to the sampling process. It may be this error that is causing some spurious turbulence in the sampled data to appear in the denoised data, but a more extensive analysis is required to confirm the hypothesis.}\\

\paragraph*{Local-slope analysis:} The local scaling exponent $\zeta_p$ characterizes the power-law scaling of  $S_\tau^{(p)}(v_i)$ and provides a stringent diagnostic test for the distributions that we obtain from our diffusion model [cf. Ref.~\cite{li2024generative}]. This exponent is defined as 
\begin{equation}
    \zeta_p(\tau)\equiv\frac{d\log S_\tau^{(p)}}{d\log\tau}\,.
    \label{eq:zetap}
\end{equation}
It is often advantageous to use the \textit{extended-self-similarity} (ESS) procedure~\cite{benzi1993extended,pandit2009statistical,li2024generative} to obtain the exponent ratio
\begin{equation}
    \frac{\zeta_p}{\zeta_2}=\frac{d\log S_p}{d\log S_2}\,,
    \label{eq:ESS}
\end{equation}
because, if we plot $S_\tau^{(p)}$ versus $S_\tau^{(2)}$, the range over which power-law scaling is observed is extended.

We present $\frac{d\log S_p}{d\log S_2}$ vs $\tau$ in Fig.~\ref{fig:structure_fn_flatness_local_slope}(c); again, solid lines indicate data from our DNSs for $\mathcal{D}_{12}=0.24$ (first column), $\mathcal{D}_{12}=0.33$ (second column), and $\mathcal{D}_{12}=0.37$ (third column). At low $\tau$, this ratio yields 
$\frac{\zeta_p}{\zeta_2}=p/2$, a result that follows from a low-$\tau$ Taylor expansion of the velocity increments. As $\tau$ increases, this ratio shows  pronounced maxima, at intermediate values of $\tau$, especially for $\mathcal{D}_{12}=0.33$ and $0.37$; the maxima are not as prominent for $\mathcal{D}_{12}=0.24$. We now check if these features in the ground-truth DNS plots of $\frac{d\log S_p}{d\log S_2}$ vs $\tau$ are reproduced by an ESS analysis with synthetic Lagrangian trajectories that we obtain using DDPM-800 and DDIM (with 100, 50, 25 denoising steps). Consider first $\mathcal{D}_{12}=0.24$ for which all these diffusion models show deviations from the DNS results, and display maxima that are more pronounced than their DNS counterparts. By contrast, if $\mathcal{D}_{12}=0.33$ or $0.37$, the DDPM-$800$ results match their DNS counterparts closely;
and the DDIM sampler struggles to match the DNS plots as the number of denoising steps are reduced from 100 to 25.

We end this Section with some qualitative observations. As a tracer enters, circulates within, or leaves a coherent structure, its velocity direction can change rapidly, producing intermittent and unusually large velocity increments. This provides a natural physical connection between the population of rare, strongly curved trajectories and the observed overshoot of the local ESS exponent ratios $\zeta_p/\zeta_2$ [see, e.g., Ref.~\cite{biferale2004multifractal,xu2007curvature,hengster2024effects}]. At very short time lags, tracer motion remains approximately ballistic, and $\zeta_p/\zeta_2=p/2$. At intermediate time lags, the influence of coherent vortical structures becomes strongest. The resulting intermittent trajectory bending enhances the contribution of rare events to high-order structure functions. Our plots are similar to those obtained for fluid~\cite{li2024generative} and bacterial turbulence~\cite{kiran2023irreversibility} except for the peak at intermediate values of $\tau$; this peak is a signature of NRCHNS turbulence.


\subsection{Irreversibility in the NRCHNS system}
\label{subsec:irrev_analysis}

If we were to play a movie, both forward and backward, of Lagrangian particles in a turbulent flow, we would be hard put to tell the forward- and backward-running movies from each other merely by visual inspection. To uncover the irreversibility of turbulent flows, we must use the asymmetry of the Lagrangian energy increments $W(\tau)$ and the Lagrangian power $p_L(t)$. Such irreversibility has been explored
for fluid turbulence~\cite{xu2014flight,bhatnagar2018heavy}, for a hydrodynamical model for bacterial turbulence~\cite{kiran2023irreversibility}, and superfluid turbulence~\cite{shukla2023inertial}. We now present a study of the irreversibility of NRCHNS turbulence; and we also show how this can be captured by the diffusion models we use.  Specifically, we evaluate
\begin{eqnarray}
    W(t,\tau)&\equiv&E(t+\tau)-E(t)\,,\\
    {\rm and}\quad p_L(t)&\equiv&\frac{dE(t)}{dt}\,,
\end{eqnarray}
where $E(t)$ is the kinetic energy of a tracer particle at time $t$. We also measure the skewness of the energy increments
\begin{equation}
    \mathcal{W}_{Sk}(\tau)\equiv\frac{\langle W^3(\tau)\rangle}{\langle W^2(\tau)\rangle^{3/2}}\,.
\end{equation}
In Figs.~\ref{fig:irreversibility} (a), (b), (c), and (d) we present, respectively, PDFs of $p_L(t)$, PDFs of $W(t,\tau)$, the skewness 
$\mathcal{W}_{Sk}(\tau)$, and illustrative trajectories of the normalized energy $E(t)/\langle E \rangle$ versus time $t$ (each separated by 3 units along the vertical axis); the plots in rows one, two, and three are for $\mathcal{D}_{12}=0.24$, $\mathcal{D}_{12}=0.33$, and $\mathcal{D}_{12}=0.37$, respectively.
The DNS ground-truth plots in Figs.~\ref{fig:irreversibility} (a)-(c) are indicated by solid lines. From these plots we see that The Lagrangian dynamics of the non-reciprocal turbulent state exhibit a distinct temporal asymmetry. The skewness of the Lagrangian energy increments is close to zero at short time lags but becomes increasingly negative as the time lag increases. Negative skewness indicates an asymmetric distribution with a pronounced tail towards intense energy-loss events, i.e., on average, particles gain gain energy slowly but lose it rapidly as in conventional fluid turbulence~\cite{xu2014flight,bhatnagar2018heavy}, where  such rapid energy-loss events have been associated preferentially with strain-dominated regions of the flow \cite{bhatnagar2018heavy}; this behavior is distinct from that observed in bacterial turbulence~\cite{kiran2023irreversibility}.

Next we compare our DNS ground-truth results with those that we obtain with synthetic trajectories from our diffusion models. We observe that DDPM-800 aligns the best with the DNS plots for $\mathcal{P}(p_L)$ and $\mathcal{P}(-p_L)$, followed by DDIM-100, DDIM-50, and DDIM-25. This result is consistent for  all values of $\mathcal{D}_{12}$, but is more prominent for $\mathcal{D}_{12}=0.33$ and $0.37$. If $\mathcal{D}_{12}=0.24$, we find that the difference between DDPM-800 and DDIM-100 is negligible, but DDIM-50 and DDIM-25 show larger deviations.
This deviation from the true PDF with decreasing number of steps is also present in $\mathcal{P}(W(\tau=0.05))$ and $\mathcal{W}_{Sk}(\tau)$. Figure~\ref{fig:irreversibility}(c) reveals that $\mathcal{W}_{Sk}(\tau)$ is extremely sensitive to the number of denoising steps. DDPM-800 maintains a reasonable accuracy compared to the DNS ground truth for all three values of $\mathcal{D}_{12}$; DDIM-100 performs best for $\mathcal{D}_{12}=0.33$ compared to $\mathcal{D}_{12}=0.37$, but is erroneous for $\mathcal{D}_{12}=0.24$. DDIM-50 and DDIM-25 also show the same performance trends, but with worse results compared to those from DDPM-800 and DDIM-100. Our analysis brings out clearly that $\mathcal{W}_{Sk}(\tau)$ is a stringent metric for the evaluation of diffusion models for the trajectories of Lagrangian particles in turbulent flows.

\subsection{Model Transferability with $\mathcal{D}_{12}$}
\label{subsec:model_transferability}

It is important to examine the feasibility of transfer learning via model fine-tuning across different values of $\mathcal{D}_{12}$ values. Specifically, we fine-tune a model, which has already been trained on an initial dataset $\mathfrak{D}_{initial}$ for $3.5\times10^5$ iterations, on a target dataset $\mathfrak{D}_{target}$ for $10^5$ iterations; we then compare its performance against the model trained on $\mathfrak{D}_{target}$ from scratch (for $3.5\times10^5$ iterations) with different statistical measures, namely, the velocity-increment PDFs, the structure functions and flatness, the exponent ratios $\zeta_p/\zeta_2$, and the PDFs of energy increments and their skewnesses. Fine-tuning helps in reducing the training time by utilizing a pre-trained model for other related datasets. In particular, we find that the diffusion model, trained with $\mathcal{D}_{12}=0.37$, gives accurate results when fine-tuned for $\mathcal{D}_{12}=0.33$. However, we find that fine-tuning completely fails to match the accuracy of the ground-truth DNS data for the flatness and the exponent ratios when the target dataset has $\mathcal{D}_{12}=0.24$; the model performs slightly better when pre-trained on $\mathcal{D}_{12}=0.33$ instead of $\mathcal{D}_{12}=0.37$. 
These results can be attributed to the data distributions being close for $\mathcal{D}_{12}=0.33$ and $\mathcal{D}_{12}=0.37$. Similarly, the data distribution of $\mathcal{D}_{12}=0.24$ is closer to that of $\mathcal{D}_{12}=0.33$ as compared to that of $\mathcal{D}_{12}=0.37$.
The analysis in this subsection is based on 10,240 samples obtained from the fine-tuned diffusion models; all plots related to fine-tuning are presented in the Appendix~\ref{subsec:appendix:finetuning_comparison}.

\section{Discussion and Conclusions}
\label{sec:Discussions}

We have presented the first investigation of the statistical properties of Lagrangian-tracer particles that are advected by non-reciprocal binary-fluid flows, as modeled by the NRCHNS framework. Our study brings out several remarkable results that we recapitulate briefly. We find that the PDFs for $a_\parallel$ and $a_\perp$ are qualitatively similar to their fluid-turbulence counterparts~\cite{Biferale_DM_2024}, in as much as they are close to Guassian PDFs near their peaks, but they have fat tails. By contrast, the velocity-component PDFs and JPDFs  are bimodal, unlike their 2D-fluid-turbulence counterparts~\cite{perlekar2009statistically,pandit2017overview}. This bimodality is a consequence of lane-type structures~\cite{reeves2021emergence} in pseudocolor plots of the Eulerian-velocity components; these also show bimodal PDFs. We then uncover multiscaling in the Lagrangian framework by computing Lagrangian velocity increments, their structure functions and flatnesses, and the exponent ratios $\zeta_p/\zeta_2$. Such Lagrangian studies are well-known in conventional fluid turbulence~\cite{biferale2004multifractal,arneodo2008universal,benzi2010inertial}; recently, these types of investigations have been carried out for a model for bacterial turbulence~\cite{kiran2025onset,pandit2025particles}; but they have remained unexplored, hitherto, in non-reciprocal hydrodynamics. Last, but not least, we examine the unexplored irreversibility of NRCHNS flows by quantifying the asymmetry of the Lagrangian energy increments $W(\tau)$ and the Lagrangian power $p_L(t)$ and using methods developed for fluid turbulence~\cite{xu2014flight,bhatnagar2018heavy}, a model for bacterial turbulence~\cite{kiran2023irreversibility}, and superfluid turbulence~\cite{shukla2023inertial}. 

Furthermore, we have shown how to use generative diffusion models to obtain synthetic Lagrangian trajectories for the NRCHNS system and then to assess how effectively they can emulate the Lagrangian statistics of this system. To the best of our knowledge, this is the first time that such models have been used for a non-reciprocal system~\footnote{ When evaluating the efficacy of these models, it is critical to distinguish the physical time-evolution of the system from the data generation process in diffusion models. Sampling synthetic data from diffusion models occurs via a predefined stochastic differential equation, rather than by autoregressively integrating the exact physical trajectories. Consequently, the generative process is not hindered by the fundamentally non-conservative and non-reciprocal dynamics of the NRCHNS equations.}. 
In particular, we have demonstrated that both DDPM-800 and DDIM-100 models yield synthetic Lagrangian trajectories whose statistical properties are very close to those from our DNS ground truth, as is evident from our extensive benchmarking against our DNS results for acceleration, velocity-component, and curvature PDFs and for Lagrangian velocity increments, their structure functions and flatnesses, and the exponent ratios $\zeta_p/\zeta_2$.
However these diffusion models exhibit slight inaccuracies in the case $\mathcal{D}_{12}=0.24$, revealing a potential performance gap in learning highly subtle flow features.



The presence of singularities and highly localized peaks in the target data distribution is a genuine challenge, which may arise (potentially) in modeling, via generative AI, the complicated distributions we consider, Neural networks generally struggle with such sharply defined distributions and tend to learn a smooth functional fit instead. Overcoming this limitation is an active area of research in generative AI.
Beyond these representational challenges, a major advantage of generative diffusion models lies in their capacity for highly parallelized data generation across Graphics Processing Units (GPUs). Given adequate computational infrastructure, this enables massive acceleration of data synthesis compared to traditional Direct Numerical Simulations (DNS), which face fundamental bottlenecks because of their sequential time-stepping. Large-scale GPU-based computations are resource-intensive, so a full realization of this advantage requires the development of efficient model architectures that minimize the training overhead. Equally important is the acceleration of the inference phase, e.g., by the reduction of the Number of Function Evaluations (NFEs) via fewer denoising steps. There are proposals for  multiple accelerated sampling schemes and more efficient modeling approaches in the literature, such as Denoising Diffusion Implicit Models (DDIMs), flow matching~\cite{lipman2023flow}, denoising diffusion generative adversarial networks (DDGANs)~\cite{xiao2022tackling}, and consistency models~\cite{song2023consistency}; these techniques still need to be tested extensively across all statistical benchmarks to confirm their ability to model, with high accuracy, the physical systems we consider. Our empirical observations indicate that reducing denoising steps (e.g., via a DDIM sampler) can significantly degrade the accuracy, in certain sensitive statistical measures, as compared to a full 800-step DDPM baseline. Therefore, an important direction for future research is to conduct extensive tests of existing approaches against various statistical benchmarks and develop data-targeted modifications of these accelerated samplers. This should ensure that a reduction in sampling steps does not compromise the statistical fidelity of these models.


\section{Acknowledgments}
   We thank N.B. Padhan and K.V. Kiran for discussions, the Anusandhan National Research Foundation (ANRF), the Science and Engineering Research Board (SERB), for support,  and the Supercomputer Education and Research Centre (IISc), for computational resources. 
\section{Appendix}
\subsection{Non-dimensional Parameters}\label{subsec:appendix:non_dim_parameters}
\begin{table}[h]
    \centering
\renewcommand{\arraystretch}{1.3} 
\setlength{\tabcolsep}{6pt} 
\begin{tabular}{|c|c|c|c|c|c|c|} 
 \hline
  $\mathcal{D}_{12}$ & $\nu$  & $We$ & $Cn$ & $Pe$ & $Re$  \\ 
 \hline
 0.24 & $10^{-4}$    & 0.1480 & 0.024& 35.50  &  5786.7 \\ 
 \hline
 0.33 & $10^{-4}$    & 0.3948 & 0.020 & 63.45  & 10341.2 \\ 
 \hline
 
 0.37 & $10^{-4}$   & 0.5440 & 0.0197 & 75.72 &  12340.1 \\ 
 \hline
 
\end{tabular}
\caption{Table of the values of the non-dimensional parameters, for different values of $\mathcal{D}_{12}$ [column 1]: Cahn number $Cn_1 \equiv \epsilon_1/L $, $Cn_2 \equiv \epsilon_2/L $, with $Cn=Cn_1=Cn_2$, Weber number $We_1 \equiv L u_{rms}^2/\sigma_1$, $We_2 \equiv L u_{rms}^2/\sigma_2$, with $We=We_1=We_2$, P\'eclet number $ Pe_1 \equiv Lu_{rms} \epsilon_1/M_1\sigma_1$, $ Pe_2 \equiv Lu_{rms} \epsilon_2/M_2\sigma_2$, with $Pe=Pe_1=Pe_2$  and friction $\alpha'=\alpha L /u_{rms}$, Reynolds number $Re \equiv L u_{rms}/{\nu}$, where $u_{rms}$ is the root-mean-square velocity, $L$ the integral length scale. We use $\alpha = 0$.}
 \label{tab:param}
\end{table}
\begin{algorithm}[htbp]
\caption{U-Net Forward Pass $\epsilon_\theta(\mathcal{V}_n, n)$}
\label{alg:unet_forward}
\begin{algorithmic}[1]
\Require Noisy trajectory $\mathcal{V}_n \in \mathbb{R}^{2 \times L_{max}}$, discrete diffusion step $n \in \mathbb{Z}^+$.
\State $n_{emb} \gets W_{n2}\varphi(W_{n1}\text{PE}(n)+b_{n1})+b_{n2}$
\State $h \gets W_{init} *_1^3 \mathcal{V}_n$
\State $S_{kip} \gets \text{Empty Stack}$
\State $S_{kip}\text{.push}(h)$
\Statex
\State \textbf{// Encoder}
\For{$i = 1$ \textbf{to} $N_{stages}$}
    \For{$j = 1$ \textbf{to} $M_{res}$}
        \State $h \gets \text{ResBlock}_{enc,i,j}(h, n_{emb})$
        \If{stage $i$ requires attention}
            \State $h \gets \text{Attn}_{enc,i,j}(h)$
        \EndIf
        \State $S_{kip}\text{.push}(h)$
    \EndFor
    \If{$i < N_{stages}$}
        \State $h \gets D_{sample_i}(h)$ \Comment{Strided Downsampling}
        \State $S_{kip}\text{.push}(h)$
    \EndIf
\EndFor
\Statex
\State \textbf{// Bottleneck}
\State $h \gets \text{ResBlock}_{bottleneck,1}(h, n_{emb})$
\State $h \gets \text{Attn}_{bottleneck}(h)$
\State $h \gets \text{ResBlock}_{bottleneck,2}(h, n_{emb})$
\Statex
\State \textbf{// Decoder}
\For{$i = N_{stages}$ \textbf{down to} $1$}
    \For{$j = 1$ \textbf{to} $M_{res} + 1$}
        \State $s \gets S_{kip}\text{.pop}()$
        \State $h \gets [h, s]$ \Comment{Concatenate along channel dimension}
        \State $h \gets \text{ResBlock}_{dec,i,j}(h, n_{emb})$
        \If{stage $i$ requires attention}
            \State $h \gets \text{Attn}_{dec,i,j}(h)$
        \EndIf
        \If{$i > 1$ \textbf{and} $j = M_{res} + 1$}
            \State $h \gets U_{sample_i}(h)$ \Comment{Nearest-Neighbor Upsampling}
        \EndIf
    \EndFor
\EndFor
\Statex
\State \textbf{// Final Output Projection}
\State $h \gets \varphi(\text{GN}(h))$
\State $\epsilon_\theta \gets W_{final} *_1^3 h$
\State \Return $\epsilon_\theta$
\end{algorithmic}
\end{algorithm}

\begin{figure*}[htbp]
    \centering
    \textbf{DDIM 100 Steps:} \\
    \vspace{0.1cm}
    \includegraphics[width=0.9\linewidth]{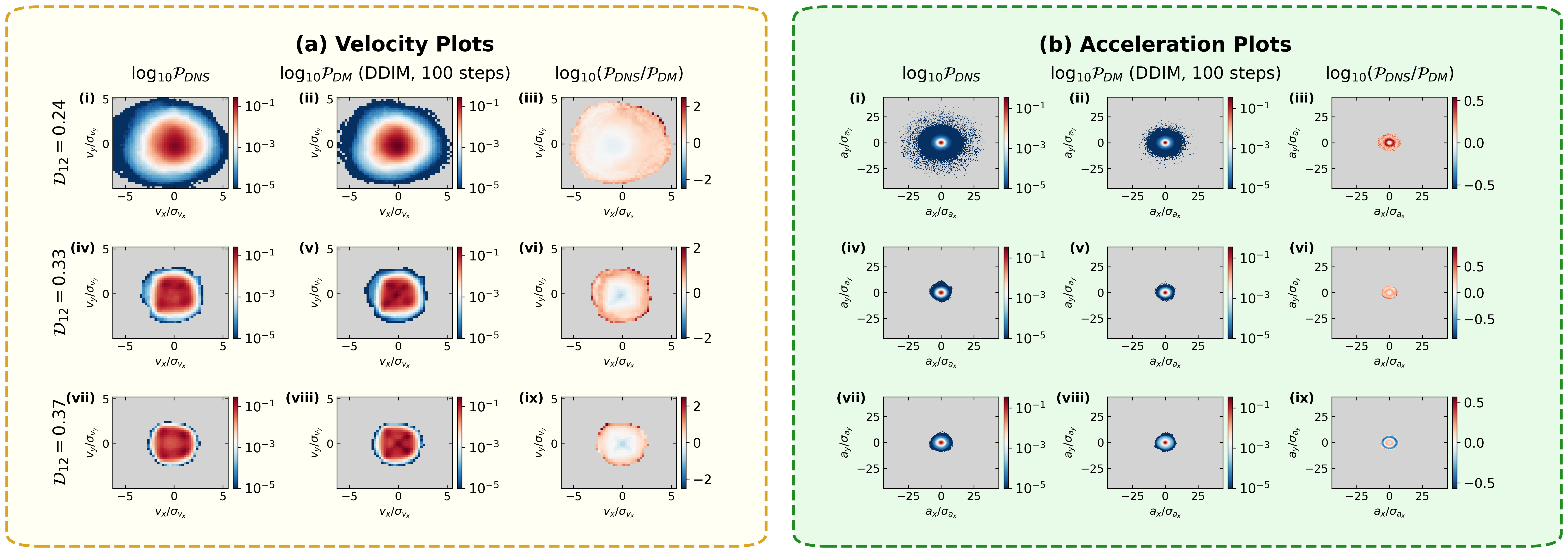}
    
    \vspace{0.4cm}
    
    \textbf{DDIM 50 Steps:} \\
    \vspace{0.1cm}
    \includegraphics[width=0.9\linewidth]{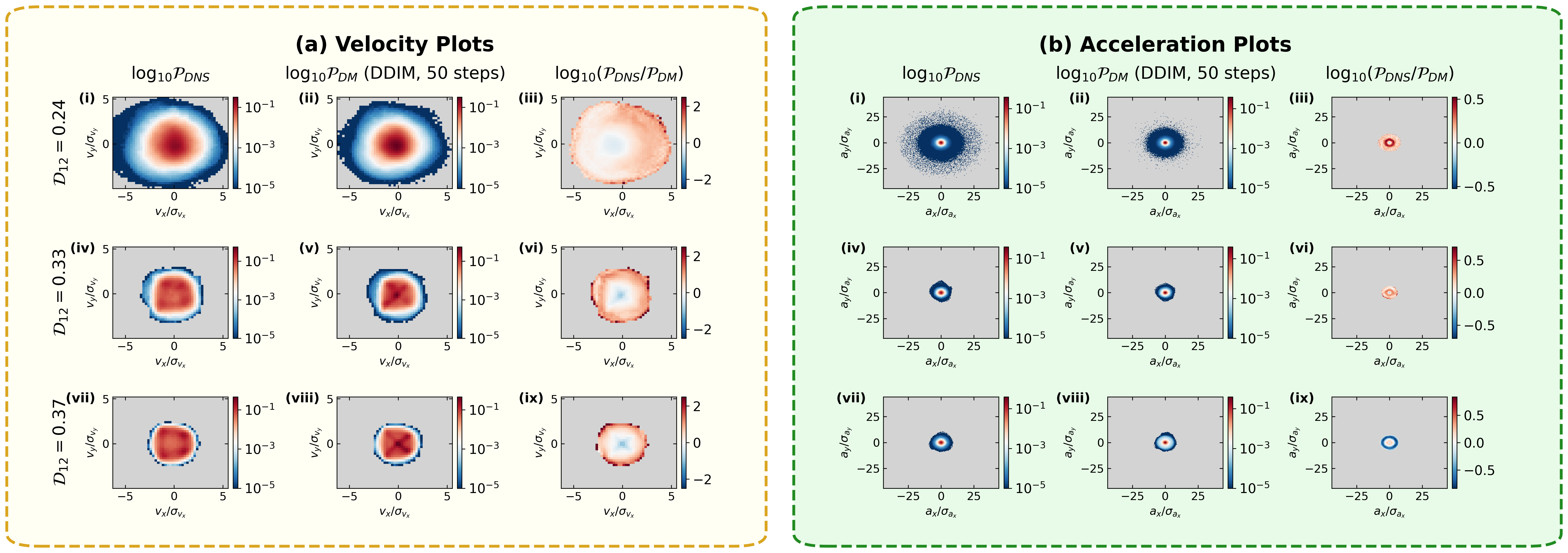}
    
    \caption{\textbf{(a) - Left} Comparison of the joint probability density functions (JPDFs) of normalized velocity components $(v_x/\sigma_{v_x}, v_y/\sigma_{v_y})$ between the ground-truth DNS data and synthetic trajectories generated by the diffusion model (DDIM sampler with 100 denoising steps: DDIM-100, and DDIM sampler with 50 denoising steps: DDIM-50).
    \textbf{(b) - Right} Comparison of the JPDFs of normalized acceleration components $(a_x/\sigma_{a_x}, a_y/\sigma_{a_y})$ from DNS data versus diffusion model (DDIM-100 and DDIM-50). The residuals, plotted as $\log_{10}(\mathcal{P}_{DNS}/\mathcal{P}_{DM})$, provide a quantitative assessment of the model's fidelity in reliably reproducing the target distribution.}
    \label{fig:ddim_higher_steps}
\end{figure*}

\begin{figure*}[htbp]
    \centering
    \textbf{DDIM 25 Steps:} \\
    \vspace{0.1cm}
    \includegraphics[width=0.85\linewidth]{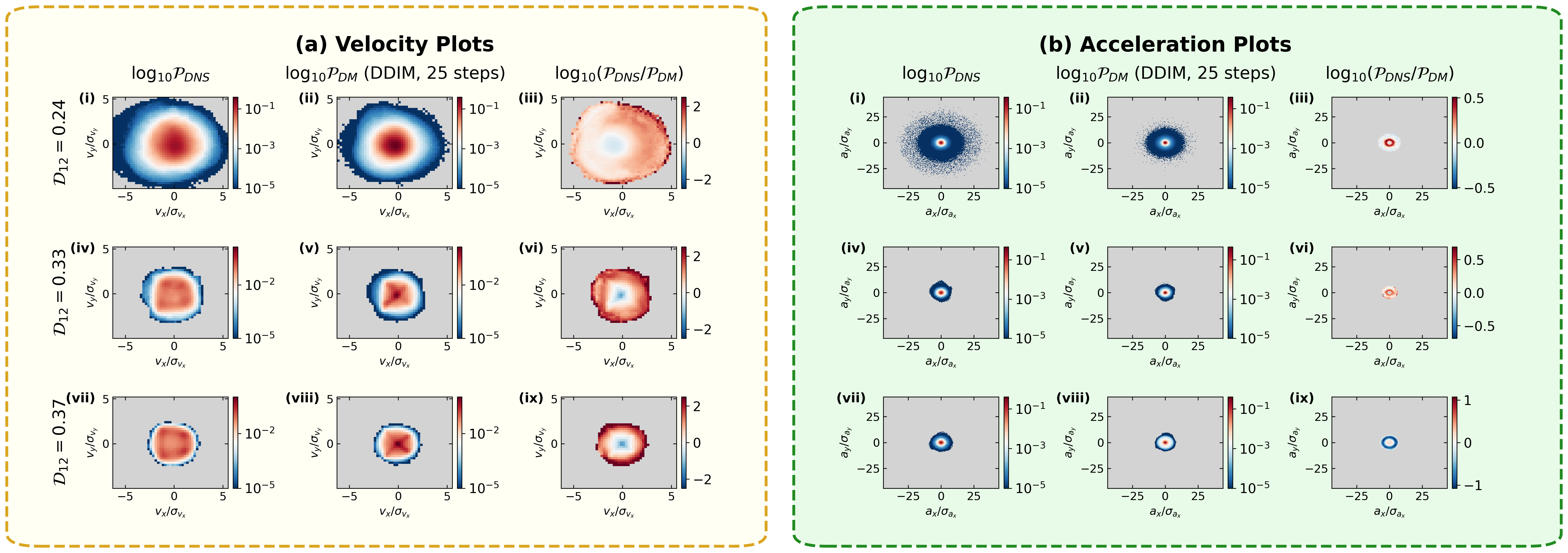}

    \vspace{0.3cm}

    \textbf{DDIM 10 steps:} \\
    \vspace{0.1cm}
    \includegraphics[width=0.85\linewidth]{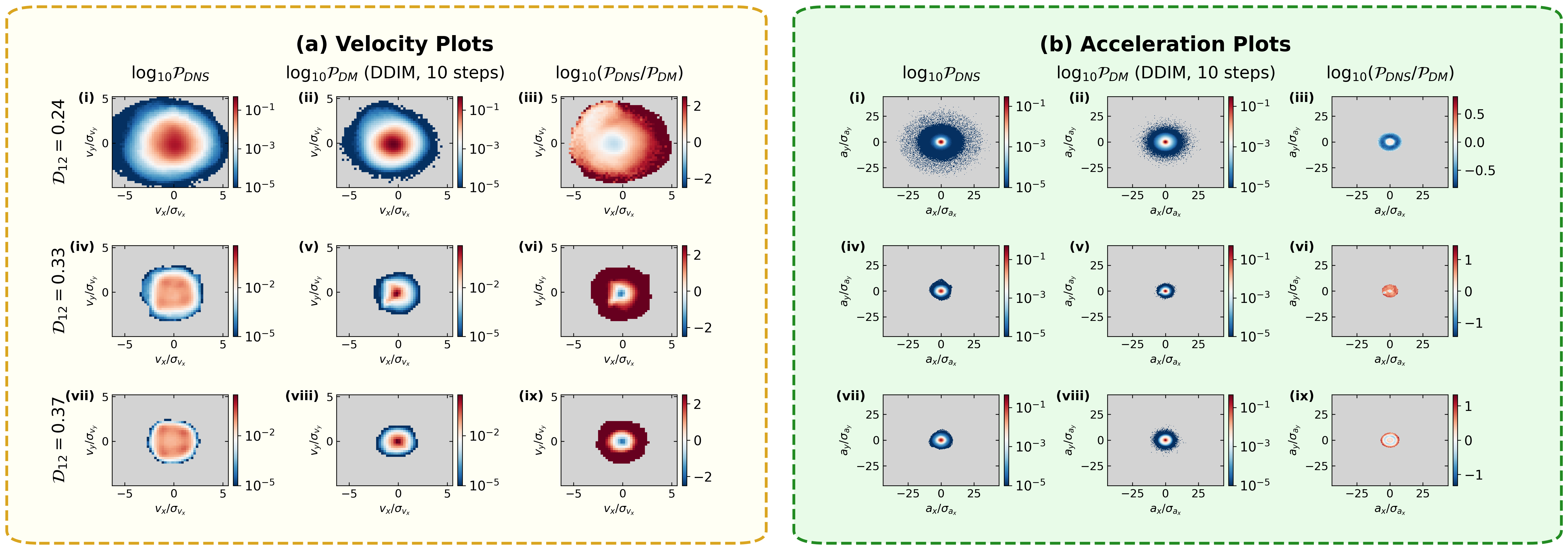}
    
    \vspace{0.3cm}

    \textbf{DDIM 5 steps:} \\
    \vspace{0.1cm}
    \includegraphics[width=0.85\linewidth]{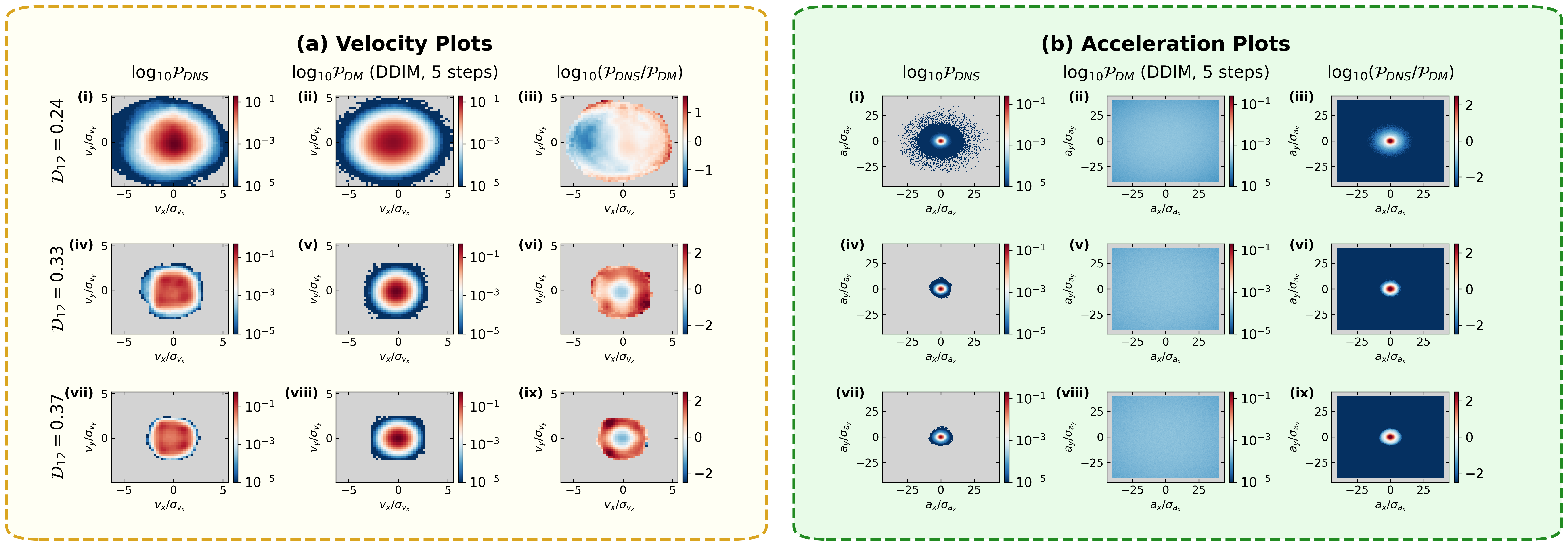}
    \caption{\textbf{(a) - Left} Comparison of the joint probability density functions (JPDFs) of normalized velocity components $(v_x/\sigma_{v_x}, v_y/\sigma_{v_y})$ between the ground-truth DNS data and synthetic trajectories generated by the diffusion model (DDIM-25, DDIM-10 and DDIM-5).
    \textbf{(b) - Right} Comparison of the JPDFs of normalized acceleration components $(a_x/\sigma_{a_x}, a_y/\sigma_{a_y})$ from DNS data versus diffusion model (DDIM-25, DDIM-10 and DDIM-5). The residuals, plotted as $\log_{10}(\mathcal{P}_{DNS}/\mathcal{P}_{DM})$, provide a quantitative assessment of the model's fidelity in reliably reproducing the target distribution.}
    \label{fig:ddim_lower_steps}
\end{figure*}

\begin{figure*}[htbp]
    \centering
    \includegraphics[width=0.9\linewidth, height=9cm]{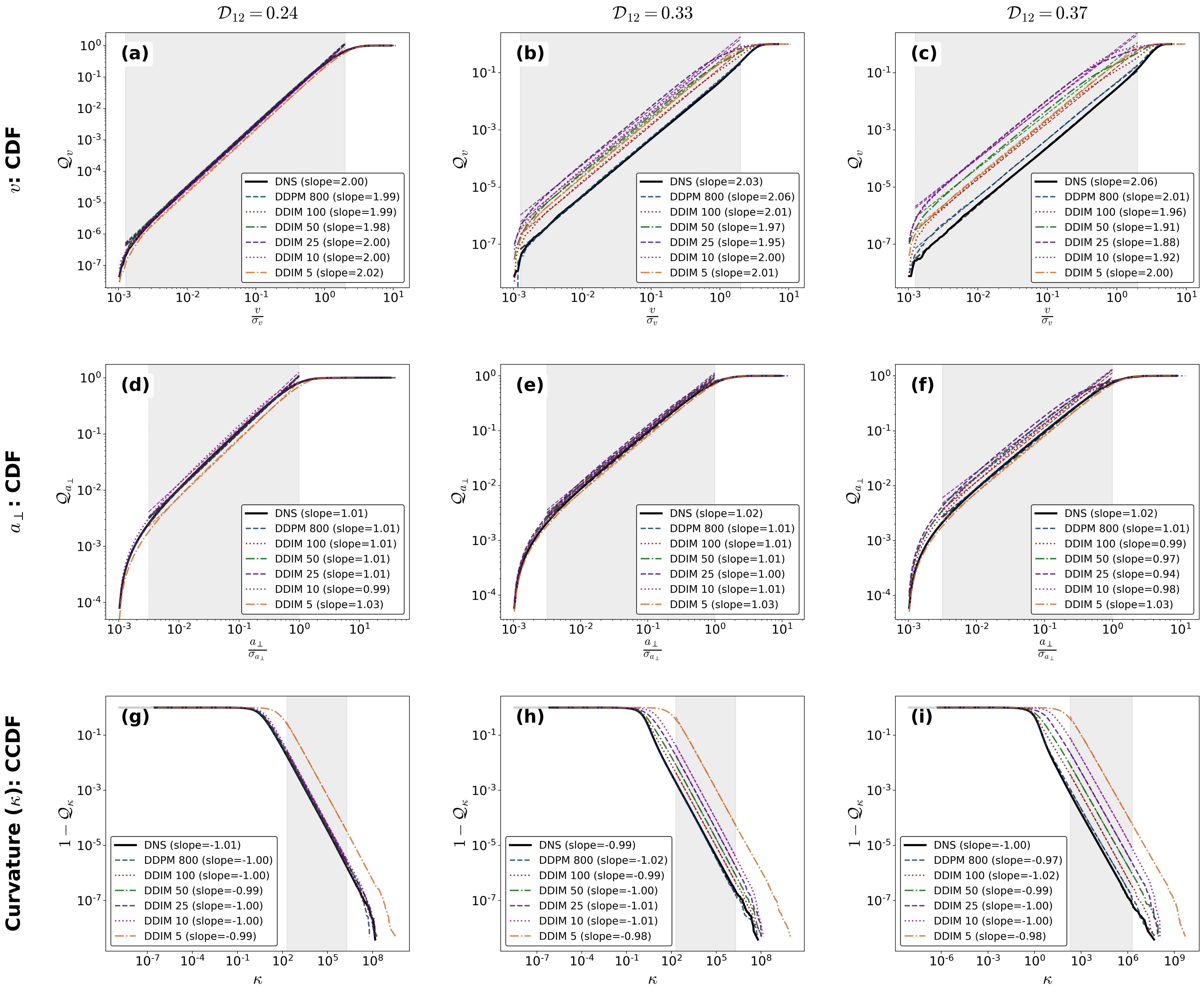}
    \caption{(a)-(c): Normalized velocity magnitude CDFs ($\mathcal{Q}_v$), (d)-(f): Normalized perpendicular acceleration magnitude CDFs ($\mathcal{Q}_{a_\perp}$), and (g)-(i): trajectory curvature CCDFs ($1-\mathcal{Q}_\kappa$), for DNS and DDPM-800, DDIM-100, DDIM-50, DDIM-25, DDIM-10, DDIM-5. Red dashed lines indicate the linear fits within the shaded asymptotic regimes.}
    \label{fig:curvature_comparison_dns_vs_dm}
\end{figure*}

\begin{figure*}[htbp]
    \centering
    \includegraphics[width=0.7\linewidth, height=8cm]{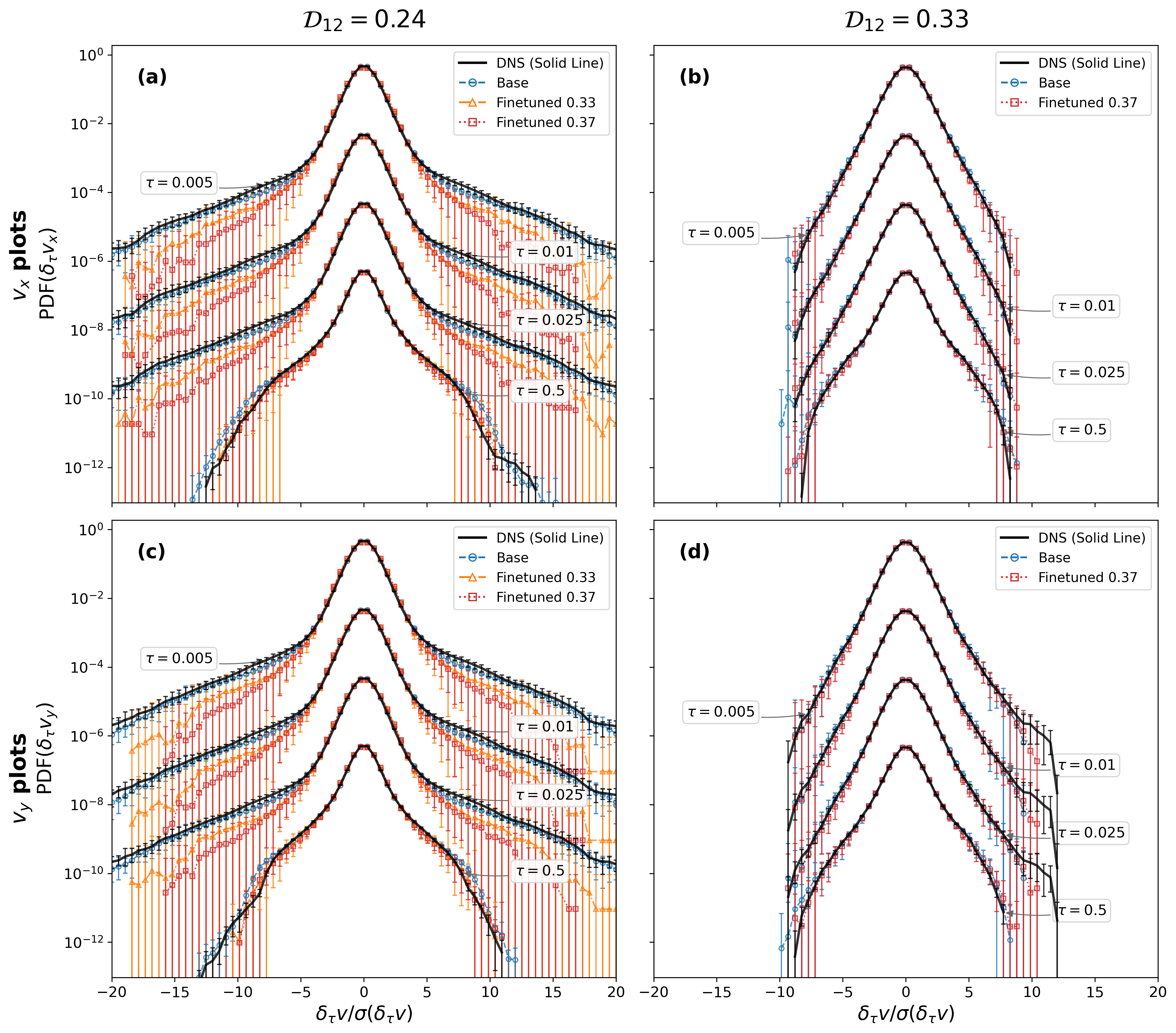}
    \caption{Probability density functions (PDFs) of Lagrangian velocity increments $\delta_{\tau}v_{i}(t)$ for the $x$ and $y$ components. ``Finetuned 0.33'' denotes that the U-Net is pre-trained on $\mathcal{D}_{12}=0.33$. Similarly ``Finetuned 0.37'' means that the U-Net is pre-trained on $\mathcal{D}_{12}=0.37$. ``Base'' denotes that the U-Net was trained on that $\mathcal{D}_{12}$ value from scratch. Training from scratch performs better than fine-tuning for $\mathcal{D}_{12}=0.24$, with the fine-tuning performing slightly better when pre-trained on $\mathcal{D}_{12}=0.33$ than on $\mathcal{D}_{12}=0.37$. Fine-tuning on $\mathcal{D}_{12}=0.37$ data performs equally good to training from scratch for $\mathcal{D}_{12}=0.33$ in reproducing the velocity increment PDF. Error bars represent the minimum and maximum values obtained after dividing the dataset into $10$ smaller sub-batches.}
    \label{fig:vel_increments_finetune}
\end{figure*}
\begin{figure*}[htbp]
    \centering
    \begin{subfigure}{0.48\textwidth}
        \centering
        \includegraphics[width=\linewidth, height=9cm]{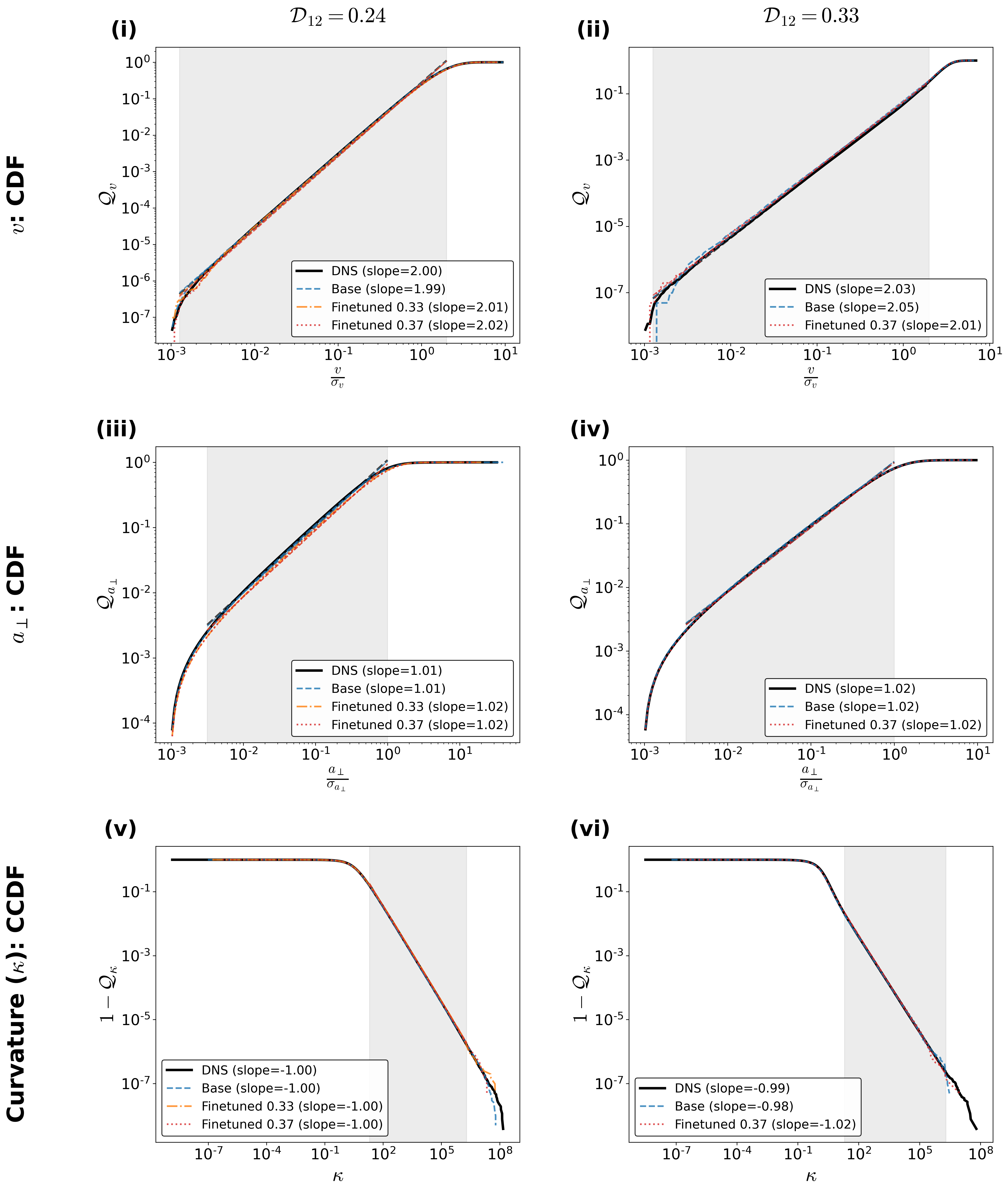}
        \caption{}
        \label{fig:finetune_kinematics}
    \end{subfigure}\hfill
    \begin{subfigure}{0.48\textwidth}
        \centering
        \includegraphics[width=\linewidth, height=9cm]{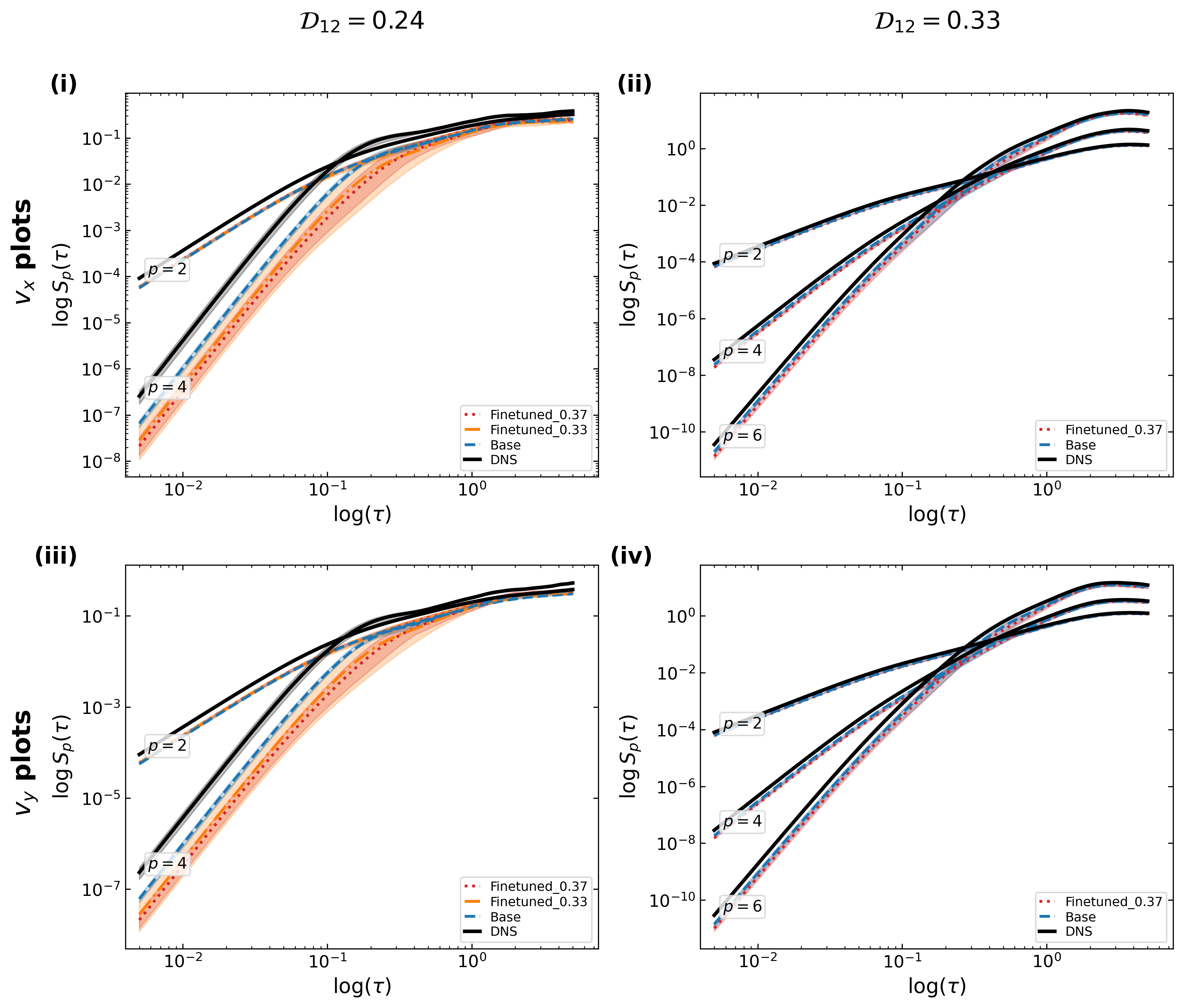}
        \caption{}
        \label{fig:finetune_structure}
    \end{subfigure}
    
    \vspace{0.3cm} 
    
    \begin{subfigure}{0.48\textwidth}
        \centering
        \includegraphics[width=\linewidth, height=9cm]{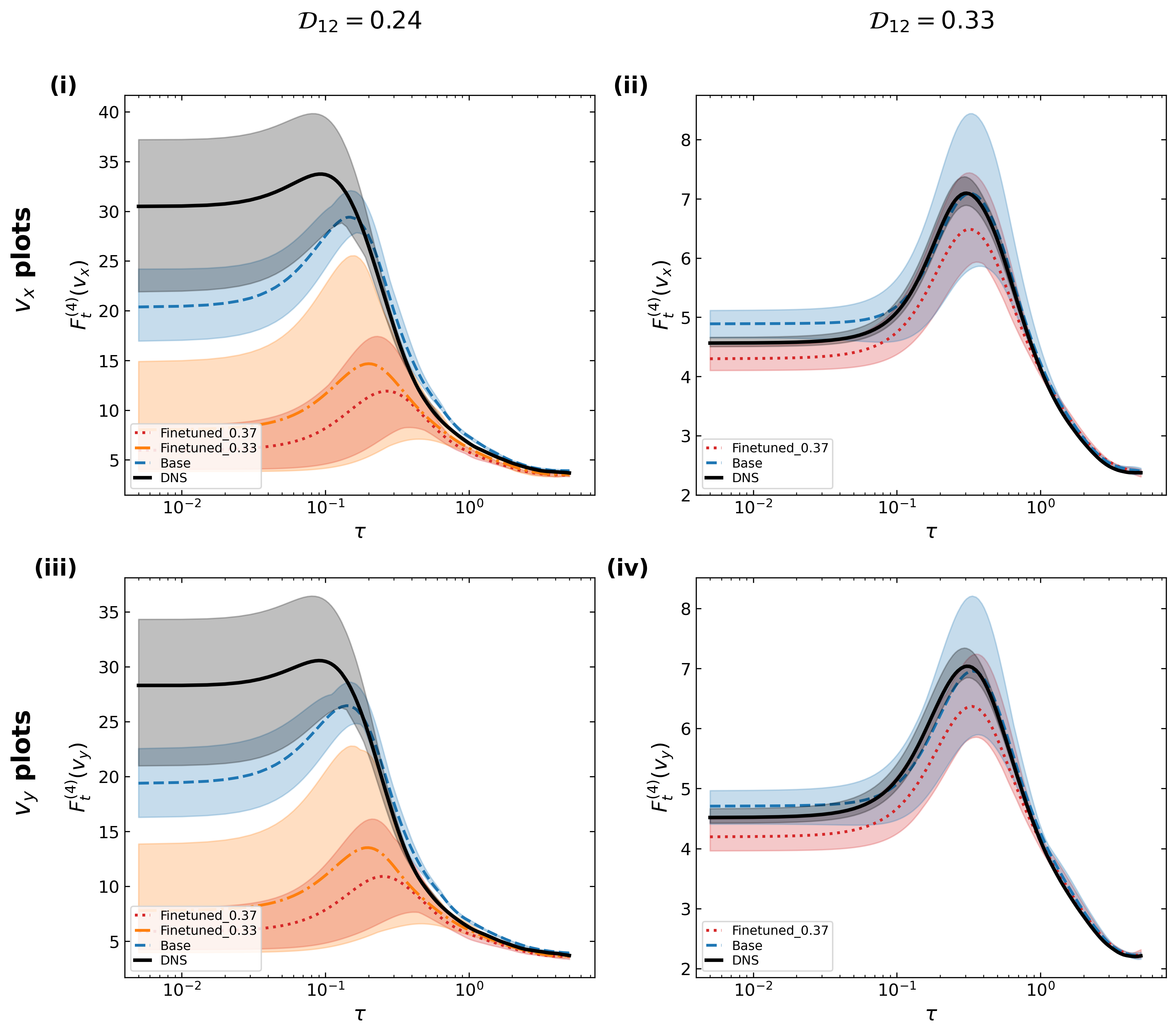}
        \caption{}
        \label{fig:finetune_flatness}
    \end{subfigure}\hfill
    \begin{subfigure}{0.48\textwidth}
        \centering
        \includegraphics[width=\linewidth, height=9cm]{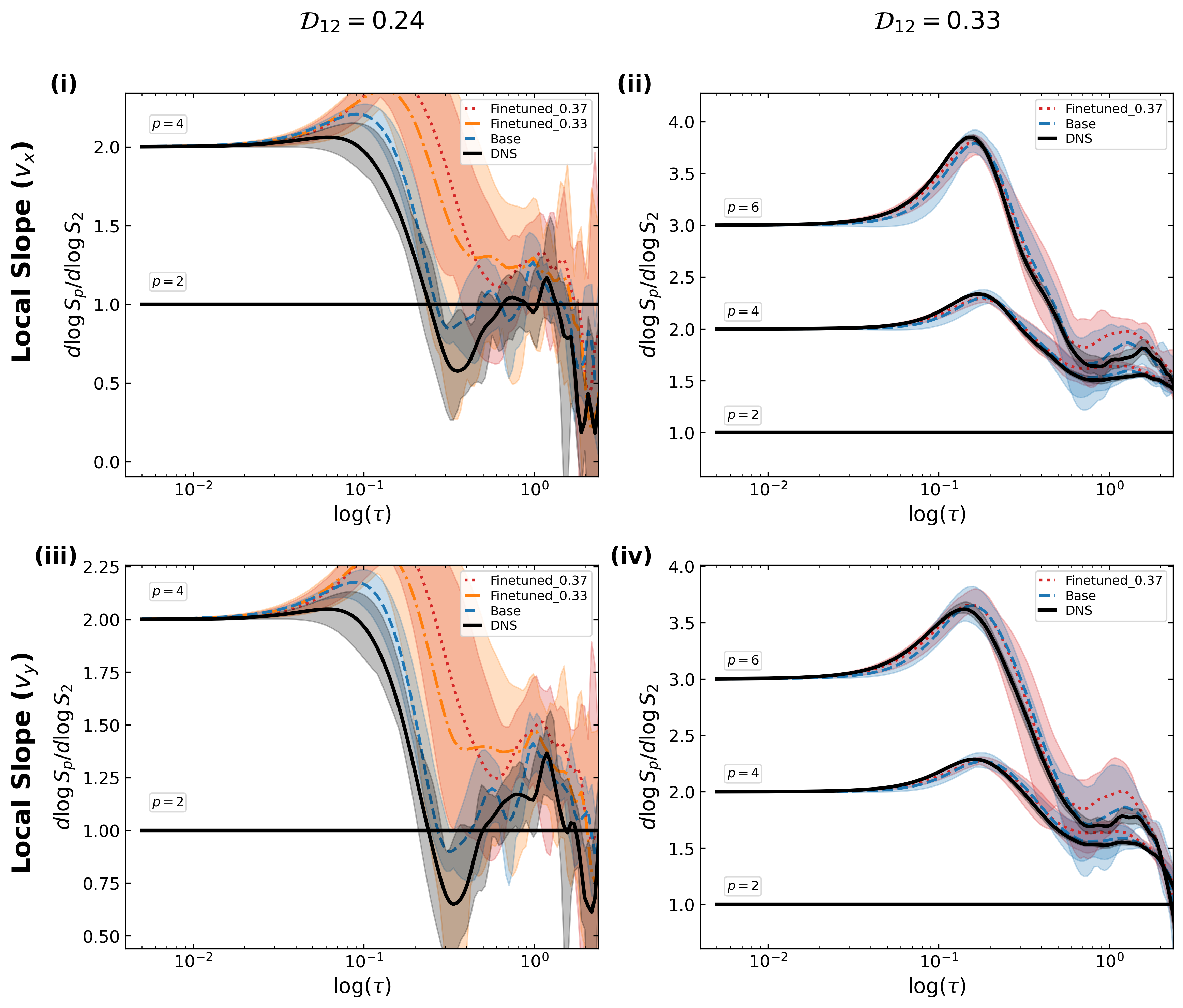}
        \caption{}
        \label{fig:finetune_slope}
    \end{subfigure}
    
    \caption{Comparison of fine-tuned models with the ``Base'' model and DNS: (a) curvature statistics: velocity magnitude CDF, perpendicular acceleration magnitude CDF, and trajectory curvature CCDF, (b) structure functions, (c) flatness, and (d) Local slope, for both velocity components, for $\mathcal{D}_{12}=0.24$ and $\mathcal{D}_{12}=0.33$. ``Finetuned 0.33'' denotes that the U-Net is pre-trained on $\mathcal{D}_{12}=0.33$. Similarly ``Finetuned 0.37'' means that the U-Net is pre-trained on $\mathcal{D}_{12}=0.37$. ``Base'' denotes that the U-Net was trained on that $\mathcal{D}_{12}$ value from scratch. The shaded regions represent the error, which are the minimum and maximum values obtained for a particular statistic after dividing the dataset into $10$ smaller sub-batches.}
    \label{fig:curv_strFn_flatness_localSlope_finetune}
\end{figure*}

\begin{figure*}[htbp]
    \centering
    \begin{tikzpicture}
        \node[anchor=south west,inner sep=0] (image) at (0,0) {\includegraphics[width=1.0\linewidth]{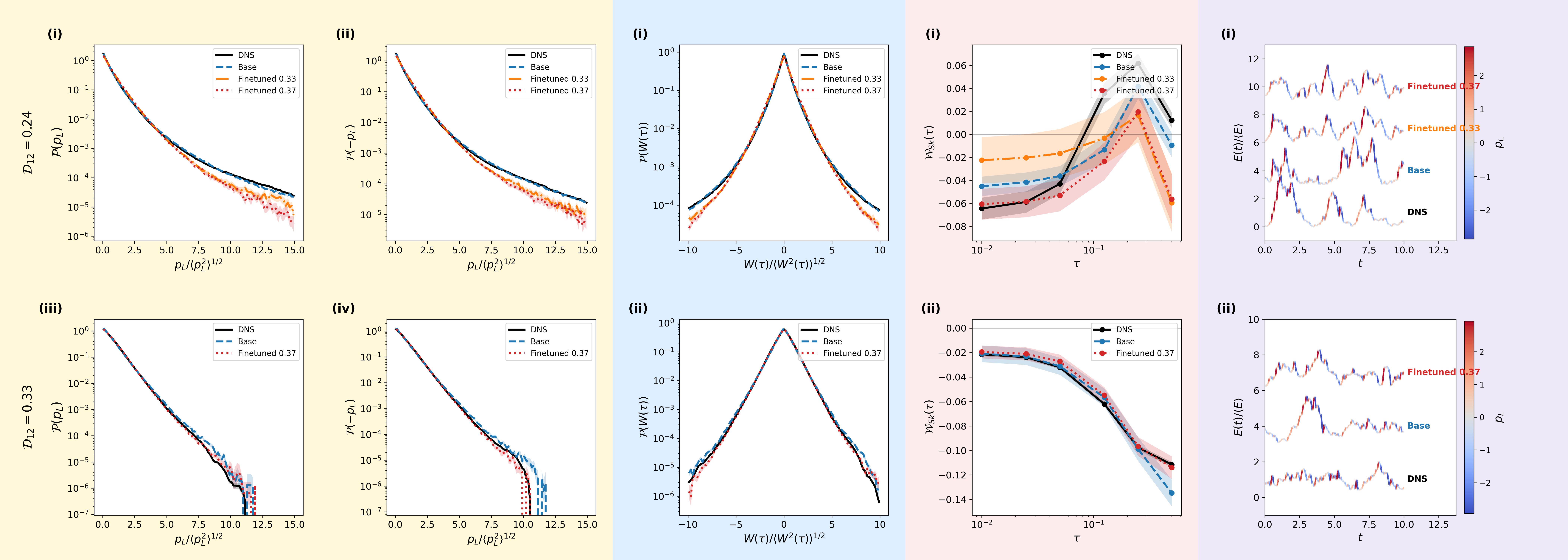}};
        
        \begin{scope}[x={(image.south east)},y={(image.north west)}, every node/.style={font=\tiny}]
            
            \node[anchor=north west] at (0.001, 0.99) {\textbf{(a)}};
            \node[anchor=north west] at (0.386, 0.99) {\textbf{(b)}};
            \node[anchor=north west] at (0.57, 0.99) {\textbf{(c)}};
            \node[anchor=north west] at (0.76, 0.99) {\textbf{(d)}};

        \end{scope}
    \end{tikzpicture}

    \caption{Comparing the performance of the fine-tuned models for $\mathcal{D}_{12}=0.24$ and $\mathcal{D}_{12}=0.33$ with DNS (the ground truth), and the respective models trained from scratch. The figure is divided into four groups: \textbf{(a)} the PDFs $\mathcal{P}(p_L)$ and $\mathcal{P}(-p_L)$ of power $p_L$, which represents the rate of change of energy; \textbf{(b)} show the energy increment PDFs $\mathcal{P}(W(\tau))$ for an energy increment $W(\tau)$ at a time delay of $\tau = 0.05$; \textbf{(c)} skewness of energy increments $\mathcal{W}_{Sk}(\tau)$ across various time delays $\tau$, used to measure the asymmetry in how particles gain or lose energy; and \textbf{(d)} sample trajectories of the normalized energy $E(t)/\langle E \rangle$ over time $t$, each separated by 3 units on the vertical axis. ``Finetuned 0.33'' denotes that the U-Net is pre-trained on $\mathcal{D}_{12}=0.33$. Similarly ``Finetuned 0.37'' means that the U-Net is pre-trained on $\mathcal{D}_{12}=0.37$. ``Base'' denotes that the U-Net was trained on that $\mathcal{D}_{12}$ value from scratch. }
    \label{fig:irreversibility_finetune}
\end{figure*}

\subsection{U-Net Architecture: Mathematical Formulation}
\label{subsec:appendix:unet_math}

We provide mathematical formulations of the terms used to describe our U-Net in Fig.~\ref{fig:unet}. Let $\mathcal{H}\in\mathbb{R}^{C\times L}$ denote an arbitrary intermediate feature sequence with $C$ channels and spatial length $L$. The network relies on a discrete diffusion step $n\in\mathbb{Z}^+$. We implement our model using the PyTorch  library~\cite{pytorch}.

\subsubsection{Fundamental Operations}

\paragraph{Diffusion Step Embedding ($n_{emb}$):}
The discrete step $n$ is mapped to a continuous embedding using fixed sinusoidal positional encodings of dimension $\mathscr{D}$. For $k\in\{0,\dots,\mathscr{D}/2-1\}$, the vector components of the positional embedding $\text{PE}(n)$ is given below:
\begin{eqnarray}
    \text{PE}(n)_{k} &=& \cos(n\,\omega_k)\,; \label{eq:PE_cos}\\
    \text{PE}(n)_{k+\mathscr{D}/2} &=& \sin(n\,\omega_k)\,; \label{eq:PE_sin}\\
    \omega_k &=& 10000^{-2k/\mathscr{D}}\,.
    \label{eq:PE_w}
\end{eqnarray}
This representation is passed through a two-layer Multi-Layer Perceptron (MLP) to generate $n_{emb}$:
\begin{eqnarray}
    n_{emb} &=& W_{n2}\varphi(W_{n1}\text{PE}(n)+b_{n1})+b_{n2}\,,
\end{eqnarray}
where $W$ and $b$ are the learnable weight matrices and bias vectors, respectively, and $\varphi$ is the SiLU activation function $\varphi(z)=z\cdot(1+e^{-z})^{-1}$.\\

\paragraph{Group Normalization (GN):}
Group Normalization (GN) partitions the $C$ channels into $G=32$ groups of size $m=C/G$. For a feature $\mathcal{H}_{c,l}$ in channel $c$ and position $l$, the corresponding group index is $g(c)=\lceil c/m \rceil$. The group mean $\mu_g$ and variance $\sigma_g^2$ are:
\begin{eqnarray}
    \mu_g &=& \frac{1}{m L}\sum_{c'=m(g-1)+1}^{mg}\sum_{l'=1}^{L}\mathcal{H}_{c',l'}\,; \\
    \sigma_g^2 &=& \frac{1}{m L}\sum_{c'=m(g-1)+1}^{mg}\sum_{l'=1}^{L}(\mathcal{H}_{c',l'}-\mu_g)^2\,.
\end{eqnarray}
The normalized output is scaled and shifted by learnable parameters $\gamma_c$ and $\beta_c$ for each channel $c$:
\begin{eqnarray}
    \text{GN}(\mathcal{H})_{c,l} &=& \gamma_c\left(\frac{\mathcal{H}_{c,l}-\mu_g}{\sqrt{\sigma_g^2+{\epsilon_{GN}}}}\right)+\beta_c\,.
\end{eqnarray}
We take the PyTorch default value of $\epsilon_{GN}=10^{-5}$.
\paragraph{1D Convolution (Kernel $\mathcal{K}$, Padding $P$, Stride $S$):}
For $\mathcal{H}\in\mathbb{R}^{C\times L}$, we define the zero-padded sequence $\tilde{\mathcal{H}}\in\mathbb{R}^{C\times(L+2P)}$ by $\tilde{\mathcal{H}}_{c,l}=\mathcal{H}_{c,l-P}$, for $P<l\le L+P$, and $\tilde{\mathcal{H}}_{c,l}=0$ otherwise. For the learnable weights $W\in\mathbb{R}^{C_{out}\times C_{in}\times \mathcal{K}}$ and bias $b\in\mathbb{R}^{C_{out}}$, the discrete cross-correlation output is \footnote{This cross-correlation is termed as convolution in the PyTorch library used in our code. We follow the same convention in our paper. Ref.\cite{Goodfellow2016} discusses it in detail in Sec. 9.1, page 329.}:
\clearpage
\begin{eqnarray}
    (W*_S^\mathcal{K}\mathcal{H})_{c,l} &=& b_c+\sum_{c'=1}^{C_{in}}\sum_{j=1}^{\mathcal{K}}W_{c,c',j}\,\tilde{\mathcal{H}}_{c',\,S(l-1)+j},
\end{eqnarray}
for $l\in\left\{1,\dots,\lfloor\frac{L+2P-K}{S}\rfloor+1\right\}$, where $*_S^\mathcal{K}$ denotes a convolution of kernel size $\mathcal{K}$ and stride S. In our U-Net, we use $(K,P)=(3,1)$, which preserves the spatial length $L$ when the stride $S=1$ and scales the length with 2 [i.e., $L\mapsto L/2$] when $S=2$; and $(K,P)=(1,0)$ at $S=1$.\\

\paragraph{Nearest-Neighbor Upsampling:}
Spatial upsampling is performed using the nearest-neighbor interpolation. For a sequence $h$ of length $L$, the interpolated sequence of length $2L$ is defined as:
\begin{eqnarray}
    \text{Interpolate}(h)_l &=& h_{\lfloor(l-1)/2\rfloor+1}\,, \nonumber \\
    && (\text{for } l\in\{1,\dots,2L\})\,.
\end{eqnarray}

\subsubsection{U-Net Blocks}

\paragraph{Spatial Resampling:}
Downsampling ($D_{sample}$) is achieved with a 1D convolution of stride $S=2$. Upsampling ($U_{sample}$) uses nearest-neighbor interpolation followed by a 1D convolution ($S=1$):
\begin{eqnarray}
    D_{sample}(\mathcal{H}) &=& W_{down}*_2^3\mathcal{H}\,; \\
    U_{sample}(\mathcal{H}) &=& W_{up}*_1^3\text{Interpolate}(\mathcal{H})\,.
\end{eqnarray}

\paragraph{Residual Block ($\text{ResBlock}$):}
The input to the residual block consists of $n_{emb}$ and $\mathcal{H}$. The input $\mathcal{H}$ undergoes group normalization, activation ($\varphi$), and convolution. The step embedding $n_{emb}$ is projected to the required channel dimension and broadcast spatially (added as a spatially uniform $C$-vector to the $C\times L$ tensor), followed by a second sequence of normalization, activation, dropout, and convolution:
\begin{eqnarray}
    h_1 &=& W_{conv1}*_1^3\varphi(\text{GN}(\mathcal{H}))\,; \\
    (h_2)_{c,l} &=& (h_1)_{c,l}+\left(W_{emb}\varphi(n_{emb})+b_{emb}\right)_c\,; \\
    h_3 &=& W_{conv2}*_1^3\varphi(\text{GN}(h_2))\,.
\end{eqnarray}
The final output uses a skip connection: $\text{ResBlock}(\mathcal{H},n_{emb})=\text{Skip}(\mathcal{H})+h_3$. The function $\text{Skip}(\mathcal{H})$ acts as the identity mapping $\mathcal{H}$ if the channel dimensions remain unchanged, or a $1\times 1$ convolution ($W_{skip}*_1^1\mathcal{H}$) if a channel projection is required.\\


\paragraph{Attention Block ($\text{Attn}$):}
This block first performs group normalization to the input of $C$ channels, followed by a $1\times 1$ convolution that projects the features into an intermediate tensor $Z$ with $3C$ output channels:
\begin{eqnarray}
    Z &=& W_{qkv}*_1^1\text{GN}(\mathcal{H})\,.
\end{eqnarray}
To reflect the specific channel-ordering implemented in the codebase, the $3C$ channels of $Z$ are first partitioned into $N_h$ heads of width $d_k=C/N_h$. We use $N_h=4$ for our U-Net. For a given head $a \in \{1, \dots, N_h\}$, its assigned block of $3d_k$ channels is subsequently split into $d_k$-dimensional Queries ($Q^{(a)}$), Keys ($K^{(a)}$), and Values ($V^{(a)}$). Using bracketed slice notation for the channel indices, the features for head $a$ are extracted as follows:
\begin{eqnarray}
    Q^{(a)} &=& Z_{[(a-1)3d_k + 1 \,\,:\,\, (a-1)3d_k + d_k],\, :}\,; \\
    K^{(a)} &=& Z_{[(a-1)3d_k + d_k + 1 \,\,:\,\, (a-1)3d_k + 2d_k],\, :}\,; \\
    V^{(a)} &=& Z_{[(a-1)3d_k + 2d_k + 1 \,\,:\,\, a(3d_k)],\, :}\,.
\end{eqnarray}
For a given head $a$, the attention weights $\alpha^{(a)}$ and the corresponding output $A^{(a)}$ are computed by applying the softmax function over the spatial dimension $l'$:
\begin{eqnarray}
    \alpha^{(a)}_{l,l'} &=& \frac{\exp\left(\langle Q^{(a)}_{:,l},\,K^{(a)}_{:,l'}\rangle/\sqrt{d_k}\right)}{\sum_{l''=1}^{L}\exp\left(\langle Q^{(a)}_{:,l},\,K^{(a)}_{:,l''}\rangle/\sqrt{d_k}\right)}\,; \\
    A^{(a)}_{:,l} &=& \sum_{l'=1}^{L}\alpha^{(a)}_{l,l'} V^{(a)}_{:,l'}\,.
\end{eqnarray}
where $\langle \cdot, \cdot \rangle$ represents inner product, and the subscript ($:$) indicates the extraction of the full $d_k$-dimensional channel vector at spatial position $l$. The channel-wise concatenation of all $A^{(a)}$ forms the full multi-head output $A$. Finally, the output of \text{Attn}($\mathcal{H}$) is given as:
\begin{eqnarray}
    \text{Attn}(\mathcal{H}) &=& \mathcal{H}+(W_{proj\_out}*_1^1 A)\,.
\end{eqnarray}

\subsubsection{Network Configuration and Constants}
The architecture relies on several fixed hyperparameters:
\begin{itemize}
    \item \textbf{Base Channels:} The first convolution step maps the input trajectory $\mathcal{V}_n$ to $C_{base}$ channels. It is also the dimension $\mathscr{D}=C_{base}$ for the sinusoidal positional embedding calculation in Eqs.~\eqref{eq:PE_cos}--~\eqref{eq:PE_w}. We take $C_{base}=128$ in our U-Net implementation.
    \item \textbf{Embedding Dimension:} The continuous step embedding $n_{emb}\in\mathbb{R}^{4C_{base}}$.
    \item \textbf{Zero-Initialization:} The trainable parameters are zero-initialized. This ensures that every residual branch and the initial network output evaluate to zero at the start of training.
\end{itemize}

\subsubsection{Network Forward Pass Algorithm}
The U-Net takes the noisy trajectory $\mathcal{V}_n$ and the diffusion step $n$ to provide an estimate of the noise $\epsilon_{\theta}$. The outputs of the initial convolution, every residual block (including the attention layer, wherever present), and every downsampling layer is appended to a stack $S_{kip}$. We recursively pop from $S_{kip}$ in the decoder to implement the skip connection. Given $N_{stages}$ resolution levels and $M_{res}$ residual blocks per stage, the stack holds exactly $1+N_{stages}M_{res}+(N_{stages}-1)$ tensors, which is the same as the number of pop operations performed in the decoder. In our U-Net, $N_{stages}=5$ and $M_{res}=3$. The feature channel dimension at stage $i \in \{1, \dots, 5\}$ is $C_i = \text{mult}_i \cdot C_{base}$, using the multiplier tuple $\text{mult} = (1, 1, 2, 3, 4)$. The pseudo-code for the forward pass is presented in Algorithm~\ref{alg:unet_forward}.

\subsection{Velocity and Acceleration JPDFs: DNS vs Diffusion Models}
As discussed in Sec.~\ref{subsec:vel_accel_pdfs}, we present the comparison of DDIM samplers with denoising step sizes 100, 50, and 25, 10 and 5 with DNS in Fig.~\ref{fig:ddim_higher_steps} and Fig.~\ref{fig:ddim_lower_steps}, respectively.

A counterintuitive observation arises in the velocity JPDF for $\mathcal{D}_{12}=0.24$, where the residual is low at merely 5 denoising steps. This occurs because the ground-truth velocity JPDF is isotropic and Gaussian-like and 5 denoising steps are insufficient for the DDIM sampler to remove the initial Gaussian noise. The acceleration JPDF for $\mathcal{D}_{12}=0.24$ reveals this failure, where a prominent residual confirms the inadequacy of 5-step DDIM sampling.

\subsection{Derivation for asymptotic scaling in the curvature PDF}
\label{subsec:appendix:kappa_derivation}
In the 2D NRCHNS system, Fig.~\ref{fig:vel_accel_stats} shows that $a_\perp$ does not strictly follow a Gaussian distribution, nor does $v$ follow a Rayleigh distribution. 
For the NRCHNS system, the observed values of $\rho = 0$ and $\Upsilon = 1$ in Fig.~\ref{fig:residual_plot} are obtained by a linear fit to the log-log plots of cumulative distribution functions (CDFs), $\mathcal{Q}_{a_\perp}(a_\perp)$ and $\mathcal{Q}_v(v)$. This allows for $\mathcal{P}(\kappa) \sim \kappa^{-2}$ as $\kappa\rightarrow\infty$. Substituting $\mathcal{P}(a_\perp) = h(a_\perp)$ and $\mathcal{P}(v) = v g(v)$ along with Eq.~\eqref{eq:independence} into Eq.~\eqref{eq:p_kappa_integral} yields:
\begin{eqnarray}
    \mathcal{P}_\kappa(\kappa) = \int_{0}^{\infty} dv \int_{-\infty}^{\infty}& da_\perp h(a_\perp) v g(v) \nonumber \\
    & \times \delta\left(\kappa - \frac{a_\perp}{v^2}\right) \label{eq:p_kappa_sub}
\end{eqnarray}
where $\delta$ is the Dirac delta function. 

We can evaluate the inner integral over $a_\perp$ by utilizing the scaling and the sifting properties of the Dirac delta function. By evaluating this at $a_\perp = \kappa v^2$, we extract a factor of $v^2$:
\begin{eqnarray}
    \mathcal{P}_\kappa(\kappa) &=& \int_{0}^{\infty} v^3 h(\kappa v^2) g(v) dv \label{eq:p_kappa_v}\,.
\end{eqnarray}
To extract the asymptotic behavior as $\kappa \to \infty$, we perform a change of variables, setting $y = \kappa v^2$. This substitution gives $v = \sqrt{y/\kappa}$ and $dv = \frac{1}{2\sqrt{\kappa y}} dy$.
Substituting these into the integral we get
\begin{eqnarray}
    \mathcal{P}_\kappa(\kappa) &=& \int_{0}^{\infty} \left(\frac{y}{\kappa}\right)^{3/2} h(y) g\left(\sqrt{\frac{y}{\kappa}}\right) \frac{dy}{2\sqrt{\kappa y}} \nonumber \\
    &=& \frac{1}{2\kappa^2} \int_{0}^{\infty} y h(y) g\left(\sqrt{\frac{y}{\kappa}}\right) dy \,.\label{eq:p_kappa_y}
\end{eqnarray}

As $\kappa \to \infty$, the argument in the function $g$ approaches zero ($\sqrt{y/\kappa} \to 0$). As observed in Fig.~\ref{fig:vel_accel_stats}, $g(v)$ is non-zero at the origin. This simplifies the integral to:
\begin{eqnarray}
    \mathcal{P}_\kappa(\kappa) &\approx& \frac{g(0)}{2\kappa^2} \int_{0}^{\infty} y h(y) dy \,.\label{eq:p_kappa_approx}
\end{eqnarray}
In Eq.~\eqref{eq:p_kappa_approx}, the integral $\int_{0}^{\infty} y h(y) dy$ is a finite constant, and $\mathcal{P}_\kappa(\kappa) \sim \kappa^{-2}$ as $\kappa \to \infty$ [Eq.~\eqref{eq:p_kappa_scale}].
Consequently, the Complementary Cumulative Distribution Function (CCDF), represented as $1-\mathcal{Q}_\kappa(\kappa)$, for the curvature obtained by integrating the PDF from $\kappa$ to infinity exhibits the following power-law scaling:
\begin{equation}
    1-\mathcal{Q}_\kappa(\kappa) 
    \sim \int_{\kappa}^{\infty} (\kappa')^{-2} d\kappa' \sim \kappa^{-1} \,. \label{eq:ccdf_scale_result}
\end{equation}
Thus, the $\kappa^{-1}$ scaling of the curvature's CCDF is preserved in the 2D NRCHNS system because of Eqs.~\eqref{eq:independence},~\eqref{eq:p_v_gen} and~\eqref{eq:p_an_gen}, where $\rho=0$ and $\Upsilon=1$ for the 2D NRCHNS system.

\subsection{Velocity Magnitude, Perpendicular Acceleration Magnitude and Curvature PDFs: Diffusion Models vs DNS}
\label{subsec:appendix:curvature_comparison}
As mentioned in Sec.~\ref{subsec:curvature_analysis}, we present the comparison of velocity and acceleration JPDFs of diffusion models with the 
following samplers: DDPM-800, DDIM-100, DDIM-50, DDIM-25, DDIM-10 and DDIM-5 with our DNS results in Fig.~\ref{fig:curvature_comparison_dns_vs_dm}.

\FloatBarrier 

\subsection{Fine-tuning Comparison}
\label{subsec:appendix:finetuning_comparison}
\label{subsec:appendix:vel_accel_jpdf}
This section contains all the comparative plots evaluating the model transferability mentioned in Sec.~\ref{subsec:model_transferability}: the velocity-increment PDFs [Fig.~\ref{fig:vel_increments_finetune}]; curvature analysis [Fig.~\ref{fig:curv_strFn_flatness_localSlope_finetune}(a)]; structure functions [Fig.~\ref{fig:curv_strFn_flatness_localSlope_finetune}(b)]; flatness [Fig.~\ref{fig:curv_strFn_flatness_localSlope_finetune}(c)]; local slope [Fig.~\ref{fig:curv_strFn_flatness_localSlope_finetune}(d)]; and irreversibility analysis [Fig.~\ref{fig:irreversibility_finetune}]. Specifically, we evaluate three transfer learning cases: fine-tuning a model pre-trained on $\mathcal{D}_{12}=0.37$ for target datasets with $\mathcal{D}_{12}=0.33$ and $\mathcal{D}_{12}=0.24$; and fine-tuning a model pre-trained on $\mathcal{D}_{12}=0.33$ for a target dataset with $\mathcal{D}_{12}=0.24$. In each case, the fine-tuned model is compared against a model trained from scratch on the target distribution.

\FloatBarrier
\bibliography{main}
\end{document}